\documentclass[aps,prb]{revtex4-2}
\usepackage{amsmath,amssymb}
\usepackage{graphicx}
\usepackage{dcolumn}
\usepackage{bm}
\usepackage{color}
\usepackage{amsmath}
\DeclareMathAlphabet{\mathbbmsl}{U}{bbm}{m}{sl}
\makeatletter
\newsavebox{\@brx}
\newcommand{\llangle}[1][]{\savebox{\@brx}{\(\m@th{#1\langle}\)}%
\mathopen{\copy\@brx\kern-0.5\wd\@brx\usebox{\@brx}}}
\newcommand{\rrangle}[1][]{\savebox{\@brx}{\(\m@th{#1\rangle}\)}%
\mathclose{\copy\@brx\kern-0.5\wd\@brx\usebox{\@brx}}}
\makeatother

\begin{document}
\draft

\title{Self-consistent quantum-kinetic theory for interacting drifting electrons and force-driven phonons in a 1D system}

\author{Xuejun Lu$^{1}$ and Danhong Huang$^{2}\footnote{E-mail contact:\ danhong.huang@us.af.mil}$}
\address{$^{1}$Department of Electrical and Computer Engineering, Francis College of Engineering, University of Massachusetts Lowell, Lowell, Massachusetts 01854, USA}
\address{$^{2}$US Air Force Research Laboratory, Space Vehicles Directorate (AFRL$/$RVSU),\\
Kirtland Air Force Base, New Mexico 87117, USA}

\date{\today}

\begin{abstract}
A self-consistent quantum-kinetic model is developed for studying strong-field nonlinear electron transport interacting with force-driven phonons within a quantum-wire system.
For this model, phonons can be dragged into motion through strong electron-phonon scattering by fast-moving 
electrons along the opposite direction of the DC electric field. 
Meanwhile, the DC-field induced charge current of electrons can be either enhanced or reduced by the same electron-phonon scattering, depending on the relative direction of a DC field with respect to that of an applied temperature gradient for driving phonons. 
By making use of this quantum-kinetic model beyond the relaxation-time approximation, neither electron nor phonon temperature is required for describing ultrafast electron-phonon scattering  and their correlated transports in this 1D electronic-lattice system. 
\end{abstract}
\pacs{PACS:}
\maketitle

\section{Introduction}
\label{sec-1}

Historically, for a quasi-thermal-equilibrium case, Boltzmann transport equation\,\cite{te-6,te-9} has been extensively utilized to study both electron and phonon transports, respectively, where position-dependent temperatures are introduced through their thermal-equilibrium occupation functions for electrons and phonons. For non-thermal-equilibrium occupation function of either electrons or phonons, on the other hand, no such thermal-temperature concept can be defined and used for exploring these thermodynamic processes. In these cases, however, transports of electrons and phonons for electrical and thermal conduction remain and their physical descriptions require a novel quantum-kinetic theory\,\cite{te-7,te-10} beyond a semi-classical Boltzmann transport equation without involving a thermal-equilibrium based temperature concept. 
\medskip

Electron drifting motion can be initiated and controlled by applying a DC electric field along a quantum wire, 
while thermal conduction of phonons occurs under a temperature drop between two external thermal-equilibrium  reservoirs.\,\cite{ziman} Due to strong electron-phonon coupling, drifting electrons can drag phonon into motion along the crystal chain by transferring their directional momentum to phonons, leading to simultaneously heating/cooling at two ends of the chain.\,\cite{new-7} Similarly, force-driven phonons along a chain also enable transferring their directional momentum to electrons through the same electron-phonon interaction,\,\cite{bookdhh} leading to a net electric current under a zero bias.
\medskip

If only a weak temperature difference is applied to a lattice system, phonon transport is usually described by classical heat-diffusion equation in combination with Fourier's law, which depends on the gradient of a temperature distribution within a sample.\,\cite{heat}
The thermoelectric effect\,\cite{te-4,te-5,te-8} usually refers to a direct conversion of temperature differences $\Delta T=T_R-T_L$ between right $(R)$ and left $(L)$ reservoirs to electric voltage and vice versa. On the other hand, as a voltage is applied to the sample, heat is transferred from one side to the other, leading to a temperature difference. 
\medskip

To account for rich thermal-dynamic features resulting from interacting non-equilibrium electrons and phonons as well as their mutual dragging effects, we include strong surface-roughness scattering of phonons within a chain, in addition to three-phonon anharmonic interaction,\,\cite{r1,r21,r25,r26,r27} in current quantum-kinetic theory for describing non-thermal phonon properties. Meanwhile, by taking into account coulomb and phonon scatterings of electrons, we employ a quantum-kinetic theory for dealing with nonlinear transport of electrons which are coupled strongly to non-thermal phonons in the same system. Such a quantum-kinetic description for ultrafast dynamics in a mutually-dragged non-thermal electron-phonon system enables to reveal new physics mechanism behind both electron-diffusion power and phonon-drag thermoelectric power under extreme conditions, such as a very-high bias, a very-large temperature gradient, and even an incidence of a blast wave. In particular, one can choose different external-force turning-on times for driving either electrons or phonons independently. 
\medskip

In early thermoelectric studies,\,\cite{r3,new-2,new-7} coupled Boltzmann transport equations are often employed for describing dynamics of electrons and phonons. However, these two Boltzmann transport equations are commonly solved by applying the well-known relaxation-time approximation (RTA)\,\cite{iop,add-3,add-4,add-31,add-32,add-33} under the assumption that non-equilibrium occupation functions of electrons and phonons are only slightly deviated from their initial thermal-equilibrium ones. Later, by going beyond the RTA, non-equilibrium parts of electron occupation and phonon distribution for their scatterings are linearized by eliminating nonlinear scattering contributions,\,\cite{ziman,kohn,lyo,new-1,new-3,new-4} which holds only for small deviations from corresponding thermal-equilibrium states. Very recently, a quantum-kinetic theory\,\cite{te-10,add-1,add-2,add-30} has been proposed for treating electron scatterings, and furthermore, it has been solved exactly for a semiconductor-wire and a quantum-dot systems, where one does not define a thermal temperature within the considered system. \medskip

In Sec.\,\ref{sec-2}, we start by introducing three-phonon anharmonic interactions and quantum-kinetic models for both electrons and phonons in a 1D electronic-lattice system, which is followed by surface-roughness scattering of longitudinal phonons. Based on numerically computed non-equilibrium occupation functions for both electrons and phonons, we evaluate time evolution of both electron drift velocity and phonon heat-current densities. Meantime, induced electron-diffusion and phonon-drag thermoelectric powers are also studied along with thermal conductivity of longitudinal phonons. Section\,\ref{sec-3} are devoted to presenting calculated results for transient electron drift velocity, time evolution of non-thermal occupation functions for electrons and longitudinal phonons, transient electric and heat currents for drifting electrons and force-driven phonons, as well as Doppler effect on 
electron-phonon interaction, {\em etc\/}. Finally, a summary and remarks are given in Sec.\,\ref{sec-4}.

\section{Model and Theory}
\label{sec-2}

In Sec.\,\ref{sec-2}, we begin with the discussion in Subsection\,\ref{sec-2-3} on surface-roughness scattering of phonons within a chain. After this, we introduce in Subsection\,\ref{sec-2-4} a three-phonon anharmonic interaction by going beyond commonly used harmonic approximation. Meanwhile, we also take into account the electron-phonon couplings in Subsection\,\ref{sec-2-4} for both longitudinal-optical and longitudinal-acoustic phonon modes in this lattice-electronic system. Moreover, we propose in Subsection\,\ref{add-1} a quantum-kinetic transport equation for electrons in a corresponding quantum-wire system. 
\medskip

Based on acquired exact solution for non-equilibrium occupation function of electrons, we further discuss in Subsection\,\ref{add-2} an effective thermal temperature of non-equilibrium drifting electrons. This is followed in Subsection\,\ref{sec-2-5} by another analysis of an effective thermal temperature for non-equilibrium longitudinal phonons. Finally, in Subsection\,\ref{sec-2-6}, we finish our discussions by presenting an analysis for effects of mutual drags on electron-diffusion and phonon-drag thermoelectric powers. 

\subsection{Surface-Roughness Scattering of Longitudinal Phonons}
\label{sec-2-3}

For surface-roughness scattering of phonons, let us consider a cylindrical-shape crystal chain along the elastic-wave propagation direction. In this case, the total $\gamma$-phonon creation ${\cal C}^{(s)}_{Q.\gamma}$ and annihilation ${\cal D}^{(s)}_{Q.\gamma}$ rates related to surface roughness are given by 

\begin{eqnarray}
\label{rough-0}
{\cal C}^{(s)}_{Q,\gamma}(t)&=&\sum\limits_{Q'}\,{\cal P}_\gamma^s(Q,Q')\,N_{Q'}^{\gamma}(t)=\frac{{\cal L}}{2\pi}\int\limits_{-\pi/d}^{\pi/d} dQ'\,{\cal P}_\gamma^s(Q,Q')\,N_{Q'}^{\gamma}(t)\ ,\\
\label{rough-1}
{\cal D}^{(s)}_{Q,\gamma}(t)&=&\sum\limits_{Q'}\,{\cal P}_\gamma^s(Q',Q)\left[1+N_{Q'}^{\gamma}(t)\right]=\frac{{\cal L}}{2\pi}\int\limits_{-\pi/d}^{\pi/d} dQ'\,{\cal P}_\gamma^s(Q',Q)\left[1+N_{Q'}^{\gamma}(t)\right]\ ,
\end{eqnarray}
where ${\cal L}$ represents the chain length, $d=2a$ is the lattice period, and a summation over all possible wave numbers $Q'$ of $\gamma$ phonons is introduced. By utilizing quantum-mechanics first-order time-dependent perturbation theory, the elastic scattering probability per unit time, {\em i.e.\/} ${\cal P}_\gamma^s(Q',Q)={\cal P}_\gamma^s(Q,Q')$ employed in Eqs.\,(\ref{rough-0}) and (\ref{rough-1}) for surface scattering between $Q$ and $Q'$ states of $\gamma$ phonons,\,\cite{surfscat,surfscat-2} takes the form 

\begin{equation}
{\cal P}_\gamma^s(Q',Q)=\frac{2\pi}{\hbar}\,\Big|\langle Q'\vert\hat{\cal H}'_\gamma\vert Q\rangle\Big|^2\,{\cal L}_0(\hbar\omega_{Q',\gamma}-\hbar\omega_{Q,\gamma},\,\hbar\Gamma^\gamma_{\rm ph})\ ,
\label{rough-2}
\end{equation}
where 

\begin{equation}
{\cal L}_0(x,\,\eta)=\frac{1}{\pi}\,\left(\frac{\eta}{x^2+\eta^2}\right)
\end{equation}
is the Lorentz line-shape function and $\hbar\Gamma^\gamma_{\rm ph}$ is a homogeneous lifetime broadening for $\gamma$-phonons. 
The integrals with repect to $Q'$ in Eqs.\,\eqref{rough-0}--\eqref{rough-1} are numerically evaluated by taking $dQ'\sim\Gamma^\gamma_{\rm ph}/v_{\rm LA}$ with phase velocity $v_{\rm LA}$ for longitudinal-acoustic phonons in the system. 
In Eq.\,\eqref{rough-2}, the perturbation matrix element for such a $\gamma$-phonon surface-scattering event is found to be\,\cite{prl}

\begin{equation}
\Big|\langle Q'\vert\hat{\cal H}'_\gamma\vert Q\rangle\Big|^2=\frac{|\xi_{Q',\gamma}\,\xi_{Q,\gamma}|}{2\pi{\cal R}{\cal L}}\,\left(\hbar\omega_{Q',\gamma}\,\hbar\omega_{Q,\gamma}\right)\Delta_{\rm sr}(|Q'-Q|)\ ,
\label{rough-3}
\end{equation}
where $\hbar\omega_{Q',\gamma}$ and $\hbar\omega_{Q,\gamma}$ are phonon energies,\,\cite{callaway} ${\cal R}$ is the radius of a cylindrical-shape crystal, $2\pi {\cal R}{\cal L}$ is the lateral surface area of the cylinder,
and $\xi_{Q,\gamma}$ is determined by $\xi_{Q,\gamma}=(1/\bar{\Delta}_0)\,\Delta\omega_{Q,\gamma}/\omega_{Q,\gamma}$ with $\bar{\Delta}_0$ as the root-mean-square of surface fluctuations. 
\medskip

For longitudinal-acoustic phonons and $Qa\ll 1$, we get from Eq.\,\eqref{rough-7} that $\xi_{Q,LA}=1/{\cal R}$.
For longitudinal-optical phonons, on the other hand, we find from Eq.\,\eqref{rough-7} that $\xi_{Q,LO}=-b_Q/{\cal R}$,
where $b_Q=(Qa)^2[m_1m_2/(m_1+m_2)^2]\sim\omega_{Q,LA}^2$ which is dimensionless. As a result, we have $(\xi_{Q,LO}\,\omega_{Q,LO}/\xi_{Q,LA}\,\omega_{Q,LA})^2=b^2_Q/(\omega_{Q,LA}^2)\sim b_Q\ll 1$, and therefore, we expect $\tau^{(s)}_{Q,LO}\gg\tau^{(s)}_{Q,LA}$ as long as $Qa\ll 1$, where relaxation times $\tau^{(s)}_{Q,LO},\,\tau^{(s)}_{Q,LA}$ are defined in Eq.\,\eqref{rough-6} below.
\medskip

More specifically, for surface scattering of longitudinal phonons, we get that\,\cite{callaway}

\begin{equation}
\xi_{Q,\gamma}=\mp\frac{1}{{\cal R}}\left[\frac{2m_1m_2}{(m_1+m_2)^2}\right]\left\{\frac{(Qa)\,\sin(Qa)}{{\cal G}_0(Qa)[1\pm{\cal G}_0(Qa)]}\right\}\ ,
\label{rough-7}
\end{equation}
where the structural factor ${\cal G}_0(Qa)$ is given by

\begin{equation}
{\cal G}_0(Qa)=\sqrt{1-\frac{4m_1m_2}{(m_1+m_2)^2}\,\sin^2(Qa)}\ ,
\label{rough-8}
\end{equation}
and $\gamma=LO$ ($\gamma=LA$) correspond to upper (lower) signs, respectively. Meanwhile, by assuming a Gaussian-form correlation function for two surface fluctuations\,\cite{lyo} as $\llangle\Delta_{\rm sr}(\mbox{\boldmath$\hat{r}$}_\|)\,\Delta_{\rm sr}(\mbox{\boldmath$\hat{r}$}')_\|\rrangle_{\rm av}=\bar{\Delta}_0^2
\exp(-|\mbox{\boldmath$\hat{r}$}_\|-\mbox{\boldmath$\hat{r}$}'_\||^2/\Lambda_0^2)$, and its Fourier transform $\Delta_{\rm sr}(|Q'-Q|)$, introduced in Eq.\,\eqref{rough-3}, is calculated as 

\begin{equation}
\Delta_{\rm sr}(|Q'-Q|)=\pi\Lambda_0^2\,\bar{\Delta}_0^2\exp\left[-\frac{|Q'-Q|^2\Lambda_0^2}{4}\right]\ ,
\label{rough-4}
\end{equation}
where $\Lambda_0$ denotes the correlation length for a surface-roughness distribution which characterizes the interaction range of surface fluctuations, whereas $\bar{\Delta}_0$ denotes the average magnitude of surface fluctuations.
\medskip

Formally, we could write

\begin{equation}
{\cal C}^{(s)}_{Q,\gamma}(t)\left[1+N_Q^{\gamma}(t)\right]-{\cal D}^{(s)}_{Q,\gamma}(t)\,N_Q^{\gamma}(t)
\approx\frac{N_{Q,\gamma}^{(0)}-N_Q^{\gamma}(t)}{\tau^{(s)}_{Q,\gamma}(t)}\ ,
\label{rough-5}
\end{equation}
where $1/\tau^{(s)}_{Q,\gamma}(t)\approx{\cal D}^{(s)}_{Q,\gamma}(t)-{\cal C}^{(s)}_{Q,\gamma}(t)$ is the relaxation rate for surface-roughness scattering of $\gamma$ phonons, and $N_{Q,\gamma}^{(0)}\approx {\cal C}^{(s)}_{Q,\gamma}(t)/[{\cal D}^{(s)}_{Q,\gamma}(t)-{\cal C}^{(s)}_{Q,\gamma}(t)]$ is the initial
thermal-equilibrium distribution of $\gamma$ phonons. From Eqs.\,\eqref{rough-0} and \eqref{rough-1}, we further find 

\begin{equation}
{\cal D}^{(s)}_{Q,\gamma}(t)-{\cal C}^{(s)}_{Q,\gamma}(t)
=\sum\limits_{Q'}\,{\cal P}_\gamma^s(Q',Q)=\frac{{\cal L}}{2\pi}\int\limits_{-\pi/d}^{\pi/d} dQ'\,{\cal P}_\gamma^s(Q',Q)\equiv\frac{1}{\tau^{(s)}_{Q,\gamma}}\ ,
\label{rough-6}
\end{equation}
which becomes independent of time $t$, where $\tau^{(s)}_{Q,\gamma}>0$ is the so-called phnon energy-relaxation time and independent of time $t$.

\begin{table}[htbp]
\centering
\caption{Four Dominant Contributions from $A_1$-$A_8$ for the $A$-Process}
\vspace{0.3cm}
\begin{tabular}{cccc}
\hline\hline
$\gamma$\ \ \ \ \ \   &   $\gamma'$\ \ \ \ \ \    &    $\gamma^{''}$\ \ \ \ \ \   &    Symbol\\
\hline
LA-in\ \ \ \ \ \      &   LA-out\ \ \ \ \ \       &    LA-out\ \ \ \ \ \ & $A_1$\\
LO-in\ \ \ \ \ \      &   LA-out\ \ \ \ \ \       &    LA-out\ \ \ \ \ \ & $A_2$\\
LO-in\ \ \ \ \ \      &   LA-out\ \ \ \ \ \       &    LO-out\ \ \ \ \ \ &      $A_3$\\
LO-in\ \ \ \ \ \      &   LO-out\ \ \ \ \ \       &    LA-out\ \ \ \ \ \ &      $A_4$\\
\hline\hline
\end{tabular}
\label{tab-a}
\end{table}

\begin{table}[htbp]
\centering
\caption{Four Dominant Contributions from $B_1$-$B_8$ for the $B$-Process}
\vspace{0.3cm}
\begin{tabular}{cccc}
\hline\hline
$\gamma$\ \ \ \ \ \   &   $\gamma'$\ \ \ \ \ \    &    $\gamma^{''}$\ \ \ \ \ \   &    Symbol\\
\hline
LA-in\ \ \ \ \ \      &   LA-in\ \ \ \ \ \        &    LA-out\ \ \ \ \ \ & $B_1$\\
LA-in\ \ \ \ \ \      &   LA-in\ \ \ \ \ \        &    LO-out\ \ \ \ \ \ & $B_2$\\
LA-in\ \ \ \ \ \      &   LO-in\ \ \ \ \ \        &    LO-out\ \ \ \ \ \ & $B_3$\\
LO-in\ \ \ \ \ \      &   LA-in\ \ \ \ \ \        &    LO-out\ \ \ \ \ \ & $B_4$\\
\hline\hline
\end{tabular}
\label{tab-b}
\end{table}

\subsection{Three-Phonon Anharmonic Interaction and Electron-Phonon Coupling}
\label{sec-2-4}

\begin{figure}[htbp]
\centering
\includegraphics[width=0.45\textwidth]{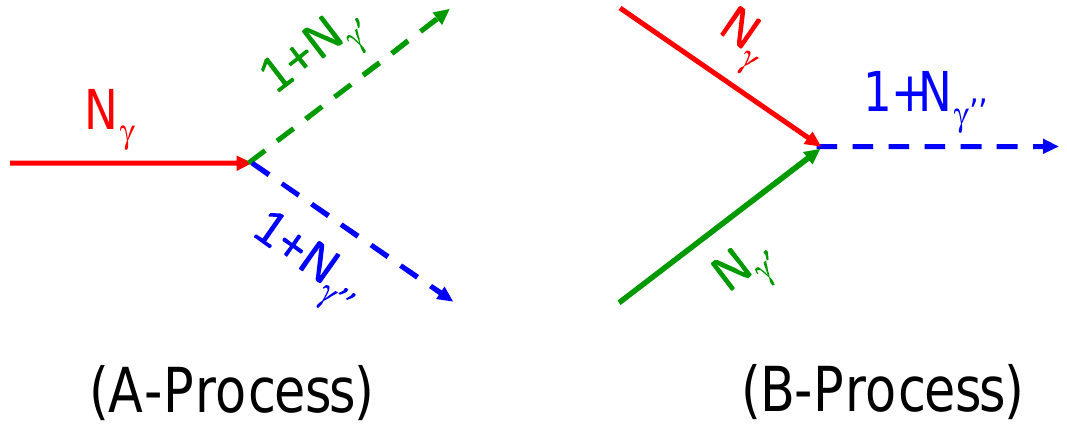}
\caption{Illustrations for the $A$-process (left) with annihilation of one phonon and creation of two phonons,
as well as for the $B$-process (right) with annihilation of two phonons and creation of one phonon.
Here, three different phonon modes are represented by colors (red, blue, green) for their creations (dashed) and annihilations (solid).}
\label{fig1}
\end{figure}

By going beyond a two-atom harmonic-interaction model, we turn to discussions on three-phonon anharmonic interactions\,\cite{r1,r21,r10} in an atomic chain. From the perspective of energy-momentum conservation for three-phonon anharmonic interactions, there exist only two distinct processes as schematically shown in Fig.\,\ref{fig1}, where either one phonon ($A$-process) or two phonons ($B$-process) are annihilated, while the other processes involving either creating or annihilating all three phonons can be excluded in our current study. If only one longitudinal-acoustic (LA) and one longitudinal-optical (LO) phonon modes\,\cite{callaway} are considered, there will be totally eight contributions for each process. For process-$A$, four dominant contributions to three-phonon anharmonic interaction are listed in Table\,\ref{tab-a}. Meanwhile,
the other four significant contributions are indicated in Table\,\ref{tab-b} for process-$B$.
\medskip

In our quantum-kinetic model, we will turn on the phonon and coulomb scatterings and phonon-electron interaction right after the initial time $t=0$, which is accompanied by gradually turning on a DC electric field for electron drifting at $t=\tau_0$. Meanwhile, we also turn on the thermal and pressure driving forces to a lattice at $t=\tau_0$. The thermal effect resulting from laser-plasmon induced lattice heating, while the lattice expansion was discussed by Bragas, {\em et al\/}.\,\cite{expansion} under high-temperature limit.
\medskip

By following this scheme and using the non-interacting phonon states as base functions for density-matrix operator of phonons, 
the quantum-kinetic equations, including external driving forces to lattice, for non-equilibrium  distributions $N_{Q}^{\gamma}(t)$ of both $\gamma=$LA and $\gamma=$LO phonons
take the form

\begin{eqnarray}
\nonumber
&&\frac{d\Delta N_{Q}^{\gamma}(t)}{dt}-\left[\frac{{\cal F}_{\rm th}^\gamma(t)}{\hbar}\right]
\frac{\partial N_{Q}^{\gamma}(t)}{\partial Q}
-\left[\frac{Q{\cal P}^{(\tau_0)}_{\rm b}(t){\cal A}d}{2\hbar}\right]\,\frac{\partial N_{Q}^{\gamma}(t)}{\partial Q}\\
&=&\left.\frac{\partial\Delta N_{Q}^{\gamma}(t)}{\partial t}\right|^{(A)}_{\rm ph-ph}
+\left.\frac{\partial\Delta N_{Q}^{\gamma}(t)}{\partial t}\right|^{(B)}_{\rm ph-ph}
+\left.\frac{\partial\Delta N_{Q}^{\gamma}(t)}{\partial t}\right|_{\rm ph-sr}
+\left.\frac{\partial\Delta N_{Q}^{\gamma}(t)}{\partial t}\right|_{\rm ph-el}\ ,\ \ \ \
\label{qdeph}
\end{eqnarray}
where ${\cal F}_{\rm th}^\gamma(t)=\alpha_0C_0^\gamma\,\Theta^{(M)}_{\tau_0}(t)$
plays the role of a static thermal-driving force 
acting on lattice vibrations and directly leading to a thermal-conduction process, 
$\alpha_0=\Delta T/{\cal L}\equiv(T_R-T_L)/{\cal L}$ represents the spatially-uniform temperature gradient applied to lattices by two external thermal-equilibrium reservoirs with their fixed temperatures $T_L<T_R$ and separation (or chain length) ${\cal L}$, 
$C_0^\gamma=[\overline{U}_{\gamma}^{(0)}(T_R)/n^{\gamma}_0(T_R)-\overline{U}_{\gamma}^{(0)}(T_L)/n^{\gamma}_0(T_L)]/\Delta T$ is the thermal-equilibrium heat capacity per reservoir thermal-equilibrium phonon, $\overline{U}_{\gamma}^{(0)}(T)$ is given by Eq.\,\eqref{efft3} below for $\gamma={\rm LA}$ or by Eq.\,\eqref{efft4} for $\gamma={\rm LO}$, 
$\Delta N_Q^\gamma(t)\equiv N_Q^\gamma(t)-N^{(0)}_{Q,\gamma}(T_0)$ stands for the non-equilibrium part of phonon distribution, $N^{(0)}_{Q,\gamma}=\{\exp[\hbar\omega_{Q}^{\gamma}/(k_BT_0)]-1\}^{-1}$ is the thermal-equilibrium phonon distribution function, $T_0=(T_R+T_L)/2$ is initial environmental temperature,
and $n^{\gamma}_0(T_{L,R})$ is the thermal-equilibrium phonon number in reservoirs under the harmonic approximation for $\gamma={\rm LA},\,{\rm LO}$ and determined by Eqs.\,\eqref{num-1} and \eqref{num-2} below, respectively. Here, for $\gamma=LA$, its minimum energy $\hbar\omega_Q^\gamma$ at $Q=0$ is cut off by $\hbar\Gamma^\gamma_{\rm ph}$ due to a finite lifetime $1/\Gamma^\gamma_{\rm ph}$ of $LA$ phonons. 
In Eq.\,\eqref{qdeph}, $\tau_0$ stands for a turning-on time for thermal and blasting forces on the lattice chain. Moreover, the slowly turning-on (step) function $\Theta^{(M)}_{\tau_0}(t)$ inexplicitly introduced in Eq.\,\eqref{qdeph} for $\tau_0>0$ can be formally written as a multi-step function, {\em i.e.\/},

\begin{equation}
\Theta^{(M)}_{\tau_0}(t)=\frac{1}{M}\,\sum\limits_{j=1}^M\,\left[\frac{1}{2}+\frac{1}{\pi}\,\tan^{-1}\left(\frac{t-t_j}{\gamma_\tau}\right)\right]\ ,
\label{step}
\end{equation}
where $M$ represents the chosen total number of sub-steps, $t_j=j\Delta t$ designates different sub-step positions with $j=1,\,2,\,\cdots,\,M$, $\Delta t=\tau_0/M$ is the sub-step width, and $0<\gamma_\tau\ll\Delta t$ quantifies the sub-step broadening.
The physical effect of the thermal force ${\cal F}_{\rm th}^\gamma(t)$ is decreasing the phonon distribution $N_{Q}^{\gamma}(t)$ in the $Q\gtrapprox 0$ region while increasing that in the $Q\lessapprox 0$ region at the same time, giving rise to a net heat-current density for $LA$ phonons from the right $(T_R)$ to the left $(T_L)$ end of an atomic chain as expected. 
Moreover, as shown in Eq.\,\eqref{drag} below, the temperature gradient $\alpha_0=\Delta T/{\cal L}$ plays a similar role as an electrical voltage to a charge current as to a heat-current density ${\cal Q}^\gamma_{\rm ph}(t)$. Correspondingly, we are also able to calculate the transient electrical current $I_{\rm e}(t)=-n_{\rm e}\,ev_{\rm d}(t)$ by using the drift velocity $v_{\rm d}(t)$ determined numerically from Eq.\,\eqref{boltz-2}. 
\medskip

On the other hand, the second dynamic force term $\sim Q{\cal P}^{(\tau_0)}_{\rm b}(t){\cal A}d/2$ occurring in Eq.\,\eqref{qdeph} has been fully derived and explained in Appendix\,\ref{sec-2.3}.  
The physical influence of such a mechanical force is pushing the very-high phonon distribution $N_{Q}^{\gamma}(t)$ at $Q=0$ towards both $Q>0$ and $Q<0$ regions, simultaneously and equally. 
In such a situation, no flow of heat-current density is expected under the condition of $\alpha_0=0$. However, the effective temperature $T^{\rm LA}_{\rm eff}(t)$ of $LA$ phonons will still grow, giving rise to a lattice-heating effect. 
\medskip

\begin{figure}[htbp]
\centering
\includegraphics[width=0.65\textwidth]{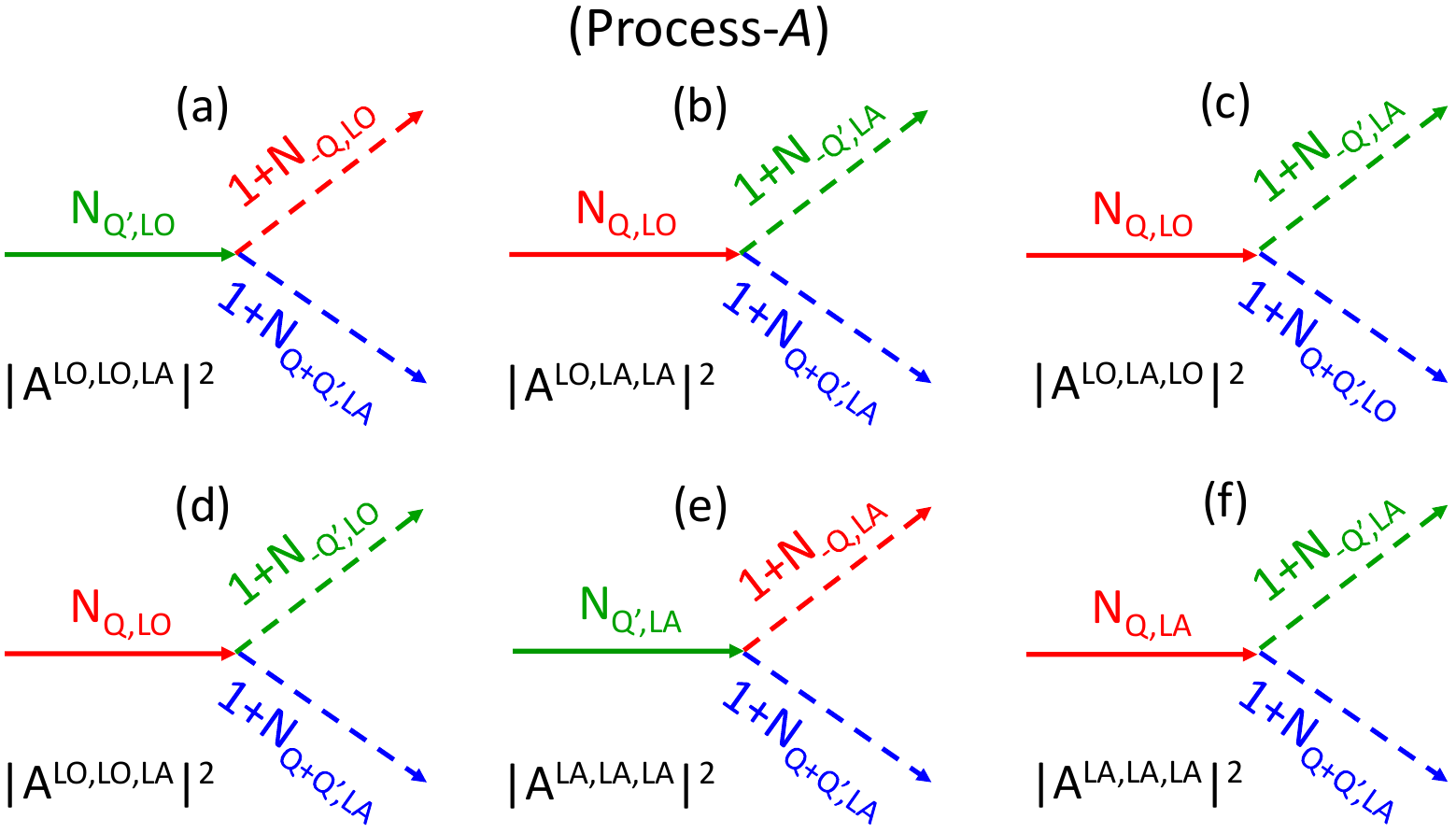}
\caption{Schematic representations of six different decays [(a)-(f)] for the $A$-process, where
$\Big|A^{\gamma,\gamma^{'},\gamma^{''}}_{{\bf Q},{\bf Q}^{'},{\bf Q}^{''}}\Big|^2$ represents the square of
anharmonic interaction among three phonons, and other notations are the same as those in Fig.\,\ref{fig1}.}
\label{fig14}
\end{figure}

The third term in four different scattering contributions on the right-hand side of Eq.\eqref{qdeph}, {\em i.e.\/}

\begin{equation}
\label{qdeph-1}
\left.\frac{\partial\Delta N_{Q}^{\gamma}(t)}{\partial t}\right|_{\rm ph-sr}={\cal F}_{\rm ph-sr}[N_{Q}^{\gamma}(t)]-{\cal F}_{\rm ph-sr}[N^{(0)}_{Q,\gamma}(T_0)]
\end{equation}
with $N_{Q}^{\gamma}(t)=\Delta N_{Q}^{\gamma}(t)+N^{(0)}_{Q,\gamma}(T_0)$, and

\begin{equation}
{\cal F}_{\rm ph-sr}[N_{Q}^{\gamma}(t)]\equiv{\cal C}^{(s)}_{Q,\gamma}(t)\left[N_{Q}^\gamma(t)+1\right]-{\cal D}^{(s)}_{Q,\gamma}(t)\,N_{Q}^\gamma(t)
\end{equation}
is attributed to the surface-roughness scattering of longitudinal $\gamma$ phonons in a crystal cylinder, ${\cal C}^{(s)}_{Q,\gamma}(t)$ is a creation rate for longitudinal $\gamma$ phonons given by Eq.\,\eqref{rough-0}, while ${\cal D}^{(s)}_{Q,\gamma}(t)$ is an annihilation rate defined in Eq.\,\eqref{rough-1} for longitudinal $\gamma$ phonons. 
Moreover, $N^{(0)}_{Q,\gamma}(T_0)$ in Eq.\,\eqref{qdeph-1} is a thermal-equilibrium occupation function for $\gamma$ phonons at their initial given environmental temperature $T_0$. 
If we adopt a harmonic approximation for lattice vibrations within an atomic chain, both $A$- and $B$ anharmonic-interaction terms in Eq.\,\eqref{qdeph} due to three-phonon scattering can be ignored. Therefore, only the surface-roughness scattering of phonons will be retained in this case so as to enable a process for energy relaxation of phonons.
\medskip

If the cross-sectional area of a chain is small enough, significant energy gap between transverse and longitudinal phonon modes occur.\,\cite{ando} In this case, we need consider only low-energy longitudinal-acoustic $(LA)$ and longitudinal-optical $(LO)$ phonons in the system while neglecting all high-energy transverse phonon modes at the same time.
As a result, the first term for the second scattering contribution on the right-hand side of Eq.\,\eqref{qdeph}, as illustrated in the left panel of Fig.\,\ref{fig1} for the three-phonon $A$-process, is calculated according to\,\cite{r1}

\begin{equation}
\label{decayloa}
\left.\frac{\partial\Delta N_{Q}^{\rm LO}(t)}{\partial t} \right|_{\rm ph-ph}^{(A)}={\cal F}^{(A)}_{\rm ph-ph}[N_Q^{\rm LO}(t)]-{\cal F}^{(A)}_{\rm ph-ph}[N^{(0)}_{Q,LO}(T_0)]
\end{equation}
with $N_{Q}^{\rm LO}(t)=\Delta N_{Q}^{\rm LO}(t)+N^{(0)}_{Q,LO}(T_0)$ and 

\begin{eqnarray}
\nonumber
&&{\cal F}^{({\rm A})}_{\rm ph-ph}[N_Q^{\rm LO}(t)]\equiv
{\cal C}^{(A)}_{Q,LO}(t)\left[N_{-Q}^{\rm LO}(t)+1\right]-{\cal D}^{(A)}_{Q,LO}(t)\,N_{Q}^{\rm LO}(t)\\
\nonumber
&=&\frac{d}{2\hbar}\int\limits_{-\pi/d}^{\pi/d} dQ'\,\Big(\left[N_{-Q}^{\rm LO}(t)+1\right]\Big\{2\left|A_{Q,Q',Q+Q'}^{LO,LO,LA} \right|^2N_{Q'}^{\rm LO}(t)\left[N_{Q+Q'}^{\rm LA}(t)+1\right]\,{\cal L}_0\left(\hbar\omega_{Q}^{\rm LO}-\hbar\omega_{Q'}^{\rm LO}+\hbar\omega_{Q+Q'}^{\rm LA},\,\hbar\Gamma^\gamma_{\rm ph}\right)\Big\}\\
\nonumber
&-&N_{Q}^{\rm LO}(t)\Big\{\left|A_{Q,Q',Q+Q'}^{LO,LA,LA}\right|^2\left[N_{-Q'}^{\rm LA}(t)+1\right]\left[N_{Q+Q'}^{\rm LA}(t)+1\right]\,
{\cal L}_0\left(\hbar\omega_{Q}^{\rm LO}-\hbar\omega_{Q'}^{\rm LA}-\hbar\omega_{Q+Q'}^{\rm LA},\,\hbar\Gamma^\gamma_{\rm ph}\right)\\
\nonumber
&+&\left|A_{Q,Q',Q+Q'}^{LO,LA,LO}\right|^2\left[N_{-Q'}^{\rm LA}(t)+1\right]\left[N_{Q+Q'}^{\rm LO}(t)+1\right]\,{\cal L}_0
\left(\hbar\omega_{Q}^{\rm LO}-\hbar\omega_{Q'}^{\rm LA}-\hbar\omega_{Q+Q'}^{\rm LO},\,\hbar\Gamma^\gamma_{\rm ph}\right)\\
\label{decayloa-2}
&+&\left|A_{Q,Q',Q+Q'}^{LO,LO,LA}\right|^2\left[N_{-Q'}^{\rm LO}(t)+1\right]\left[N_{Q+Q'}^{\rm LA}(t)+1\right]\,
{\cal L}_0\left(\hbar\omega_{Q}^{\rm LO}-\hbar\omega_{Q'}^{\rm LO}-\hbar\omega_{Q+Q'}^{\rm LA},\,\hbar\Gamma^\gamma_{\rm ph}\right)\Big\}\Big)\ ,
\end{eqnarray}
while

\begin{equation}
\label{decaylaa}
\left.\frac{\partial\Delta N_{Q}^{\rm LA}(t)}{\partial t} \right|_{\rm ph-ph}^{(A)}={\cal F}^{(A)}_{\rm ph-ph}[N_Q^{\rm LA}(t)]-{\cal F}^{(A)}_{\rm ph-ph}[N^{(0)}_{Q,LA}(T_0)]
\end{equation}
with $N_{Q}^{\rm LA}(t)=\Delta N_{Q}^{\rm LA}(t)+N^{(0)}_{Q,LA}(T_0)$ and  

\begin{eqnarray}
\nonumber
&&{\cal F}^{({\rm A})}_{\rm ph-ph}[N_Q^{\rm LA}(t)]\equiv{\cal C}^{(A)}_{Q,LA}(t)\left[N_{-Q}^{\rm LA}(t)+1\right]-{\cal D}^{(A)}_{Q,LA}(t)\,N_{Q}^{\rm LA}(t)\\
\nonumber
&=&\frac{d}{2\hbar}\int\limits_{-\pi/d}^{\pi/d} dQ'\,\Big(\left[N_{-Q}^{\rm LA}(t)+1\right]\Big\{2\left|A_{Q,Q',Q+Q'}^{LA,LA,LA}\right|^2N_{Q'}^{\rm LA}(t)\left[N_{Q+Q'}^{\rm LA}(t)+1\right]\,
{\cal L}_0\left(\hbar\omega_{Q}^{\rm LA}-\hbar\omega_{Q'}^{\rm LA}+\hbar\omega_{Q+Q'}^{\rm LA},\,\hbar\Gamma^\gamma_{\rm ph}\right)\Big\}\\
\label{decaylaa-2}
&-&N_{Q}^{\rm LA}(t)\Big\{\left|A_{Q,Q',Q+Q'}^{LA,LA,LA}\right|^2\left[N_{-Q'}^{\rm LA}(t)+1\right]\left[N_{Q+Q'}^{\rm LA}(t)+1\right]\,
{\cal L}_0\left(\hbar\omega_{Q}^{\rm LA}-\hbar\omega_{Q'}^{\rm LA}-\hbar\omega_{Q+Q'}^{\rm LA},\,\hbar\Gamma^\gamma_{\rm ph}\right)\Big\}\Big)\ .
\end{eqnarray}
Here, $d$ in Eqs.\,\eqref{decayloa-2} and \eqref{decaylaa-2} is the lattice period, $\omega_{Q}^{\gamma}$ the frequency of $\gamma$ phonons, and four terms in Eq.\,\eqref{decayloa-2} and two terms in Eq.\,\eqref{decaylaa-2} are all depicted in Fig.\,\ref{fig14}.
Moreover, in Eqs.\,(\ref{decayloa-2}) and (\ref{decaylaa-2}), a so-called Umklapp-scattering process\,\cite{umklapp} can occur whenever the final scattering wave number $Q+Q'$ goes out of the first Brillouin zone $[-G_0/2,\,G_0/2]$, where $G_0=2\pi/d$ is a reciprocal-lattice vector. In order to include this Umklapp-scattering effect, we must replace $Q+Q'$ by $Q+Q'\pm G_0$
to ensure that the new final scattering wave number $Q+Q'\pm G_0$ always remains within the first Brillouin zone.
\medskip

\begin{figure}[htbp]
\centering
\includegraphics[width=0.65\textwidth]{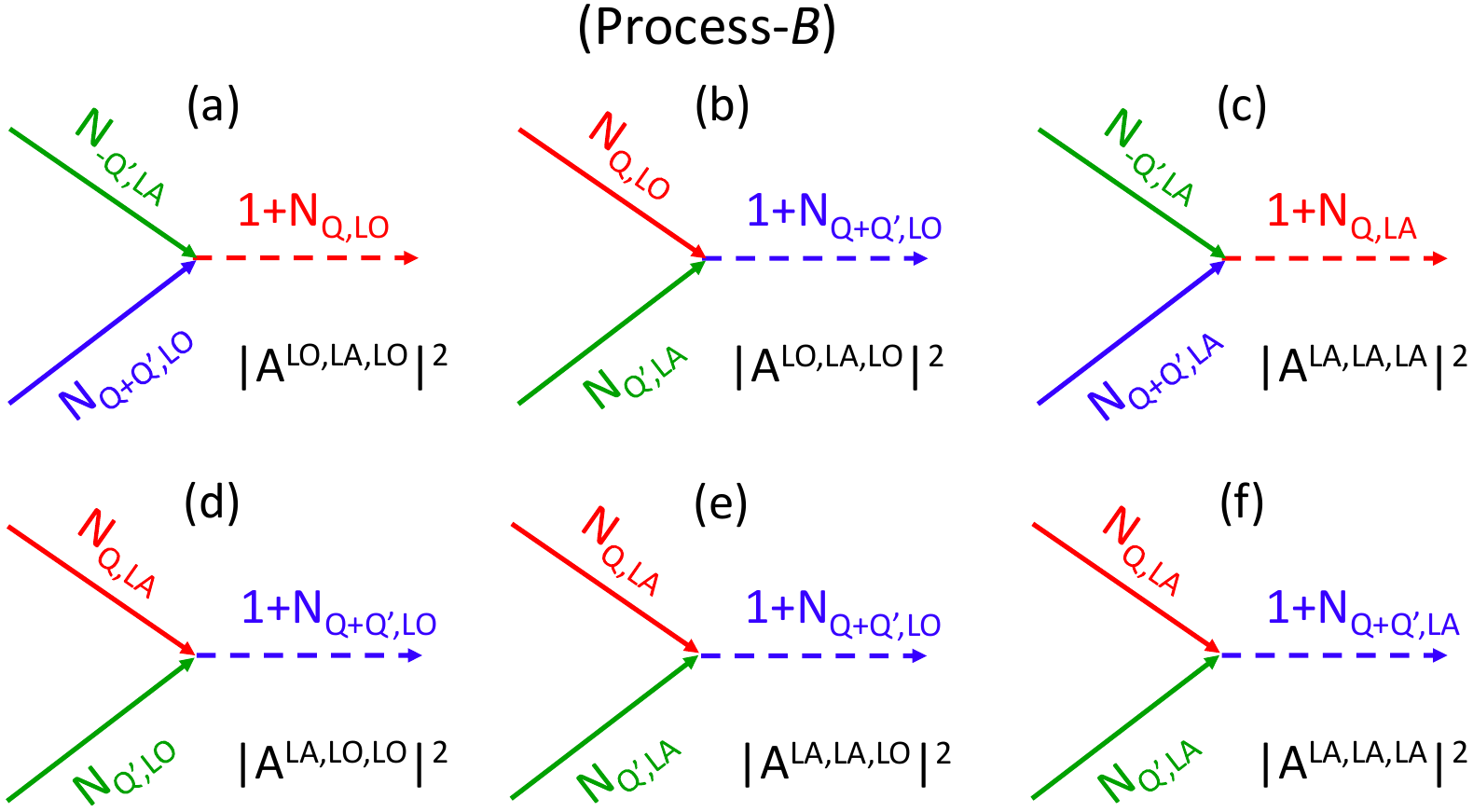}
\caption{Schematic representations of six different annihilations [(a)-(f)] for the $B$-process, where
$\Big|A^{\gamma,\gamma^{'},\gamma^{''}}_{{\bf Q},{\bf Q}^{'},{\bf Q}^{''}}\Big|^2$ represents the square of
anharmonic interaction among three phonons, and other notations are the same as those in Fig.\,\ref{fig1}.}
\label{fig15}
\end{figure}

On the other hand, the second term for the other phonon-scattering contribution on the right-hand side of Eq.\,\eqref{qdeph}, as depicted by the right panel of Fig.\,\ref{fig1}, for the three-phonon $B$-process is given by

\begin{equation}
\label{decaylob}
\left.\frac{\partial\Delta N_{Q}^{\rm LO}(t)}{\partial t} \right|_{\rm ph-ph}^{(B)}={\cal F}^{(B)}_{\rm ph-ph}[N_Q^{\rm LO}(t)]-{\cal F}^{(B)}_{\rm ph-ph}[N^{(0)}_{Q,LO}(T_0)]
\end{equation}
with $N_{Q}^{\rm LO}(t)=\Delta N_{Q}^{\rm LO}(t)+N^{(0)}_{Q,LO}(T_0)$ and  

\begin{eqnarray}
\nonumber
&&{\cal F}^{(B)}_{\rm ph-ph}[N_Q^{\rm LO}(t)]\equiv{\cal C}^{(B)}_{Q,LO}(t)\left[N_{Q}^{\rm LO}(t)+1\right]-{\cal D}^{(B)}_{Q,LO}(t)\,N_{Q}^{\rm LO}(t)\\
\nonumber
&=&\frac{d}{2\hbar}\int\limits_{-\pi/d}^{\pi/d} dQ'\,\Big(\left[N_{Q}^{\rm LO}(t)+1\right]\Big\{\left|A_{Q,Q',Q+Q'}^{LO,LA,LO}\right|^2N_{-Q'}^{\rm LA}(t)\,N_{Q+Q'}^{\rm LO}(t)\,
{\cal L}_0\left(\hbar\omega_{Q}^{\rm LO}-\hbar\omega_{Q'}^{\rm LA}-\hbar\omega_{Q+Q'}^{\rm LO},\,\hbar\Gamma^\gamma_{\rm ph}\right)\\
\label{decaylob-2}
&-&N_{Q}^{\rm LO}(t)\Big\{2\left|A_{Q,Q',Q+Q'}^{LO,LA,LO}\right|^2N_{Q'}^{\rm LA}(t)\left[N_{Q+Q'}^{\rm LO}(t)+1\right]\,
{\cal L}_0\left(\hbar\omega_{Q}^{\rm LO}+\hbar\omega_{Q'}^{\rm LA}-\hbar\omega_{Q+Q'}^{\rm LO},\,\hbar\Gamma^\gamma_{\rm ph}\right)\Big\}\Big)\ ,
\end{eqnarray}
while

\begin{equation}
\label{decaylab}
\left.\frac{\partial\Delta N_{Q}^{\rm LA}(t)}{\partial t}\right|_{\rm ph-ph}^{(B)}={\cal F}^{(B)}_{\rm ph-ph}[N_Q^{\rm LA}(t)]-{\cal F}^{(B)}_{\rm ph-ph}[N^{(0)}_{Q,LA}(T_0)]
\end{equation}
with $N_{Q}^{\rm LA}(t)=\Delta N_{Q}^{\rm LA}(t)+N^{(0)}_{Q,LA}(T_0)$ and 

\begin{eqnarray}
\nonumber
&&{\cal F}^{(B)}_{\rm ph-ph}[N_Q^{\rm LA}(t)]\equiv{\cal C}^{(B)}_{Q,LA}(t)\left[N_{Q}^{\rm LA}(t)+1\right]-{\cal D}^{(B)}_{Q,LA}(t)\,N_{Q}^{\rm LA}(t)\\
\nonumber
&=&\frac{d}{2\hbar}\int\limits_{-\pi/d}^{\pi/d} dQ'\,\Big(\left[N_{Q}^{\rm LA}(t)+1\right]\Big\{\left|A_{Q,Q',Q+Q'}^{LA,LA,LA}\right|^2N_{-Q'}^{\rm LA}(t)\,N_{Q+Q'}^{\rm LA}(t)\,
{\cal L}_0\left(\hbar\omega_{Q}^{\rm LA}-\hbar\omega_{Q'}^{\rm LA}-\hbar\omega_{Q+Q'}^{\rm LA},\,\hbar\Gamma^\gamma_{\rm ph}\right)\Big\}\\
\nonumber
&-&N_{Q}^{\rm LA}(t)\Big\{2\left|A_{Q,Q',Q+Q'}^{LA,LO,LO}\right|^2N_{Q'}^{\rm LO}(t)\left[N_{Q+Q'}^{\rm LO}(t)+1\right]\,
{\cal L}_0\left(\hbar\omega_{Q}^{\rm LA}+\hbar\omega_{Q'}^{\rm LO}-\hbar\omega_{Q+Q'}^{\rm LO},\,\hbar\Gamma^\gamma_{\rm ph}\right)\\
\nonumber
&+&2\left|A_{Q,Q',Q+Q'}^{LA,LA,LO}\right|^2N_{Q'}^{\rm LA}(t)\left[N_{Q+Q'}^{\rm LO}(t)+1\right]\,
{\cal L}_0\left(\hbar\omega_{Q}^{\rm LA}+\hbar\omega_{Q'}^{\rm LA}-\hbar\omega_{Q+Q'}^{\rm LO},\,\hbar\Gamma^\gamma_{\rm ph}\right)\\
\label{decaylab-2}
&+&2\left|A_{Q,Q',Q+Q'}^{LA,LA,LA}\right|^2N_{Q'}^{\rm LA}(t)\left[N_{Q+Q'}^{\rm LA}(t)+1\right]\,
{\cal L}_0\left(\hbar\omega_{Q}^{\rm LA}+\hbar\omega_{Q'}^{\rm LA}-\hbar\omega_{Q+Q'}^{\rm LA},\,\hbar\Gamma^\gamma_{\rm ph}\right)\Big\}\Big)\ ,
\end{eqnarray}
where two contributions in Eq.\,\eqref{decaylob-2} and four contributions in Eq.\,\eqref{decaylab-2} are
schematically presented in Fig.\,\ref{fig15}, while the Umklapp-scattering process can be included in the same way as in Eqs.\,\eqref{decayloa-2} and \eqref{decaylaa-2}.
Additionally, ${\cal C}^{(A,B)}_{Q,\gamma}(t)$ and ${\cal D}^{(A,B)}_{Q,\gamma}(t)$ in Eqs.\,\eqref{decayloa}--\eqref{decaylab-2}
represent the creation and annihilation rates of $\gamma$-phonons with wave number $Q$ in the $A$- and $B$-process, respectively.
\medskip

For numerical computations, by using a homogeneous lifetime broadening $\hbar\Gamma^\gamma_{\rm ph}$ for phonons, 
we can define a cutoff wave number $|Q_c|=0.01\,(\Gamma^{\rm LA}_{\rm ph}/\hbar v_{\rm LA})$ for initial distribution of LA phonons so that $N^{(0)}_{Q,LA}=N^{(0)}_{|Q_c|,LA}$ if $N^{(0)}_{Q,LA}>k_BT_0/\hbar v_{\rm LA}|Q_c|$ to eliminate an infinity,
where $v_{\rm LA}$ is the sound velocity of LA phonons for $Q\to 0$ in Debye model.
Moreover, the square of anharmonic interaction $\left|A_{Q,Q',Q^{''}}^{\gamma,\gamma',\gamma''}\right|^2$ in Eqs.\,\eqref{decayloa}--\eqref{decaylab-2} for random phonon decays and collisions can be formally written in a general vector form\,\cite{r1}, yielding 

\begin{equation}
\left|A_{Q,Q',Q^{''}}^{\gamma,\gamma',\gamma''}\right|^2 = \frac{1}{a^2d}\left(\frac{\hbar}{2\rho}\right)^3\,\frac{{\cal A}^2_{\rm ph}}{\omega_{Q}^{\gamma}\,\omega_{Q'}^{\gamma'}\,\omega_{ Q^{''}}^{\gamma^{''}}}
\label{def}
\end{equation}
with $\rho$ and $d=2a$ as the volume mass density and lattice constant. The three-phonon anharmonic interaction amplitude ${\cal A}_{\rm ph}\equiv{\cal A}^{(E)}_{\rm ph}+{\cal A}^{(\mu)}_{\rm ph}+{\cal A}^{(\lambda)}_{\rm ph}$ in Eq.\,\eqref{def} is calculated generally and explicitly as

\begin{eqnarray}
\nonumber
{\cal A}^{(E)}_{\rm ph}&=&E_1\,(\mbox{\boldmath$\hat{e}$}\cdot\mbox{\boldmath$Q$})\,(\mbox{\boldmath$\hat{e}$}^{'}\cdot\mbox{\boldmath$Q$}^{'})\,
(\mbox{\boldmath$\hat{e}$}^{''}\cdot\mbox{\boldmath$Q$}^{''})\\
\nonumber
&+&E_2\,(\mbox{\boldmath$\hat{e}$}\cdot\mbox{\boldmath$Q$})\left[(\mbox{\boldmath$\hat{e}$}^{'}\cdot\mbox{\boldmath$\hat{e}$}^{''})\,
(\mbox{\boldmath$Q$}^{'}\cdot\mbox{\boldmath$Q$}^{''})+(\mbox{\boldmath$\hat{e}$}^{'}\cdot\mbox{\boldmath$Q$}^{''})\,
(\mbox{\boldmath$\hat{e}$}^{''}\cdot\mbox{\boldmath$Q$}^{'})\right]\\
\nonumber
&+&E_2\,(\mbox{\boldmath$\hat{e}$}^{'}\cdot\mbox{\boldmath$Q$}^{'})\left[(\mbox{\boldmath$\hat{e}$}\cdot\mbox{\boldmath$\hat{e}$}^{''})\,
(\mbox{\boldmath$Q$}\cdot\mbox{\boldmath$Q$}^{''})+(\mbox{\boldmath$\hat{e}$}\cdot\mbox{\boldmath$Q$}^{''})\,
(\mbox{\boldmath$\hat{e}$}^{''}\cdot\mbox{\boldmath$Q$})\right]\\
\nonumber
&+&E_2\,(\mbox{\boldmath$\hat{e}$}^{''}\cdot\mbox{\boldmath$Q$}^{''})\left[(\mbox{\boldmath$\hat{e}$}\cdot\mbox{\boldmath$\hat{e}$}^{'})\,
(\mbox{\boldmath$Q$}\cdot\mbox{\boldmath$Q$}^{'})+(\mbox{\boldmath$\hat{e}$}\cdot\mbox{\boldmath$Q$}^{'})\,
(\mbox{\boldmath$\hat{e}$}^{'}\cdot\mbox{\boldmath$Q$})\right]\\
\nonumber
&+&E_3\,(\mbox{\boldmath$\hat{e}$}\cdot\mbox{\boldmath$\hat{e}$}^{'})\left[(\mbox{\boldmath$\hat{e}$}^{''}\cdot\mbox{\boldmath$Q$})\,
(\mbox{\boldmath$Q$}^{'}\cdot\mbox{\boldmath$Q$}^{''})+(\mbox{\boldmath$Q$}\cdot\mbox{\boldmath$Q$}^{''})\,
(\mbox{\boldmath$\hat{e}$}^{''}\cdot\mbox{\boldmath$Q$}^{'})\right]\\
\nonumber
&+&E_3\,(\mbox{\boldmath$\hat{e}$}\cdot\mbox{\boldmath$Q$}^{'})\left[(\mbox{\boldmath$\hat{e}$}^{''}\cdot\mbox{\boldmath$Q$})\,
(\mbox{\boldmath$\hat{e}$}^{'}\cdot\mbox{\boldmath$Q$}^{''})+(\mbox{\boldmath$Q$}\cdot\mbox{\boldmath$Q$}^{''})\,
(\mbox{\boldmath$\hat{e}$}^{'}\cdot\mbox{\boldmath$\hat{e}$}^{''})\right]\\
\nonumber
&+&E_3\,(\mbox{\boldmath$\hat{e}$}\cdot\mbox{\boldmath$\hat{e}$}^{''})\left[(\mbox{\boldmath$\hat{e}$}^{'}\cdot\mbox{\boldmath$Q$})\,
(\mbox{\boldmath$Q$}^{'}\cdot\mbox{\boldmath$Q$}^{''})+(\mbox{\boldmath$Q$}\cdot\mbox{\boldmath$Q$}^{'})\,
(\mbox{\boldmath$\hat{e}$}^{'}\cdot\mbox{\boldmath$Q$}^{''})\right]\\
&+&E_3\,(\mbox{\boldmath$\hat{e}$}\cdot\mbox{\boldmath$Q$}^{''})\left[(\mbox{\boldmath$\hat{e}$}^{'}\cdot\mbox{\boldmath$Q$})\,
(\mbox{\boldmath$\hat{e}$}^{''}\cdot\mbox{\boldmath$Q$}^{'})+(\mbox{\boldmath$Q$}\cdot\mbox{\boldmath$Q$}^{'})\,
(\mbox{\boldmath$\hat{e}$}^{'}\cdot\mbox{\boldmath$\hat{e}$}^{''})\right]\ ,
\label{ono1}
\end{eqnarray}

\begin{eqnarray}
\nonumber
{\cal A}^{(\mu)}_{\rm ph}&=&\mu_L\left[(\mbox{\boldmath$\hat{e}$}\cdot\mbox{\boldmath$\hat{e}$}^{'})\,(\mbox{\boldmath$Q$}\cdot\mbox{\boldmath$\hat{e}$}^{''})
\,(\mbox{\boldmath$Q$}^{'}\cdot\mbox{\boldmath$Q$}^{''})
+(\mbox{\boldmath$\hat{e}$}\cdot\mbox{\boldmath$Q$}^{''})\,(\mbox{\boldmath$Q$}\cdot\mbox{\boldmath$Q$}^{'})\,
(\mbox{\boldmath$\hat{e}$}^{'}\cdot\mbox{\boldmath$\hat{e}$}^{''})\right.\\
\nonumber
&+&(\mbox{\boldmath$\hat{e}$}\cdot\mbox{\boldmath$\hat{e}$}^{'})\,(\mbox{\boldmath$Q$}\cdot\mbox{\boldmath$Q$}^{''})\,
(\mbox{\boldmath$Q$}^{'}\cdot\mbox{\boldmath$\hat{e}$}^{''})+(\mbox{\boldmath$\hat{e}$}\cdot\mbox{\boldmath$Q$}^{'})\,
(\mbox{\boldmath$Q$}\cdot\mbox{\boldmath$Q$}^{''})\,(\mbox{\boldmath$\hat{e}$}^{'}\cdot\mbox{\boldmath$\hat{e}$}^{''})\\
&+&\left.(\mbox{\boldmath$\hat{e}$}\cdot\mbox{\boldmath$\hat{e}$}^{''})\,(\mbox{\boldmath$Q$}\cdot\mbox{\boldmath$\hat{e}$}^{'})\,
(\mbox{\boldmath$Q$}^{'}\cdot\mbox{\boldmath$Q$}^{''})+(\mbox{\boldmath$\hat{e}$}\cdot\mbox{\boldmath$\hat{e}$}^{''})\,
(\mbox{\boldmath$Q$}\cdot\mbox{\boldmath$Q$}^{'})\,(\mbox{\boldmath$\hat{e}$}^{'}\cdot\mbox{\boldmath$Q$}^{''})\right]\ ,	
\label{ono2}
\end{eqnarray}

\begin{eqnarray}
\nonumber
{\cal A}^{(\lambda)}_{\rm ph}&=&\lambda_L\left[(\mbox{\boldmath$\hat{e}$}\cdot\mbox{\boldmath$Q$})\,(\mbox{\boldmath$\hat{e}$}^{'}\cdot\mbox{\boldmath$\hat{e}$}^{''})\,
(\mbox{\boldmath$Q$}^{'}\cdot\mbox{\boldmath$Q$}^{''})\right.\\
&+&\left.(\mbox{\boldmath$\hat{e}$}\cdot\mbox{\boldmath$\hat{e}$}^{''})\,(\mbox{\boldmath$Q$}\cdot\mbox{\boldmath$Q$}^{''})\,
(\mbox{\boldmath$\hat{e}$}^{'}\cdot\mbox{\boldmath$Q$}^{'})+(\mbox{\boldmath$\hat{e}$}\cdot\mbox{\boldmath$\hat{e}$}^{'})\,
(\mbox{\boldmath$Q$}\cdot\mbox{\boldmath$Q$}^{'})\,(\mbox{\boldmath$\hat{e}$}^{''}\cdot\mbox{\boldmath$Q$}^{''})\right]\ ,
\label{ono3}
\end{eqnarray}
where $\mbox{\boldmath$\hat{e}$}=\mbox{\boldmath$\hat{e}$}(\mbox{\boldmath$Q$},\gamma)$, $\mbox{\boldmath$\hat{e}$}^{'}=\mbox{\boldmath$\hat{e}$}^{'}(\mbox{\boldmath$Q$}^{'},\gamma^{'})$
and $\mbox{\boldmath$\hat{e}$}^{''}=\mbox{\boldmath$\hat{e}$}^{''}(\mbox{\boldmath$Q$}^{''},\gamma^{''})$ are three unit polarization vectors of phonons.
Moreover, $\mu_L$ and $\lambda_L$ are Lam\'e constants, and $E_1$, $E_2$ and $E_3$ are third-order elastic constants, which, for cubic crystals, are given
by\,\cite{r1}

\begin{eqnarray}
\nonumber
\mu_L&=&\frac{1}{5}\left(c_{11}-c_{12}+3\,c_{44}\right)\ ,\\
\nonumber
\lambda_L&=&\frac{1}{5}\left(c_{11}+4\,c_{12}-2\,c_{44}\right)\ ,\\
\nonumber
E_1&=&\frac{1}{35}\left(c_{111}+18\,c_{112}+16\,c_{123}-30\,c_{144}-12\,c_{166}+16\,c_{456}\right)\ ,\\
\nonumber
E_2&=&\frac{1}{35}\left(c_{111}+4\,c_{112}-5\,c_{123}+19\,c_{144}+2\,c_{166}-12\,c_{456}\right)\ ,\\
\label{coeff}
E_3&=&\frac{1}{35}\left(c_{111}-3\,c_{112}+2\,c_{123}-9\,c_{144}+9\,c_{166}+9\,c_{456}\right)\ .
\end{eqnarray}
Here, all the required elastic constants for GaAs material are listed in Tables\,\ref{tab-1} and \ref{tab-2}. 
\medskip

\begin{table}[htbp]
\centering
\caption{Second-Order Elastic Constants of GaAs}
\vspace{0.3cm}
\begin{tabular}{ccc}
\hline\hline
Elastic Constant  &   $\ \ \ \ \ \ \ \ $   &   Unit (GPa)\\
\hline
$c_{11}$   &   $\ \ \ \ \ \ \ \ $   &   $118.4$\\
$c_{12}$   &   $\ \ \ \ \ \ \ \ $   &   $53.7$\\
$c_{44}$   &   $\ \ \ \ \ \ \ \ $   &   $59.12$\\
$\mu_L$    &   $\ \ \ \ \ \ \ \ $   &   $48.4$\\
$\lambda_L$   &   $\ \ \ \ \ \ \ \ $   &   $43.0$\\
\hline\hline
\end{tabular}
\label{tab-1}
\end{table}

\begin{table}[htbp]
\centering
\caption{Third-Order Elastic Constants of GaAs}
\vspace{0.3cm}
\begin{tabular}{ccc}
\hline\hline
Elastic Constant  &   $\ \ \ \ \ \ \ \ $   &   Unit (GPa)\\
\hline
$c_{111}$  &   $\ \ \ \ \ \ \ \ $   &   $-622$\\
$c_{112}$   &   $\ \ \ \ \ \ \ \ $   &   $-387$\\
$c_{123}$   &   $\ \ \ \ \ \ \ \ $   &   $-57$\\
$c_{144}$   &   $\ \ \ \ \ \ \ \ $   &   $2$\\
$c_{166}$   &   $\ \ \ \ \ \ \ \ $   &   $-269$\\
$c_{456}$   &   $\ \ \ \ \ \ \ \ $   &   $-39$\\
$E_1$    &   $\ \ \ \ \ \ \ \ $   &   $-170$\\
$E_2$    &   $\ \ \ \ \ \ \ \ $   &   $-54.7$\\
$E_3$    &   $\ \ \ \ \ \ \ \ $   &   $-67.6$\\
\hline\hline
\end{tabular}
\label{tab-2}
\end{table}

We also note from Eqs.\,(\ref{ono1})--(\ref{ono3}) that ${\cal A}_{\rm ph}$ is symmetric with respect to a cycling exchange, {\em i.e.\/}, $(\mbox{\boldmath$\hat{e}$},\mbox{\boldmath$Q$})\leftrightarrow(\mbox{\boldmath$\hat{e}$}^{'},\mbox{\boldmath$Q$}^{'})$,
$(\mbox{\boldmath$\hat{e}$}^{'},\mbox{\boldmath$Q$}^{'})\leftrightarrow(\mbox{\boldmath$\hat{e}$}^{''},\mbox{\boldmath$Q$}^{''})$
and $(\mbox{\boldmath$\hat{e}$}^{''},\mbox{\boldmath$Q$}^{''})\leftrightarrow(\mbox{\boldmath$\hat{e}$},\mbox{\boldmath$Q$})$.
Since we consider a one-dimensional (1D) propagation case ({\em i.e.\/} $\mbox{\boldmath$Q$},\,\mbox{\boldmath$Q$}^{'},\,
\mbox{\boldmath$Q$}^{''}\|\mbox{\boldmath$\hat{e}$}_w$ with a unit vector $\mbox{\boldmath$\hat{e}$}_w$ in the chain direction)
for longitudinal-phonon modes ({\em i.e.\/}, $\mbox{\boldmath$\hat{e}$},\,\mbox{\boldmath$\hat{e}$}^{'},\,\mbox{\boldmath$\hat{e}$}^{''}\|\mbox{\boldmath$\hat{e}$}_w$)
and use the condition $Q^{''}=Q+Q^{'}$ at the same time, we can greatly simplify the result in Eqs.\,\eqref{ono1}--\eqref{ono3} and obtain

\begin{equation}
{\cal A}_{\rm ph}=(E_1+6E_2+8E_3+3\lambda_L+6\mu_L)\,QQ'(Q+Q')\ ,
\label{ampd2}
\end{equation}
which can be substituted back into Eq.\,\eqref{def} for calculating $\left|A_{Q,Q',Q^{''}}^{\gamma,\gamma',\gamma^{''}}\right|^2$
with eight combinations for $\gamma,\,\gamma'$ and $\gamma^{''}$ in total, where the Umklapp-scattering effect can also be taken into account as for Eqs.\,\eqref{decayloa}--\eqref{decaylab-2} by replacing $Q+Q'$ with $Q+Q'\pm G_0$ whenever it is required.
Physically, the mirror symmetry in \mbox{\boldmath$Q$} space, {\em i.e.\/} $N_{Q}^{\gamma}(t)=N_{-Q}^{\gamma}(t)$, should be maintained
before either a thermal or a mechanical force is turned on. As $Q,\,Q',\,Q^{''}\rightarrow 0$, we find $\left|A_{Q,Q',Q^{''}}^{\gamma,\gamma',\gamma''}\right|^2=0$, as can be verified by Eq.\,\eqref{def}.
\medskip

\begin{figure}[htbp]
\centering
\includegraphics[width=0.45\textwidth]{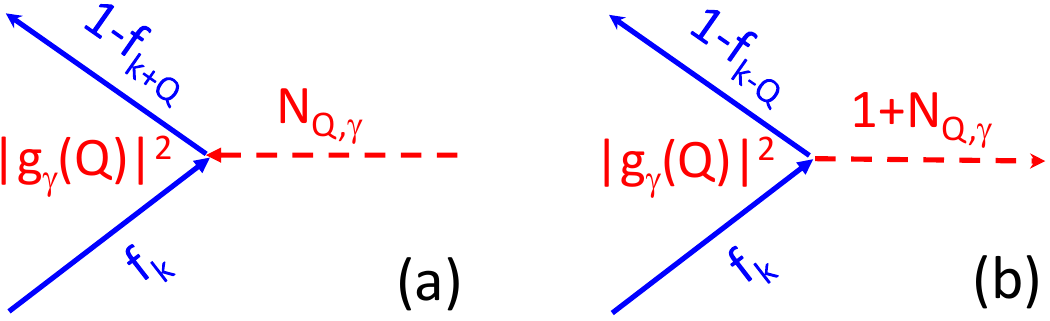}
\caption{Illustrations for (a) electron transition from initial $k$ state to final $k+Q$ state accompanied by a phonon absorption ($N^{\gamma}_{Q}$);
(b) electron transition from initial $k$ state to final $k-Q$ state with a phonon emission ($N^{\gamma}_{Q}+1$). In addition, $|g_\gamma(Q)|^2$
represents the electron-phonon coupling strength.}
\label{fig2}
\end{figure}

The third term for the electron-phonon interaction contribution on the right-hand side of Eq.\,\eqref{qdeph}, as illustrated in Figs.\,\ref{fig2}(a) and \ref{fig2}(b), can be written as\,\cite{r2}

\begin{equation}
\label{eph}
\left.\frac{\partial\Delta N_{Q}^{\gamma}(t)}{\partial t}\right|_{\rm ph-el}={\cal G}_{\rm ph-el}[N_Q^\gamma(t),\,\tilde{f}_k(t)]-{\cal G}_{\rm ph-el}[N^{(0)}_{Q,\gamma}(T_0),\,f_0(\varepsilon_k,T_0)]
\end{equation}
with $N_{Q}^{\gamma}(t)=\Delta N_{Q}^{\gamma}(t)+N^{(0)}_{Q,\gamma}(T_0)$, $\tilde{f}_k(t)=\Delta\tilde{f}_k(t)+f_0(\varepsilon_k,T_0)$ and 

\begin{eqnarray}
\nonumber
&&{\cal G}_{\rm ph-el}[N_Q^\gamma(t),\,\tilde{f}_k(t)]\equiv\frac{1}{\pi}\int\limits_{-\infty}^{\infty} dk\,\Xi^{({\rm em})}_{Q,\gamma}(k,t\,\vert\,v_{\rm d})\left[N_Q^\gamma(t)+1\right]-\frac{1}{\pi}\int\limits_{-\infty}^{\infty} dk\,\Xi^{({\rm abs})}_{Q,\gamma}(k,t\,\vert\,v_{\rm d})\,N_Q^\gamma(t)\\
\nonumber
&=&\left[N_{Q}^{\gamma}(t)+1\right]\Big\{\frac{2}{\hbar}\,\left|g_\gamma(Q)\right|^2\theta_0(\hbar\omega_{Q}^{\gamma}-\hbar Qv_{\rm d}(t))\,\frac{1}{\pi}\int\limits_{-\infty}^{\infty} dk\,\tilde{f}_k(t)\left[1-\tilde{f}_{k-Q}(t)\right]\,
{\cal L}_0\left(\varepsilon_k-\varepsilon_{k-Q}-\hbar\omega_{Q}^{\gamma}+\hbar Qv_{\rm d}(t),\,\hbar\Gamma_{\rm e}\right)\Big\}\\
\label{eph-2}
&-&N_{Q}^{\gamma}(t)\Big\{\frac{2}{\hbar}\,\left|g_\gamma(Q)\right|^2\theta_0(\hbar\omega_{Q}^{\gamma}-\hbar Qv_{\rm d}(t))\,\frac{1}{\pi}\int\limits_{-\infty}^{\infty} dk\,\tilde{f}_k(t)\left[1-\tilde{f}_{k+Q}(t)\right]\,
{\cal L}_0\left(\varepsilon_k-\varepsilon_{k+Q}+\hbar\omega_{Q}^{\gamma}-\hbar Qv_{\rm d}(t),\,\hbar\Gamma_{\rm e}\right)\Big\}\ ,\ \ \ \ \ \ \ \ 
\end{eqnarray}
where $\hbar\Gamma_{\rm e}$ is a homogeneous lifetime broadening for electrons, $\tilde{f}_k(t)$ is the full non-equilibrium occupation function of electrons with electron wave number $k$, 
while $f_0(\varepsilon_k,T_0)$ is only the thermal-equilibrium occupation function of electrons. 
In Eq.\,\eqref{eph-2}, $\Xi^{({\rm em})}_{Q,\gamma}(k,t\,\vert\,v_{\rm d})$ and $\Xi^{({\rm abs})}_{Q,\gamma}(k,t\,\vert\,v_{\rm d})$ represent emission and absorption rates of coupled $\gamma$ phonons by electrons in the $k$ state, $\varepsilon_k=\hbar^2k^2/2m^*$ is the electron kinetic energy with effective mass $m^*$, where the contributions from $Q=0$ should be excluded.
Moreover, $\theta_0(x)$ in Eq.\,\eqref{eph} is a unit step function approximated by $\theta_0(x)\approx1/2+(1/\pi)\tan^{-1}(x/\hbar\Gamma^\gamma_{\rm ph})$ with a lifetime broadening $\hbar\Gamma^\gamma_{\rm ph}$ for $\gamma$-phonons,
and the Doppler shift $\hbar Qv_{\rm d}(t)$ due to drifting electrons
has been included for the softening (hardening) of
$\gamma$-phonons energy $\hbar\omega_{Q}^{\gamma}$ for $Qv_{\rm d}(t)>0$ ($Qv_{\rm d}(t)<0$), which reflects a phonon-drag effect\,\cite{r3} from different rates for emission and absorption of phonons with respect to $Q$ and $-Q$.
\medskip

For the integral with respect to electron wave number $k$ in Eq.\,\eqref{eph},
we also numerically truncate it to $|k|\leq k_{\rm max}$ and set $\tilde{f}_k(t)=0$ for $|k|>k_{\rm max}$.
Additionally, the initial thermal-equilibrium occupation functions of LO and LA phonons in Eq.\,\eqref{qdeph} are assumed to be
$N^{(0)}_{Q,\gamma}=\{\exp[\hbar\omega_{Q}^{\gamma}/(k_BT_0)]-1\}^{-1}$ with $T_0$ as an initial environmental temperature for both $\gamma={\rm LA}$ and ${\rm LO}$ phonons.
Furthermore, we assume an initial thermal-equilibrium occupation functions for electrons by $f_{0}(\varepsilon_k,T_0)=\{1+\exp[(\varepsilon_k-u_0)/k_BT_0]\}^{-1}$ with $u_0(n_{\rm 1D},T_0)$ and $n_{\rm 1D}$ being the chemical potential and linear density of electrons, respectively.
\medskip

\begin{table}[htbp]
\centering
\caption{Electron-Phonon Coupling Parameters for GaAs}
\vspace{0.3cm}
\begin{tabular}{ccc}
\hline\hline
Parameter  &   $\ \ \ \ \ \ \ \ $   &   Value (Unit)\\
\hline
$\epsilon_{\infty}$  &   $\ \ \ \ \ \ \ \ $   &   $10.89$\\
$\epsilon_s$  &   $\ \ \ \ \ \ \ \ $   &   $12.90$\\
$D_0$  &   $\ \ \ \ \ \ \ \ $   &   $-9.3\,$eV\\
$d_{14}$ & $\ \ \ \ \ \ \ \ $   &   $-0.16\,$C/m$^2$\\
$h_{14}$ & $\ \ \ \ \ \ \ \ $   &   $1.2\times 10^9\,$V/m\\
$\rho_i$   &   $\ \ \ \ \ \ \ \ $   &   $5.3\,$g/cm$^{3}$\\
$a=d/2$   &   $\ \ \ \ \ \ \ \ $   &   $5.65\,$\AA\\
$m^*/m_0$   &   $\ \ \ \ \ \ \ \ $   &   $0.067$\\
$v_{\rm LA}$    &   $\ \ \ \ \ \ \ \ $   &   $4.78\times 10^5\,$cm/sec\\
$\bar{\Delta}_0$    &   $\ \ \ \ \ \ \ \ $   &   $2.0\,$nm\\
$\Lambda_0$    &   $\ \ \ \ \ \ \ \ $   &   $6\,$nm\\
$R_0$    &   $\ \ \ \ \ \ \ \ $   &   $50\,$nm\\
$n_{\rm 1D}$    &   $\ \ \ \ \ \ \ \ $   &   $10^6\,$cm$^{-1}$\\
\hline\hline
\end{tabular}
\label{tab-3}
\end{table}

For coupling between LA phonons and electrons in Eq.\,\eqref{eph}, we find\,\cite{r3} the coupling strength

\begin{equation}
\left|g_{\rm LA}(Q) \right|^2=\left(\frac{\hbar Q^2}{2\rho_i\omega_{Q}^{\rm LA}a^2|\epsilon_s(\vert Q\vert)|^2}\right)
\left[D_0^2+(eh_{14})^2\,\frac{9Q^2R_0^4}{(1+Q^2R_0^2)^4}\right]\ ,
\label{laph}
\end{equation}
where $D_0$ is the deformation-potential coefficient. For coupling between LO phonons and electrons, on the other hand, we use the Fl\"ohlich Hamiltonian model and get\,\cite{r2}

\begin{equation}
\left|g_{\rm LO}(Q) \right|^2 =\frac{\hbar\omega_{Q}^{\rm LO}}{2}\left(\frac{e^2}{2\pi\epsilon_0|\epsilon_s(\vert Q\vert)|^2}\right)\left(\frac{1}{\varepsilon_\infty}-\frac{1}{\varepsilon_s}\right)\,K_0(|Q|R_0)\ ,
\label{loph}
\end{equation}
where $K_0(x)$ is the modified Bessel function, while $\epsilon_\infty$ and $\epsilon_s$ are the optical and static dielectric constants, respectively. In Eq.\,\eqref{loph}, we will set $Q_{\rm min}=10^{-8}/R_0$ so that $K_0(|Q|R_0)=K_0(10^{-8})$ for $|Q|\leq Q_{\rm min}$ to eliminate an infinity, where $R_0$ is a cutting-off radius for electron-phonon interaction. Furthermore, the introduced static dielectric function $\epsilon_s(\vert Q\vert)$ in Eqs.\,\eqref{laph} and \eqref{loph} can be calculated within the random-phase approximation\,\cite{bookdhh} and given by

\begin{equation}
\epsilon_s(\vert Q\vert)=1-\frac{e^2m^*}{\epsilon_0\epsilon_r\pi^2\hbar^2|Q|}\,
\ln\left|\frac{\Omega_{-}(\vert Q\vert)}{\Omega_{+}(\vert Q\vert)}\right|K_0(|Q|R_0)\ ,
\label{rpa}
\end{equation}
where $\epsilon_r=(\epsilon_\infty+\epsilon_s)/2$, 
$\Omega_{\pm}(\vert Q\vert)=(\hbar|Q|/2m^*)\left(|Q|\pm 2k_F\right)$ with $k_F=\pi n_{\rm 1D}/2$ as the Fermi wave number and $n_{\rm 1D}$ represents the linear density of electrons. 
Here, all the required electron-phonon coupling parameters for GaAs material are listed in Table\,\ref{tab-3}. 

\subsection{Quantum-Kinetic Transport Equation for Electrons}
\label{add-1}

By choosing non-interacting electronic states as base functions for density-matrix operator, we introduce a quantum-kinetic transpor equation\,\cite{add-1} for electrons as

\begin{equation}
\label{boltz-1}
\frac{d\Delta\tilde{f}_k(t)}{dt}-\frac{e{\cal E}_{\rm DC}(t)}{\hbar}\,\frac{\partial\tilde{f}_k(t)}{\partial k}=\sum\limits_{\gamma}\,{\cal S}[\tilde{f}_k(t),N_\gamma(t)]
-\sum\limits_{\gamma}\,{\cal S}[f_0(\varepsilon_k,T_0),N^{(0)}_\gamma(T_0)]
\end{equation}
with $\tilde{f}_k(t)=\Delta\tilde{f}_k(t)+f_0(\varepsilon_k,T_0)$, $N_{Q}^{\gamma}(t)=\Delta N_{Q}^{\gamma}(t)+N^{(0)}_{Q,\gamma}(T_0)$ and 

\begin{equation}
\label{boltz-3}
{\cal S}[\tilde{f}_k(t),N_\gamma(t)]\equiv[1-\tilde{f}_k(t)]\,\Sigma^{({\rm in})}_k[\tilde{f}_k(t),N_\gamma(t)]-\tilde{f}_k(t)\,\Sigma^{({\rm out})}_k[\tilde{f}_k(t),N_\gamma(t)]\ ,
\end{equation}
where ${\cal E}_{\rm DC}(t)=\Delta{\cal E}\,\Theta^{(M)}_{\tau_0}(t)$ having a stepped strength $\Delta{\cal E}={\cal E}_0/M$. 
The initial condition is taken as $\tilde{f}_k(t=0)=f_0(\varepsilon_k,T_0)\equiv\left\{1+\exp\left[(\varepsilon_k-u_0)/k_BT_0\right]\right\}^{-1}$ at $T=T_0$ with a chemical potential $u_0(n_{\rm 1D},T_0)$, while $\Delta\tilde{f}_k(t)\equiv\tilde{f}_k(t)-f_0(\varepsilon_k,T_0)$ represents the non-equilibrium part of
occupation function for drifting electrons. At $T=0\,$K, we get $f_0(\varepsilon_k,T=0\,K)=\theta(k_F-|k|)$ with 
the Fermi wave number $k_F=\pi n_{\rm 1D}/2$ and the Fermi energy $E_F=\hbar^2k_F^2/2m^*$. 
In particular, solving Eq.\,\eqref{boltz-1} is limited by a conservation constraint in Eq.\,\eqref{constraint} below for the total number of electrons. 
Explicitly, we have 

\begin{equation}
n_{\rm 1D}=\frac{2}{\pi}\int\limits_{0}^{+\infty} \frac{dk}{1+\exp[(\varepsilon_k-u_0)/k_BT_0]}\ ,
\label{fermi}
\end{equation}
which can be used to evaluate the chemical potential $u_0(n_{\rm 1D},T_0)$ as functions of both initial temperature $T_0$ and fixed linear density $n_{\rm 1D}$ for doped electrons. Moreover, the energy-relaxation time $\tau^{(e)}_{k,\gamma}(t)$ of electrons can be calculated by using 

\begin{equation}
\frac{1}{\tau^{(e)}_{k}(t)}=\sum\limits_{\gamma}\,\Sigma^{({\rm in})}_k[\tilde{f}_k(t),N_\gamma(t)]+\sum\limits_{\gamma}\,\Sigma^{({\rm out})}_k[\tilde{f}_k(t),N_\gamma(t)]
\label{e-relax}
\end{equation}
\medskip

\begin{table}[htbp]
\centering
\caption{Chain Transport Parameters}
\vspace{0.3cm}
\begin{tabular}{ccc}
\hline\hline
Parameter  &   $\ \ \ \ \ \ \ \ $   &   Value (Unit)\\
\hline
$T_0$   &   $\ \ \ \ \ \ \ \ $   &   $150\,$K\\
$M$   &   $\ \ \ \ \ \ \ \ $   &   $5$\\
$\Delta{\cal E}$   &   $\ \ \ \ \ \ \ \ $   &   $20\,$V/cm\\
$\tau_0$   &   $\ \ \ \ \ \ \ \ $   &   $5\,$ps\\
$\gamma_\tau$   &   $\ \ \ \ \ \ \ \ $   &   $0.2\,$ps\\
$\Gamma_0$   &   $\ \ \ \ \ \ \ \ $   &   $0.5\,$ps\\
$n_{\rm 1D}$   &   $\ \ \ \ \ \ \ \ $   &   $1\times 10^6\,$cm$^{-1}$\\
${\cal R}$  &   $\ \ \ \ \ \ \ \ $   &   $5\,$nm\\
$\hbar\Gamma^{\rm LO}_{\rm ph}$   &   $\ \ \ \ \ \ \ \ $   &   $1\,$meV\\
$\hbar\Gamma_{\rm e}$   &   $\ \ \ \ \ \ \ \ $   &   $1\,$meV\\
$\hbar\Gamma^{\rm LA}_{\rm ph}$   &   $\ \ \ \ \ \ \ \ $   &   $1\,$meV\\
$\hbar\omega_0$   &   $\ \ \ \ \ \ \ \ $   &   $100\,$meV\\
$T_L$   &   $\ \ \ \ \ \ \ \ $   &   $125\,$K\\
$T_R$   &   $\ \ \ \ \ \ \ \ $   &   $175\,$K\\
$p_s$    &   $\ \ \ \ \ \ \ \ $   &   $100\,$Pa\\
$t^*$    &   $\ \ \ \ \ \ \ \ $   &   $10\,$ps\\
\hline\hline
\end{tabular}
\label{tab-4}
\end{table}

From Eqs.\,\eqref{boltz-1} and \eqref{boltz-3}, we find that the total electron scattering-in and scattering-out rates with $\tilde{f}_k(t)=\Delta\tilde{f}_k(t)+f_0(\varepsilon_k,T_0)$ are given by

\begin{eqnarray}
\nonumber
&&\sum\limits_{\gamma}\,\Sigma^{({\rm in})}_k[\tilde{f}_k(t),N_\gamma(t)]\equiv
\frac{1}{2\pi\hbar}\int\limits_{-\infty}^{\infty}\int\limits_{-\infty}^{\infty}\int\limits_{-\infty}^{\infty} dk_2\,dk_3\,dk_4\,\left|V^{\rm ee}_{k,k_2;k_3,k_4}\right|^2[1-\tilde{f}_{k_2}(t)]\tilde{f}_{k_3}(t)\tilde{f}_{k_4}(t)\, \\
\nonumber
&\times&{\cal L}_0\left(\varepsilon_{k}+\varepsilon_{k_2}-\varepsilon_{k_3}-\varepsilon_{k_4},\,\hbar\Gamma_{\rm e}\right)+\frac{d}{\hbar}\,\sum\limits_{\gamma=LA,LO}\int\limits_{-\pi/d}^{\pi/d} dQ\,\left|g_\gamma(Q)\right|^2\theta_0(\hbar\omega_Q^\gamma-\hbar Qv_d(t))\,\left\{N_Q^\gamma(t)\,\tilde{f}_{k-Q}(t)\right.\\
\label{in-rate}
&\times&\left.{\cal L}_0\left(\varepsilon_k-\varepsilon_{k-Q}-\hbar\omega_Q^\gamma+\hbar Qv_d(t),\,\hbar\Gamma_{\rm e}\right)\right.
+\left.\left[N_Q^\gamma(t)+1\right]\tilde{f}_{k+Q}(t)\,{\cal L}_0\left(\varepsilon_k-\varepsilon_{k+Q}+\hbar\omega_Q^\gamma-\hbar Qv_d(t),\,\hbar\Gamma_{\rm e}\right)\right\}\ ,\ \ \ \ 
\end{eqnarray}
and

\begin{eqnarray}
\nonumber
&&\sum\limits_{\gamma}\,\Sigma^{({\rm out})}_k[\tilde{f}_k(t),N_Q(t)]\equiv
\frac{1}{2\pi\hbar}\int\limits_{-\infty}^{\infty}\int\limits_{-\infty}^{\infty}\int\limits_{-\infty}^{\infty} dk_2\,dk_3\,dk_4\,\left|V^{\rm ee}_{k_4,k_2;k_3,k}\right|^2[1-\tilde{f}_{k_2}(t)]\tilde{f}_{k_3}(t)[1-\tilde{f}_{k_4}(t)]\,\\
\nonumber
&\times&{\cal L}_0\left(\varepsilon_{k_4}+\varepsilon_{k_2}-\varepsilon_{k_3}-\varepsilon_{k},\,\hbar\Gamma_{\rm e}\right)
+\frac{d}{\hbar}\,\sum\limits_{\gamma=LA,LO}\int\limits_{-\pi/d}^{\pi/d} dQ\,\left|g_\gamma(Q)\right|^2\theta_0(\hbar\omega_Q^\gamma-\hbar Qv_d(t))\,
\left\{N_Q^\gamma(t)\left[1-\tilde{f}_{k+Q}(t)\right]\,\right.\\
\label{out-rate}
&\times&{\cal L}_0\left(\varepsilon_k-\varepsilon_{k+Q}+\hbar\omega_Q^\gamma-\hbar Qv_d(t),\,\hbar\Gamma_{\rm e}\right)+\left.\left[N_Q^\gamma(t)+1\right]\left[1-\tilde{f}_{k-Q}(t)\right]\,
{\cal L}_0\left(\varepsilon_k-\varepsilon_{k-Q}-\hbar\omega_Q^\gamma+\hbar Qv_d(t),\,\hbar\Gamma_{\rm e}\right)\right\}\ ,\ \ \ \ \ \ 
\end{eqnarray}
where both electron-electron and electron-phonon interactions have been taken into considerations, 
while $g_{\rm LA}(Q)$ and $g_{\rm LO}(Q)$ are given by Eqs.\,\eqref{laph} and \eqref{loph}, respectively.
The integrals in Eqs.\,\eqref{in-rate} and \eqref{out-rate} exclude the terms with $Q=0$ as well as $k_2=k_3$ and $k_4=k$. 
Mathematically, we need maintain $\Delta k=\Delta Q$ so as to calculate electron-phonon interaction accurately while numerically solving Eq.\,\eqref{boltz-1} and Eq.\,\eqref{qdeph} simultaneously. 
Moreover, the element of introduced Coulomb-interaction matrix in Eqs.\,\eqref{in-rate} and \eqref{out-rate} is calculated as

\begin{equation}
\left|V^{\rm ee}_{k_1,k_2;k_3,k_4}\right|^2=\delta(k_1+k_2-k_3-k_4)\left|\frac{e^2}{2\pi\epsilon_0\epsilon_r\epsilon_s(\vert k_1-k_4\vert)}\right|^2\,K^2_0(|k_1-k_4|R_0)\ ,
\end{equation}
where $K_0(x)$ is the modified Bessel function of the third kind, $R_0$ is the cutting-off radius, and cutoff for the modified Bessel function $K_0(x)$ is set to be $x\sim 10^{-8}$.
\medskip

In addition, the drift velocity $\mbox{\boldmath$v$}_d(t)$ introduced in Eqs.\,\eqref{boltz-1}$-$\eqref{out-rate} is calculated by

\begin{equation}
v_d(t)=\frac{1}{\pi\hbar n_{\rm 1D}}\int\limits_{-\infty}^{\infty} dk\,\left(\frac{d\varepsilon_k}{dk}\right)\,\Delta\tilde{f}_k(t)\ ,
\label{boltz-2}
\end{equation}
and the transient electron mobility $\mu_{\rm e}(t)$ can be evaluated by $\mu_{\rm e}(t)=\partial v_d(t)/\partial{\cal E}_{\rm DC}(t)$ while the transient electrical current $I_{\rm e}(t)$ is given by $I_{\rm e}(t)=-n_{\rm 1D}\,ev_d(t)$. Here, all the required parameters for transport of electrons in a chain are listed in Table\,\ref{tab-4}.

\subsection{Effective Temperature of Non-Equilibrium Electrons}
\label{add-2}

The change of average thermal energy $\overline{U}_{\rm e}(t)$ per electron can be calculated from

\begin{equation}
\Delta\overline{U}_{\rm e}(t)=\frac{1}{\pi n_{\rm 1D}}\int\limits_{-\infty}^{\infty} dk\,\varepsilon_k\,\Delta\tilde{f}_k(t)\ ,
\label{ethermo}
\end{equation}
from which we define the effective temperature $T_{\rm eff}^{({\rm e})}(t)$ in the high-temperature limit for electrons through the relation

\begin{equation}
\Delta\overline{U}_{\rm e}(t)=\frac{1}{2}\,k_B\left[T_{\rm eff}^{({\rm e})}(t)-T_0\right]\ ,
\label{capacity}
\end{equation}
where $T_0$ is the initial environmental temperature.

\subsection{Effective Temperatures of Non-Equilibrium LA and LO Phonons}
\label{sec-2-5}

After having calculated the non-equilibrium part $\Delta N_{Q}^{\gamma}(t)$ of occupation functions for $\gamma$ phonons, we can write down a dynamical equation on the change of average thermal energy
$\Delta\overline{U}_{\rm LA}(t)$ for non-thermal LA phonons, {\em i.e.\/}

\begin{equation}
\frac{d\Delta\overline{U}_{\rm LA}(t)}{dt}=\frac{d}{2\pi}\int\limits_{-\pi/d}^{\pi/d} dQ\,\hbar\omega_{Q}^{\rm LA}\,\frac{d\Delta N_{Q}^{\rm LA}(t)}{dt}\ ,
\label{efft}
\end{equation}
where the rate $d\Delta N_{Q}^{\rm LA}(t)/dt$ can be evaluated from the quantum-kinetic equation in Eq.\,\eqref{qdeph}.
By using the Debye model at low temperatures, Eq.\,\eqref{efft} leads to another dynamical equation for an effective LA-phonon temperature $T^{\rm LA}_{\rm eff}(t)$, given by

\begin{equation}
\frac{d}{dt}\left\{\left[T^{\rm LA}_{\rm eff}(t)\right]^4\right\}=4\left[T^{\rm LA}_{\rm eff}(t)\right]^3\,\frac{dT^{\rm LA}_{\rm eff}(t)}{dt}
=\left[\frac{T_0^4}{\overline{U}^{(0)}_{\rm LA}(T_0)}\right]\frac{d\Delta\overline{U}_{\rm LA}(t)}{dt}=\frac{T_0^4d}{2\pi\overline{U}^{(0)}_{\rm LA}(T_0)}\,\int\limits_{-\pi/d}^{\pi/d} dQ\,\hbar\omega_{Q}^{\rm LA}\,\frac{d\Delta N_{Q}^{\rm LA}(t)}{dt}\ ,
\label{efft2}
\end{equation}
where $T_0$ is the initial lattice temperature. Moreover, the initial average thermal energy $\overline{U}^{(0)}_{\rm LA}(T_0)$ introduced in Eq.\,\eqref{efft2} is calculated as

\begin{equation}
\overline{U}^{(0)}_{\rm LA}(T_0)=\frac{d}{2\pi}\int\limits_{-\pi/d}^{\pi/d} dQ\,\hbar\omega_{Q}^{\rm LA}\,N_{Q,{\rm LA}}^{(0)}(T_0)\ ,
\label{efft3}
\end{equation}
where $N_{Q,{\rm LA}}^{(0)}(T_0)$ is the initial thermal-equilibrium distribution of LA-phonons at temperature $T_0$, and 

\begin{equation}
n_0^{\rm LA}(T_0)=\frac{d}{2\pi}\int\limits_{-\pi/d}^{\pi/d} dQ\,N_{Q,{\rm LA}}^{(0)}(T_0)
\label{num-1}
\end{equation}
is the total number of LA phonons at initial temperature $T_0$.
\medskip

In a similar way, we can write down a dynamical equation on the change of average thermal energy $\overline{U}_{\rm LO}(t)$ for non-thermal LO phonons, {\em i.e.\/}

\begin{equation}
\frac{d\Delta\overline{U}_{\rm LO}(t)}{dt}=\hbar\omega_0\left(\frac{d}{2\pi}\right)\int\limits_{-\pi/d}^{\pi/d} dQ\,\frac{d\Delta N_{Q}^{\rm LO}(t)}{dt}\ ,
\label{einst}
\end{equation}
where $\omega_0$ is a $Q$-averaged frequency with respect to a very narrow LO-phonon band,
and the rate $d\Delta N_{Q}^{\rm LO}(t)/dt$ can still be evaluated by applying the quantum-kinetic equation in Eq.\,\eqref{qdeph}.
Therefore, by using the Einstein model, we arrive at
another dynamical equation for effective LO-phonon temperature $T^{\rm LO}_{\rm eff}(t)$, yielding

\begin{eqnarray}
\nonumber
\frac{d}{dt}\left\{\exp\left[\frac{\hbar\omega_0}{k_BT_{\rm eff}^{\rm LO}(t)}\right]\right\}&=&
\frac{1}{\overline{U}^{(0)}_{\rm LO}(T_0)}\left[\exp\left(\frac{\hbar\omega_0}{k_BT_0}\right)-1\right]
\frac{d\Delta\overline{U}_{\rm LO}(t)}{dt}\\
\label{emod}
&=&\frac{\hbar\omega_0d}{2\pi \overline{U}^{(0)}_{\rm LO}(T_0)}\left[\exp\left(\frac{\hbar\omega_0}{k_BT_0}\right)-1\right]\,
\int\limits_{-\pi/d}^{\pi/d} dQ\,\frac{d\Delta N_{Q}^{\rm LO}(t)}{dt}\ ,
\end{eqnarray}
where the initial average thermal energy $\overline{U}^{(0)}_{\rm LO}(T_0)$ is given by

\begin{equation}
\overline{U}^{(0)}_{\rm LO}(T_0)=\hbar\omega_0\left(\frac{d}{2\pi}\right)\int\limits_{-\pi/d}^{\pi/d} dQ\,N_{Q,{\rm LO}}^{(0)}(T_0)\ ,
\label{efft4}
\end{equation}
$N_{Q,{\rm LO}}^{(0)}(T_0)$ is the initial thermal-equilibrium distribution of LO-phonons at temperature $T_0$, and 

\begin{equation}
n_0^{\rm LO}(T_0)=\frac{d}{2\pi}\,\int\limits_{-\pi/d}^{\pi/d} dQ\,N_{Q,{\rm LO}}^{(0)}(T_0)
\label{num-2}
\end{equation}
is the total number of LO phonons at initial temperature $T_0$.

\subsection{Electron-Diffusion and Phonon-Drag Thermoelectric Powers}
\label{sec-2-6}

In the presence of a spatially-uniform DC electric field, electrons in a chain are driven by DC field in the position space with a drifting velocity $v_{\rm d}$.
Meanwhile, the random scattering motions of these conduction electrons by lattice phonons can be described based on Boltzmann-type collisions,\,\cite{r4}
leading to a non-equilibrium occupation function $\tilde{f}_k(t)$ for drifting electrons. The conservation of electron number requires

\begin{equation}
n_{\rm 1D}=\frac{1}{\pi}\int\limits^{\infty}_{-\infty} dk\,\left\{1+\exp\left[\frac{\varepsilon_k-u_0}{k_BT_0}\right]\right\}^{-1}
=\frac{1}{\pi}\int\limits_{-\infty}^{\infty} dk\,\tilde{f}_k(t)\ ,
\label{chemi}
\end{equation}
which implies that 

\begin{equation}
\int\limits_{-\infty}^{\infty} dk\,\Delta\tilde{f}_k(t)=\Delta k\sum\limits_{j=-N}^{N}\,\Delta\tilde{f}(k_j,t)\equiv 0\ ,
\label{chemi-1}
\end{equation}
where $|k_j|\leq k_{\rm max}=k_N$ is set to cut off the range of integral, $\Delta k=k_{\max}/N$ and $k_j=j\Delta k$ with $|j|\leq N$. 
This conservation of the total number of electrons imposes a constraint for the middle point $k_0=0$, given by 

\begin{equation}
\Delta\tilde{f}(k_0,t)\equiv -\sum\limits_{j=-N}^{-1}\,\Delta\tilde{f}(k_j,t)-\sum\limits_{j=+1}^{N}\,\Delta\tilde{f}(k_j,t)\ ,
\label{constraint}
\end{equation}
which can be further utilized in numerically evaluating $d\Delta\tilde{f}(k_j,t)/dt=[\Delta\tilde{f}(k_j,t+\Delta t)-\Delta\tilde{f}(k_j,t)]/\Delta t\propto\partial\tilde{f}(k_j,t)/\partial k=[\tilde{f}(k_{j+1},t)-\tilde{f}(k_j,t)]/\Delta k$ 
for all other points $k=k_j$ in Eq.\,\eqref{boltz-1} with $-N\leq j\leq -1$ or $+1\leq j\leq N$ except for $j=0$. Meanwhile, we further set 
$\Delta\tilde{f}(k_{-(N+1)},t)=\Delta\tilde{f}(k_{N+1},t)\equiv 0$ and $f_0(k_{-(N+1)})=f_0(k_{N+1})\equiv 0$ for a large enough $k_{\rm max}$ value. As a last step, we obtain at the new time $t+\Delta t$ for $j=0$ that 

\begin{equation}
\Delta\tilde{f}(k_0,t+\Delta t)\equiv -\sum\limits_{j=-N}^{-1}\,\Delta\tilde{f}(k_j,t+\Delta t)-\sum\limits_{j=+1}^{N}\,\Delta\tilde{f}(k_j,t+\Delta t)\ .
\end{equation}
\medskip

Moreover, in Eq.\,\eqref{chemi}, $T_0$ is an electron temperature and $u_0$ is an electron chemical potential determined by the constraint for conservation of number of electrons.
Physically, if we set $T_0=0\,$ in Eq.\,\eqref{chemi}, we get $k_F=\pi n_{\rm 1D}/2$ for the Fermi wave number and $u_0$ becomes the Fermi energy $E_F=\hbar^2k_F^2/2m^*$.
\medskip

The heat-current density ${\cal Q}_{\rm d}(t)$ due to electron diffusion is given, accounting for the spin degeneracy, by\,\cite{r3}

\begin{equation}
{\cal Q}_{\rm d}(t)=\frac{1}{\pi^2\hbar{\cal R}^2}\int\limits_{-\infty}^{\infty} dk\,\left(\varepsilon_k-u_0\right)\,\frac{\partial\varepsilon_k}{\partial k}\,\Delta\tilde{f}_k(t)\ ,
\label{diff}
\end{equation}
and the electron-diffusion thermoelectric power is then calculated according to ${\cal S}_{\rm d}(t)={\cal Q}_{\rm d}(t)/T_0J_{\rm e}(t)<0$ in unit of $\mu V/K$,
where $J_{\rm e}(t)=-en_{\rm 1D}\,v_{\rm d}(t)/\pi{\cal R}^2>0$ is the electrical-current density due to DC transport of electrons along the chain, and $\sigma_{\rm e}(t)=\partial J_{\rm e}(t)/\partial{\cal E}_{\rm DC}(t)\equiv-en_{\rm 1D}\mu_{\rm e}(t)/\pi{\cal R}^2$ is the transient electrical conductivity of electrons with electron mobility $\mu_{\rm e}(t)$. 
Here, ${\cal Q}_{\rm d}(t)\neq 0$ can be maintained even with ${\cal E}_{\rm DC}(t)=0$ under the condition of ${\cal F}^\gamma_{\rm th}(t)\propto\alpha_0\neq 0$ due to phonon-electron drag or 
$\Delta\tilde{f}_k(t)\neq\Delta\tilde{f}_{-k}(t)$.  
\medskip

On the other hand, the heat-current density ${\cal Q}_{\rm ph}^\gamma(t)$ due to phonon transport is calculated as\,\cite{r3}

\begin{equation}
{\cal Q}_{\rm ph}^\gamma(t)=\frac{1}{2\pi^2{\cal R}^2}\,
\int\limits_{-\pi/d}^{\pi/d} dQ\,\hbar\omega_{Q}^{\gamma}
\left(\frac{\partial\omega_{Q}^{\gamma}}{\partial Q}\right)\,\Delta N_{Q}^{\gamma}(t)\ ,
\label{drag}
\end{equation}
and the phonon-drag thermoelectric power under the conditions of $\alpha_0\equiv\Delta T/{\cal L}=0$ and ${\cal P}^{(\tau_0)}_{\rm b}(t)=0$ can be calculated
by using the definition ${\cal S}_{\rm ph}^\gamma(t)={\cal Q}_{\rm ph}^\gamma(t)/T_0J_{\rm e}(t)$ in unit of $\mu V/K$, which is negative for $\gamma=LA$ but positive for $\gamma=LO$.
In a similar way, under the condition of ${\cal E}_{\rm DC}(t)=0$ or $J_{\rm e}(t)=0$, we get the thermal conductivity $\kappa_{\rm ph}^\gamma(t)=|{\cal Q}_{\rm ph}^\gamma(t)/\alpha_0|$ for $\gamma=LA,\,LO$, given by

\begin{equation}
\kappa_{\rm ph}^\gamma(t)=\left|\frac{{\cal Q}_{\rm ph}^\gamma(t)}{\alpha_0}\right|\ ,
\label{coeff}
\end{equation}
where $\alpha_0=\Delta T/{\cal L}=(T_R-T_L)/{\cal L}$ is the spatially-uniform temperature gradient applied to lattices by two external thermal-equilibrium reservoirs with different temperatures $T_L,\,T_R$ at two ends of a lattice chain having a length ${\cal L}$. 
For $\gamma=LA$, we expect from Eq.\,\eqref{drag} a significant heat-current density flowing from the right end of a chain towards its left end. 
For $\gamma=LO$, there also exists a heat-current density flowing from the right high-$T_R$ to the left low-$T_L$ ends. However, it becomes insignificant and can be neglected because the energy dispersion of $LO$-phonons is very small. 
Here, ${\cal Q}^\gamma_{\rm ph}(t)\neq 0$ can still be retained even with ${\cal F}^\gamma_{\rm th}(t)\propto\alpha_0=0$ under the condition of ${\cal E}_{\rm DC}(t)\neq 0$ due to electron-phonon drag or $\Delta N_{Q}^{\gamma}(t)\neq\Delta N_{-Q}^{\gamma}(t)$.  

\section{Numerical Results and Discussions}
\label{sec-3}

In Section\,\ref{sec-3}, we will present numerical results, including 
effect of surface-roughness scattering in Subsection\,\ref{sec-3-0}, 
effect of anharmonic phonon scattering in Subsection\,\ref{sec-3-1}, 
time evolution of $q$-dependent non-equilibrium phonon distribution in Subsection\,\ref{sec-3-2},
Doppler effect from moving electrons on phonon scattering in Subsections\,\ref{sec-3-3}$-$\ref{sec-3-6}, and 
heat transport of phonons and electrons in Subsection\,\ref{sec-3-7}. 
Moreover, all the parameters used in our numerical calculations are listed in Tables\,\ref{tab-1}$-$\ref{tab-4}. 
\medskip 

\subsection{Surface-Scattering Rate of Phonons}
\label{sec-3-0}

\begin{figure}[htbp]
\centering
\includegraphics[width=0.85\textwidth]{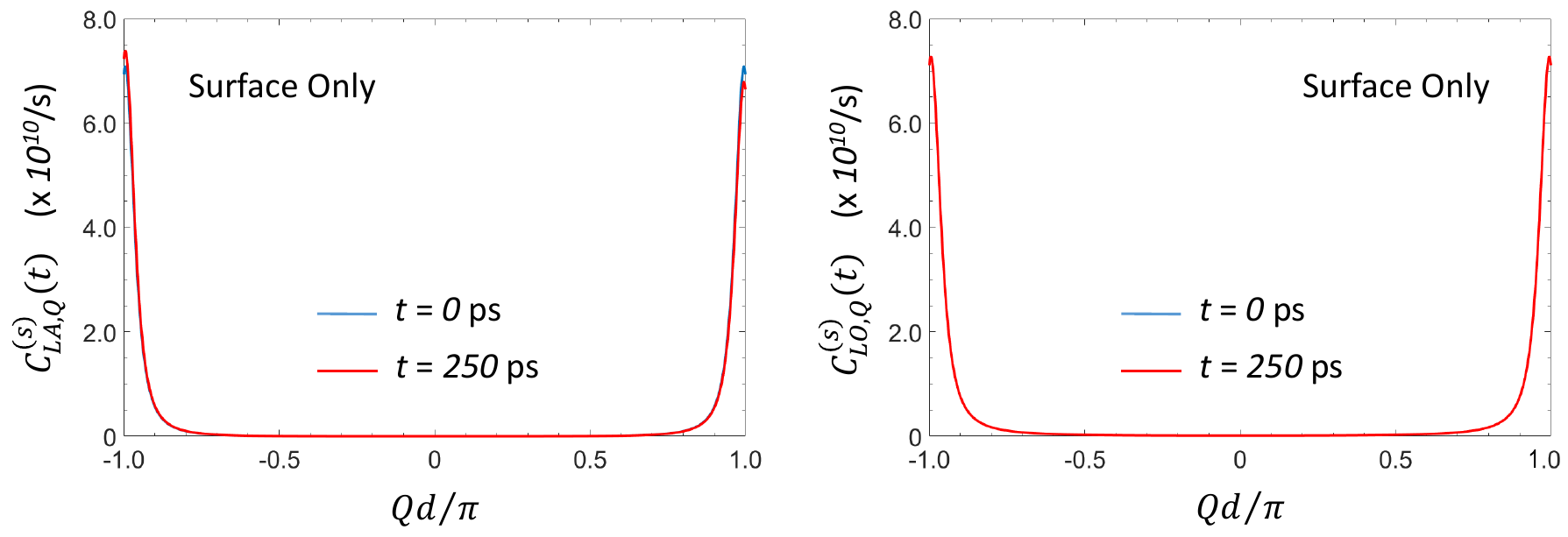}
\caption{Numerically computed creation rates $C^{(s)}_{LA,Q}(t)$ and $C^{(s)}_{LO,Q}(t)$ for surface-phonon scattering as functions of scaled wave number $Qd/\pi$ at $t=0$ and $250\,$ps in the left and right panels, respectively.}
\label{nfig1}
\end{figure}

Calculated force-driven phonon creation rates $C^{(s)}_{LA,Q}(t)$ and $C^{(s)}_{LO,Q}(t)$ 
for the surface-roughness scattering process 
at $t=0$ and $t=250\,$ps are presented in Fig.\,\ref{nfig1} for comparison. 
From Fig.\,\ref{nfig1}, we find similar $Q$-dependence for very strong surface scatterings of force-driven $LA$ and $LO$ phonons under an applied thermal force ${\cal F}^\gamma_{\rm th}(t)$ in Eq.\,\eqref{qdeph}, which is associated with phonon creations as formulated in Eq.\,\eqref{rough-0}. All features in two panels occur around the Brillouin-zone boundaries at $Q=\pm\pi/d$ for both $LA$ and $LO$ phonons within a lattice chain, which reflects the fact that strong surface scattering of phonons requires a large phonon wave number $Q$ and implies that they will play a major role in the energy-momentum relaxation of non-equilibrium $LA$ and $LO$ phonons. 
Moreover, the transient variation of surface-scattering rate for $LA$ and $LO$ phonons appears negligible by directly comparing two results at $t=0$ and $t=250\,$ps in both panels.
This is due to the small phonon-occupation changes compared to the total phonon numbers in the thermal equilibrium. 
\medskip

\subsection{$A$- and $B$-Type Anharmonic Phonon-Phonon Scattering Rates}
\label{sec-3-1}

\begin{figure}[htbp]
\centering
\includegraphics[width=0.85\textwidth]{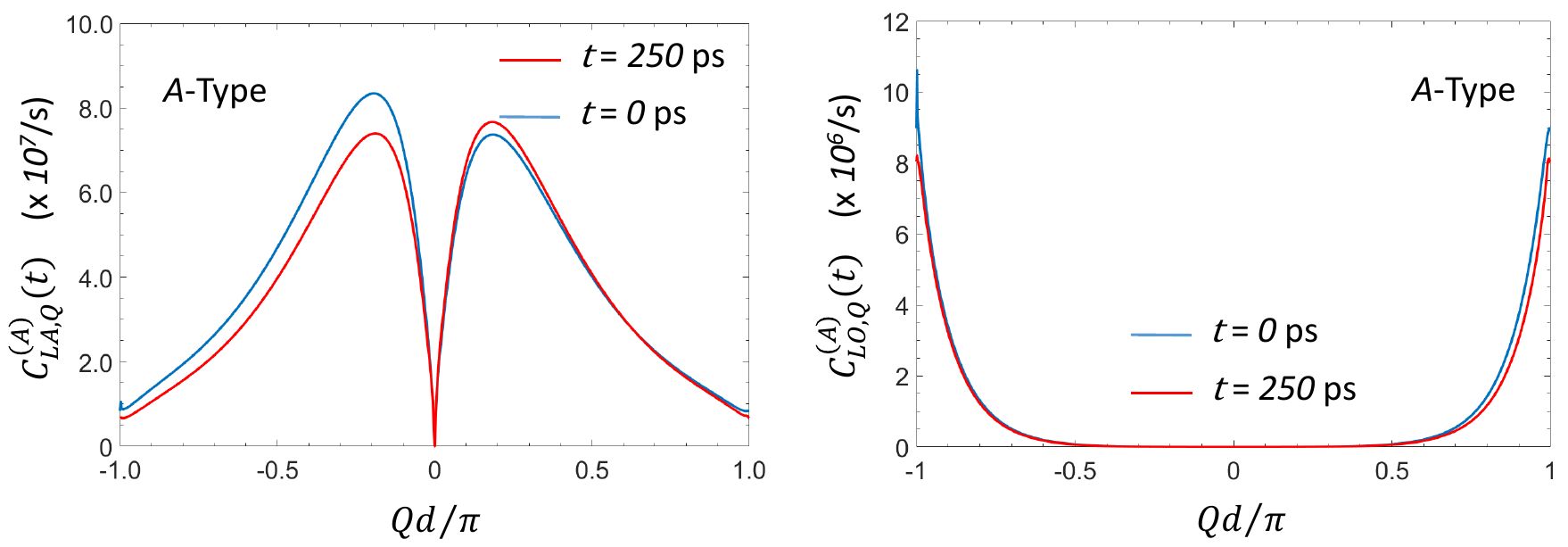}
\caption{Numerically computed creation rates $C^{(A)}_{LA,Q}(t)$ and $C^{(A)}_{LO,Q}(t)$ for $A$-type anharmonic-phonon scattering as functions of scaled wave number $Qd/\pi$ at $t=0$ and $250\,$ps in the left and right panels, respectively.}
\label{nfig2}
\end{figure}

\begin{figure}[htbp]
\centering
\includegraphics[width=0.85\textwidth]{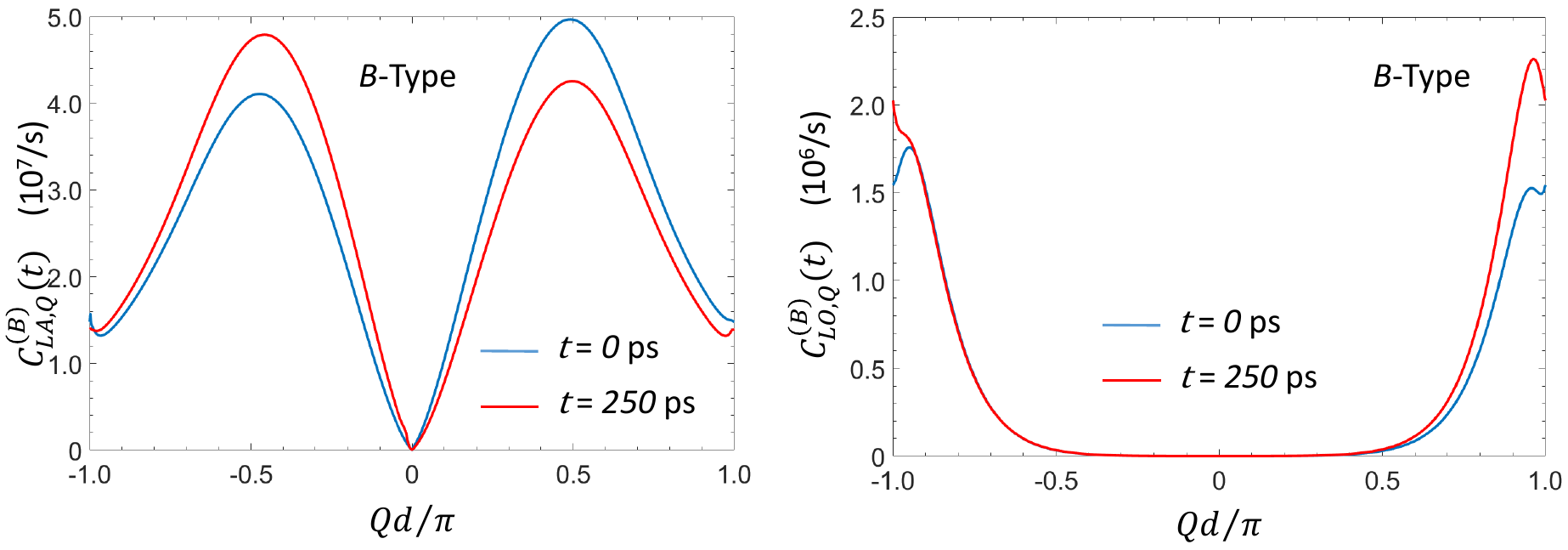}
\caption{Numerically computed creation rates $C^{(B)}_{LA,Q}(t)$ and $C^{(B)}_{LO,Q}(t)$ for $B$-type anharmonic-phonon scattering as functions of scaled wave number $Qd/\pi$ at $t=0$ and $250\,$ps in the left and right panels, respectively.}
\label{nfig3}
\end{figure}

Physically, as the lattice potential is expanded over the third-order of displacement, the three-phonon anharmonic phonon-phonon interaction shows up, and the use of harmonic approximation can no longer be justified. From the consideration of energy-momentum conservation for three-phonon anharmonic interactions, there exist only two distinct processes as illustrated in Fig.\,\ref{fig14}, where either one phonon ($A$-process) or two phonons ($B$-process) are annihilated, while other processes involving either creating or annihilating all three phonons will be ignored in this study. Moreover, if only one longitudinal-acoustic $(LA)$ and one longitudinal-optical $(LO)$ phonon modes\,\cite{callaway} are taken into consideration, there are totally eight contributions for each one of these two process. Here, the calculated three-phonon creation rates for both $A$ and $B$ processes at two time moments are presented and compared in Figs.\,\ref{nfig2} and \ref{nfig3}, respectively.
\medskip

First, for the $A$-type anharmonic scattering process, as illustrated in Fig.\,\ref{fig14} and displayed in the left panel of Fig.\,\ref{nfig2}, we observe that the obtained three-phonon creation rate $C^{(A)}_{LA,Q}(t)$ for force-driven $LA$ phonons is modified in the whole range of the lattice Brillouin zone at $t=250\,$ps in comparison with the initial thermal-equilibrium one at time $t=0$. 
In this case, $C^{(A)}_{LA,Q}(t)$ is greatly reduced in the range of $Q<0$ but is only slightly increased in the $Q>0$ range at the same time. For force-driven $LO$ phonons, on the other hand, it is one order of magnitude smaller than  $C^{(A)}_{LA,Q}(t)$ and no visible change in their creation rate $C^{(A)}_{LO,Q}(t)$ is found except very close to two Brillouin-zone boundaries at $Q=\pm\pi/d$.
\medskip

Next, for the $B$-type anharmonic scattering process, as illustrated in Fig.\,\ref{fig15} and presented in Fig.\,\ref{nfig3}, however, the opposite features in time variation of $LA$-phonon creation rate $C^{(B)}_{LA,Q}(t)$ are found by comparing it with $C^{(A)}_{LA,Q}(t)$ in Fig.\,\ref{nfig2}. Here, the enlarged variation in $C^{(B)}_{LA,Q}(t)$ appears in the whole range of $Q$. For $LO$ phonons, however, 
$C^{(B)}_{LO,Q}(t)$ is still one order of magnitude smaller than  $C^{(B)}_{LA,Q}(t)$ and no significant change in $C^{(B)}_{LO,Q}(t)$ shows up except in regions very close to two boundaries $Q=\pm\pi/d$ of the Brillouin zone.
\medskip

\subsection{Transient Non-Equilibrium Distributions of $LA$ and $LO$ Phonons}
\label{sec-3-2}

\begin{figure}[htbp]
\centering
\includegraphics[width=0.95\textwidth]{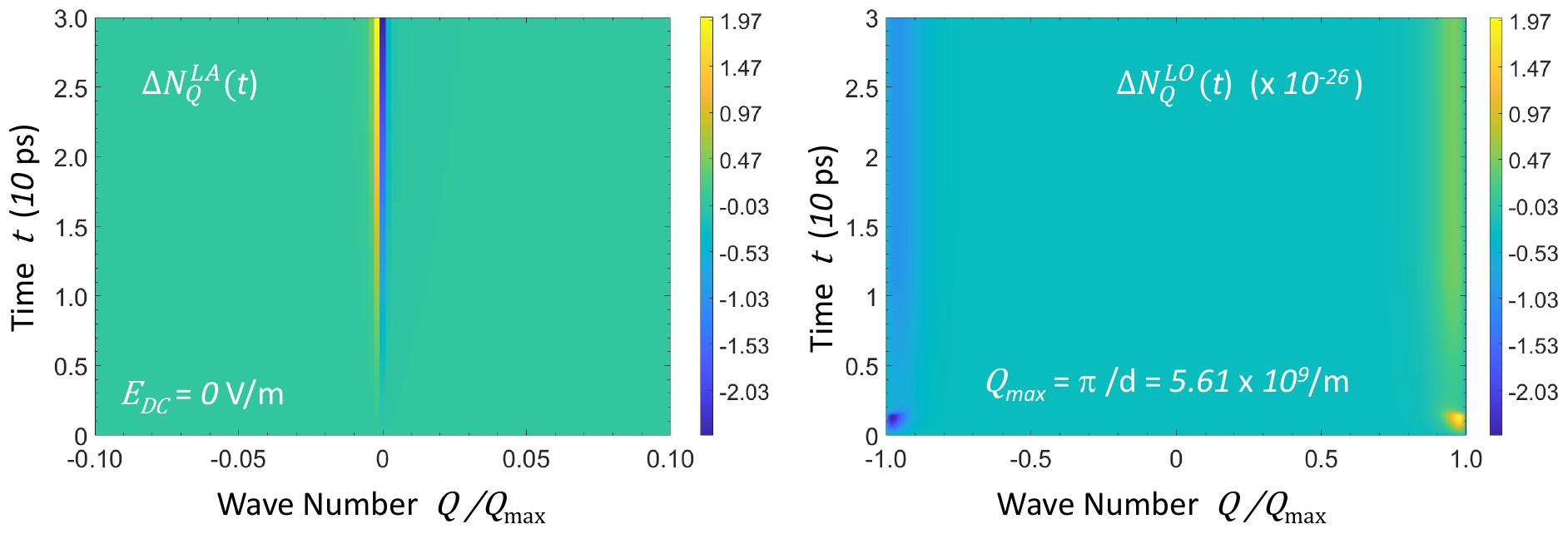}
\caption{Numerically computed time evolution of force-driven non-equilibrium parts $\Delta N^{\rm LA}_Q(t)$ for $LA$-phonons and  $\Delta N^{\rm LO}_Q(t)$ for $LO$-phonons as functions of scaled wave number $Q/Q_{\rm max}$ in the left and right panels, respectively, under $E_{\rm DC}=0$.}
\label{nfig4}
\end{figure}

\begin{figure}[htbp]
\centering
\includegraphics[width=0.85\textwidth]{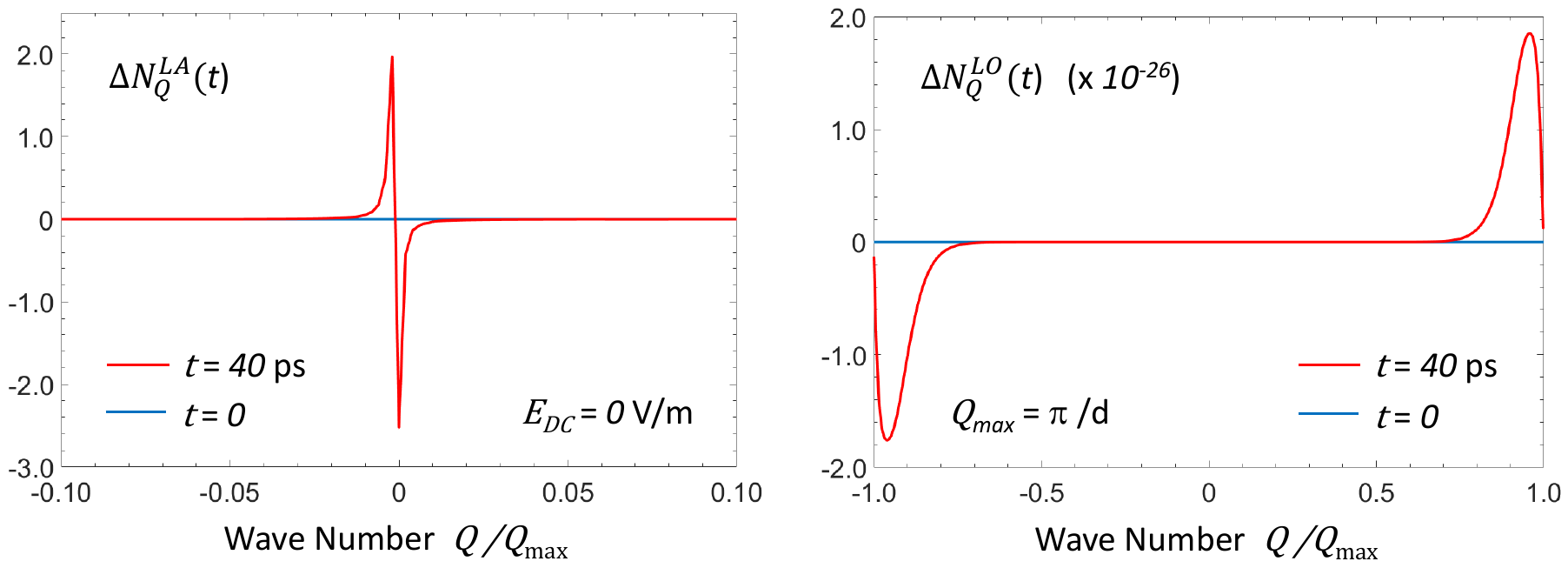}
\caption{Selected comparisons for numerically computed force-driven $\Delta N^{\rm LA}_Q(t)$ for $LA$-phonons and  $\Delta N^{\rm LO}_Q(t)$ for $LO$-phonons as functions of scaled wave number $Q/Q_{\rm max}$ in the left and right panels, respectively, at $t=0$ and $40\,$ps and under $E_{\rm DC}=0$.}
\label{nfig5}
\end{figure}

Our numerically computed time evolution of $\Delta N_Q^{\rm LA}(t)$ and $\Delta N_Q^{\rm LO}(t)$ for non-equilibrium parts of $LA$- and $LO$-phonon distributions are presented in Figs.\,\ref{nfig4} and \ref{nfig5}, respectively, under the condition of $E_{\rm DC}=0$. In particular, comparisons of $\Delta N_Q^{\rm LA}(t)$ and $\Delta N_Q^{\rm LO}(t)$ at $t=0$ and $t=40\,$ps are made in two panels of Fig.\,\ref{nfig5}. 
\medskip

As $E_{\rm DC}=0$, electrons remain in their thermal-equilibrium state except for dragging by force-driven phonons in the system. After a thermal field ${\cal F}_{\rm th}^\gamma(t)$ has been applied to phonons, as described in Eq.\,\eqref{qdeph}, a significant amount of non-equilibrium $LA$ phonons, denoted as $\Delta N_Q^{\rm LA}(t)$, develops  around the zone center at $Q=0$. This includes both phonon absorption with $\Delta N_Q^{\rm LA}(t)<0$ for $Q\gtrapprox 0$, as well as phonon emission $\Delta N_Q^{\rm LA}(t)>0$ for $Q\lessapprox 0$, as reflected in left panels of Figs.\,\ref{nfig4} and \ref{nfig5}. For $LO$ phonons, on the other hand, their finite non-equilibrium part $\Delta N_Q^{\rm LO}(t)$ is extremely small and limited only to zone edges at $Q=\pm\pi/d$, where non-equilibrium phonon absorption and emission are seen for $Q\gtrapprox -\pi/d$ and $Q\lessapprox \pi/d$, respectively. Here, two time moments at $t=0$ and $t=40\,$ps are selected for comparing non-equilibrium parts, $\Delta N_Q^{\rm LA}(t)$ and $\Delta N_Q^{\rm LO}(t)$, in two panels of Fig.\,\ref{nfig5}, where $\hbar\omega_Q^{\rm LA}\approx 0$ and a maximum $\left|d\hbar\omega_Q^{\rm LA}/dQ\right|$ are reached around the zone center, while a minimum $\hbar\omega_Q^{\rm LO}$ and $\left|d\hbar\omega_Q^{\rm LO}/dQ\right|\approx 0$ are expected around two zone boundaries. This observation implies that only a very small heat-current density ${\cal Q}_{\rm ph}^\gamma(t)$ is present for both $LA$ and $LO$ phonons in this case, as predicted by Eq.\,\eqref{drag}.
\medskip

\subsection{Doppler Effect of Slowly-Moving Electrons on Phonon Scattering}
\label{sec-3-3}

\begin{figure}[htbp]
\centering
\includegraphics[width=0.95\textwidth]{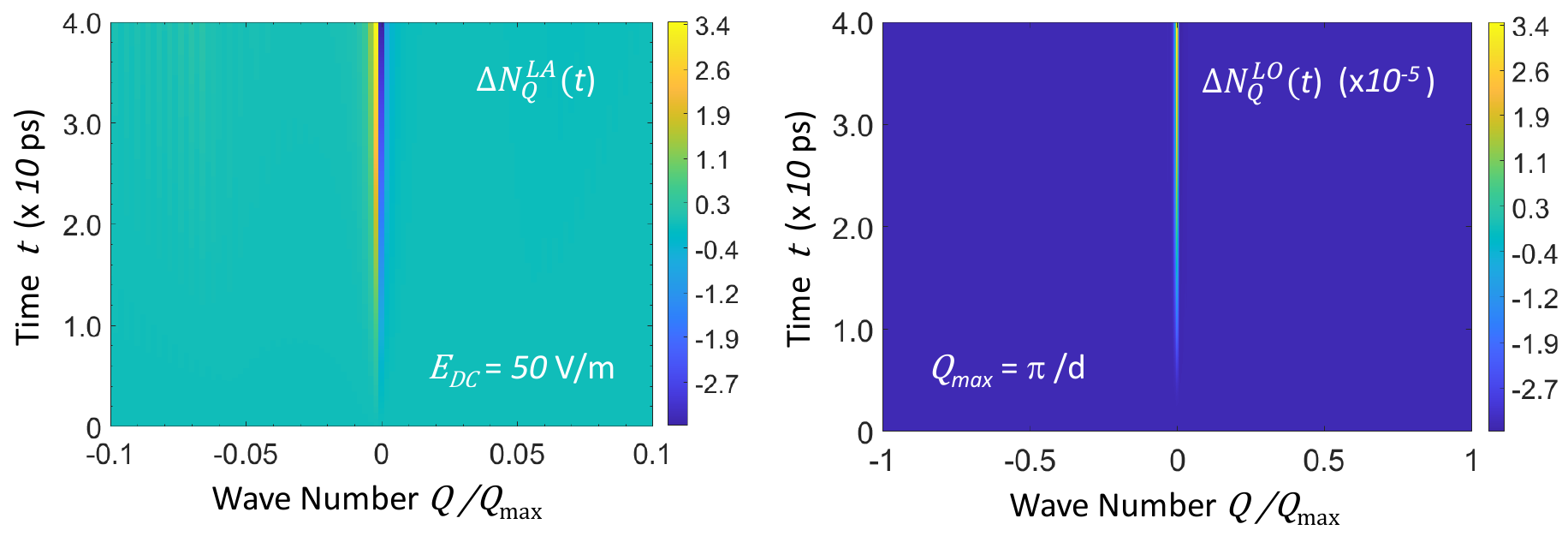}
\caption{Selected comparisons for numerically computed force-driven $\Delta N^{\rm LA}_Q(t)$ for $LA$-phonons and  $\Delta N^{\rm LO}_Q(t)$ for $LO$-phonons as functions of scaled wave number $Q/Q_{\rm max}$ in the left and right panels, respectively, at $t=0$ and $40\,$ps and under $E_{\rm DC}=50\,$V/m. Here, the circled part highlights the Doppler effect from slowly-moving electrons on a weak phonon-emission process.}
\label{nfig6}
\end{figure}

\begin{figure}[htbp]
\centering
\includegraphics[width=0.85\textwidth]{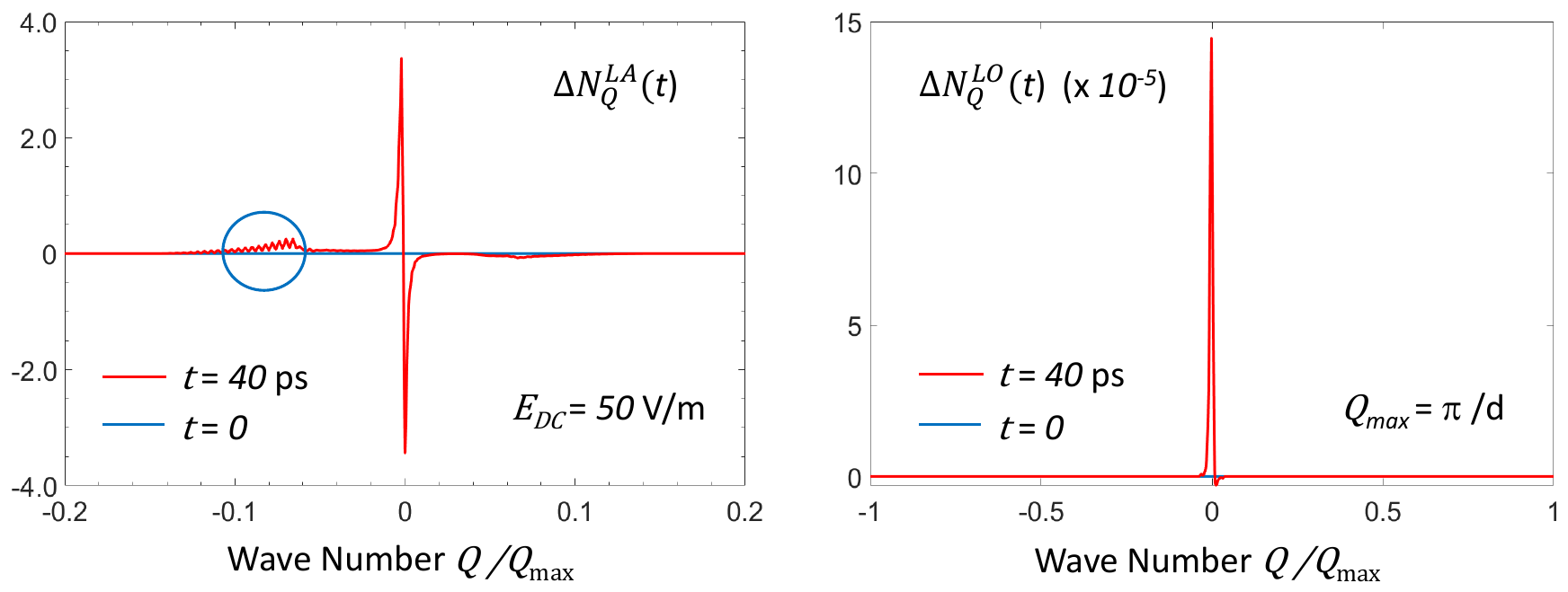}
\caption{Selected comparisons for numerically computed force-driven $\Delta N^{\rm LA}_Q(t)$ for $LA$-phonons and  $\Delta N^{\rm LO}_Q(t)$ for $LO$-phonons as functions of scaled wave number $Q/Q_{\rm max}$ in the left and right panels, respectively, at $t=0$ and $40\,$ps and under $E_{\rm DC}=50\,$V/m. Here, the circled part highlights the change due to Doppler effect from moving electrons on a phonon-emission process.}
\label{nfig7}
\end{figure}

\begin{figure}[htbp]
\centering
\includegraphics[width=0.90\textwidth]{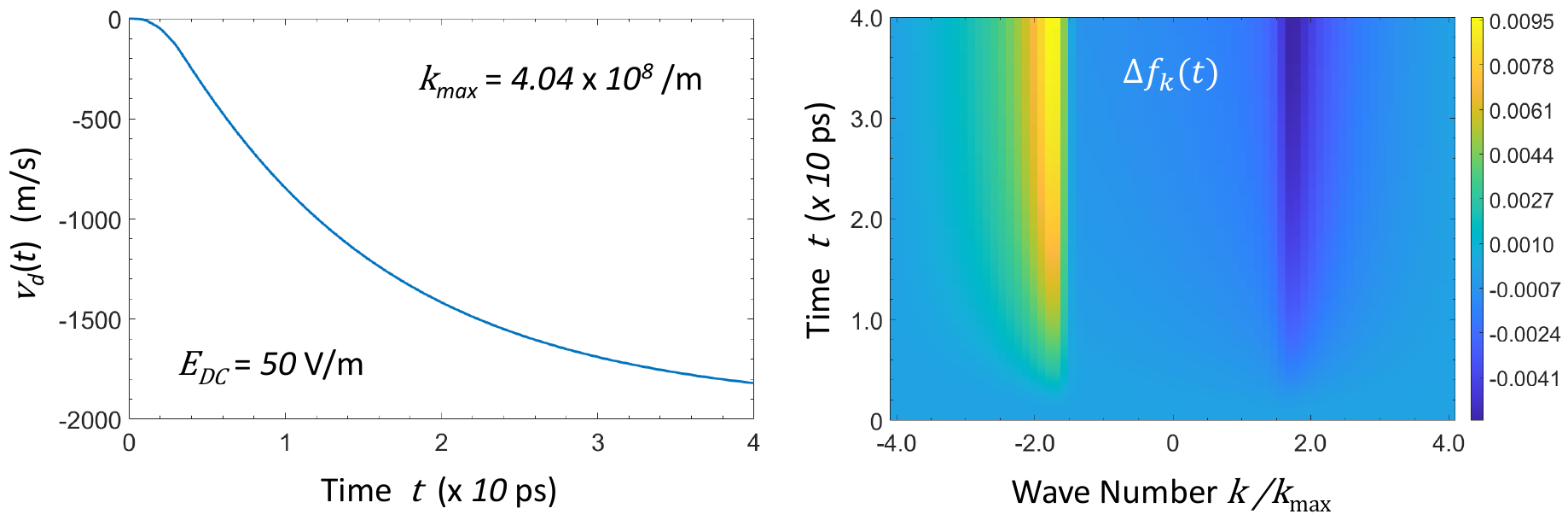}
\caption{Numerically computed time dependence of drifting velocity $\mbox{\boldmath$v$}_d(t)$ for electrons under $E_{\rm DC}=50\,$V/m in the left panel as well as the 
DC-field driven time evolution of non-equilibrium occupation part $\Delta f_k(t)$ of electrons as a function of scaled wave number $k/k_{\rm max}$ in the right panel.}
\label{nfig8}
\end{figure}

Our numerically computed time evolution for non-equilibrium parts $\Delta N_Q^{\rm LA}(t)$ and $\Delta N_Q^{\rm LO}(t)$ of both $LA$- and $LO$-phonon distributions are shown in Figs.\,\ref{nfig6} for a low DC-field strength $E_{\rm DC}=50\,$V/m. 
Meanwhile, selected comparisons for $\Delta N_Q^{\rm LA}(t)$ and $\Delta N_Q^{\rm LO}(t)$ at $t=0$ and $t=40\,$ps are made and provided in Fig.\,\ref{nfig7}. 
Moreover, calculated time evolution of non-equilibrium occupation part $\Delta\tilde{f}_k(t)$ for slowly-moving electrons under a low bias field $E_{\rm DC}=50\,$V/m is also displayed in Fig.\,\ref{nfig8}, along with the computed time dependent drift velocity $\mbox{\boldmath$v$}_d(t)$ for $E_{\rm DC}=50\,$V/m.
\medskip

Under a low DC field $E_{\rm DC}=50\,$V/m in Fig.\,\ref{nfig6}, electrons within a wire are only put into a slow drifting motion. But, according to Eq.\,\eqref{eph}, this can still affect the electron-phonon scattering, as illustrated in Fig.\,\ref{fig2}, leading to visible changes in non-equilibrium parts $\Delta N_Q^\gamma(t)$ of either $LA$ or $LO$ phonon distribution. In fact, as seen in the right panel of Fig.\,\ref{nfig6}, the time evolution of very small $\Delta N_Q^{\rm LO}(t)$ for $LO$ phonons has already moved from the Brillouin-zone boundaries at $Q=\pm\pi/d$ to the Brillouin-zone center at $Q=0$ after a weak DC field $E_{\rm DC}=50\,$V/m is applied. However, the time evolution of $\Delta N_Q^{\rm LA}(t)$ for $LA$ phonons remains largely unchanged except for its slightly increased magnitude.
\medskip

Physically speaking, as an electron is put into slow motion by an applied weak DC field $E_{\rm DC}=50\,$V/m, one expects that electron-phonon scattering starts playing a different role in modifying the non-equilibrium part $\Delta N_Q^{\gamma}(t)$ of phonon distribution, as can be verified by Eqs.\,\eqref{eph} and \eqref{eph-2}. Actually, such a modification results from a time-dependent Doppler shifted phonon frequency $\omega^{\rm LA}_Q-\mbox{\boldmath$Q$}\cdot\mbox{\boldmath$v$}_d(t)$ in Eq.\,\eqref{eph-2} for strongly dispersive $LA$ phonons, which can either increase or decrease phonon frequency $\omega^{\rm LA}_Q$, depending on the relative direction between wave vector $\mbox{\boldmath$Q$}$ of $LA$ phonons and drifting velocity $\mbox{\boldmath$v$}_d(t)$ of conducting electrons under a finite bias field $E_{\rm DC}\neq 0$. For $E_{\rm DC}>0$, pointing to the positive direction (left-to-right) of a chain in $\mbox{\boldmath$r$}$-space, $\mbox{\boldmath$v$}_d(t)$ will lie in the opposite direction of a chain due to negative charge of an electron. Therefore, if $\mbox{\boldmath$Q$}$ and $\mbox{\boldmath$v$}_d(t)$ are in the same direction, {\em i.e.\/} $Q<0$, phonon frequency $\omega^{\rm LA}_Q$ will be reduced or softened by this Doppler effect. Consequently, one finds a circled very-weak emission ($\Delta N_Q^{\rm LA}(t)>0$) peak in non-equilibrium part of $LA$ phonons for $Q<0$ in the left panel of Fig.\,\ref{nfig7}. On the other hand, the Doppler effect on $\Delta N_Q^{\rm LO}(t)$ for high-frequency $LO$ phonons also appears by moving its two peaks around the Brillouin-zone boundaries at $Q=\pm\pi/d$ to the Brillouin-zone center at $Q=0$ although $\Delta N_Q^{\rm LO}(t)$ itself is still very small in magnitude, as can be verified from the right panel of Fig.\,\ref{nfig7}.
\medskip

The presence of electron-phonon scattering in this system changes not only the non-equilibrium part $\Delta N_Q^{\rm LA}(t)$ of force-driven $LA$-phonon distribution but also the drifting velocity $\mbox{\boldmath$v$}_d(t)$ of drifting electrons within a wire, as demonstrated in Fig.\,\ref{nfig8}. From the right panel of Fig.\,\ref{nfig8}, we find that the wavevector $\mbox{\boldmath$k$}$-dependent distribution for small non-equilibrium part $\Delta\tilde{f}_k(t)$ of drifting electrons expands outwards with time $t$ in both positive and negative directions, and it behaves as an odd function with respect to electron wave number $k$. This unique feature ensures a finite drifting velocity $\mbox{\boldmath$v$}_d(t)$, as can be verified from Eq.\,\eqref{boltz-2}, since the electron group velocity $v_k=(1/\hbar)\,d\varepsilon_k/dk$ itself is an odd function of $k$.
\medskip

\subsection{Doppler Effect of Moving Electrons on Phonon Scattering}
\label{sec-3-4}

\begin{figure}[htbp]
\centering
\includegraphics[width=0.85\textwidth]{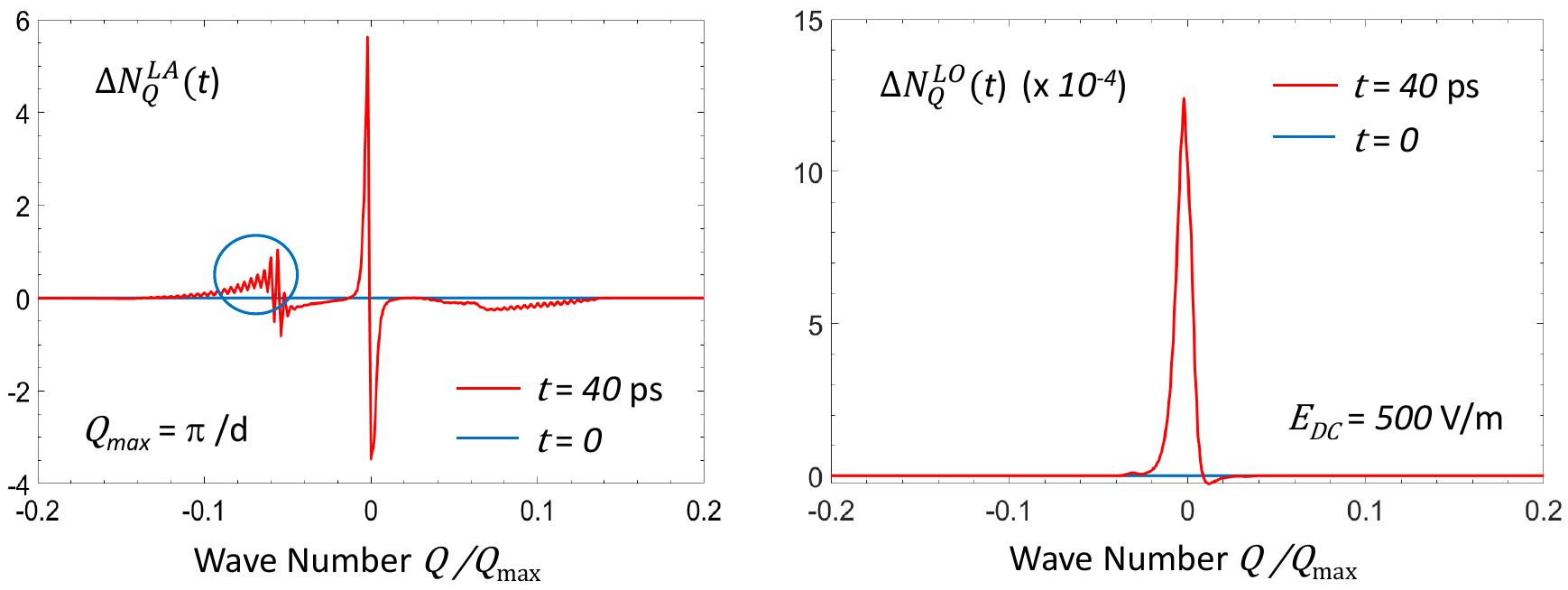}
\caption{Selected comparisons for numerically computed force-driven $\Delta N^{\rm LA}_Q(t)$ for $LA$-phonons and  $\Delta N^{\rm LO}_Q(t)$ for $LO$-phonons as functions of scaled wave number $Q/Q_{\rm max}$ in the left and right panels, respectively, at $t=0$ and $40\,$ps and under $E_{\rm DC}=500\,$V/m. Here, the circled part highlights the change due to Doppler effect from moving electrons on combined phonon-emission and -absorption processes.}
\label{nfig10}
\end{figure}

\begin{figure}[htbp]
\centering
\includegraphics[width=0.95\textwidth]{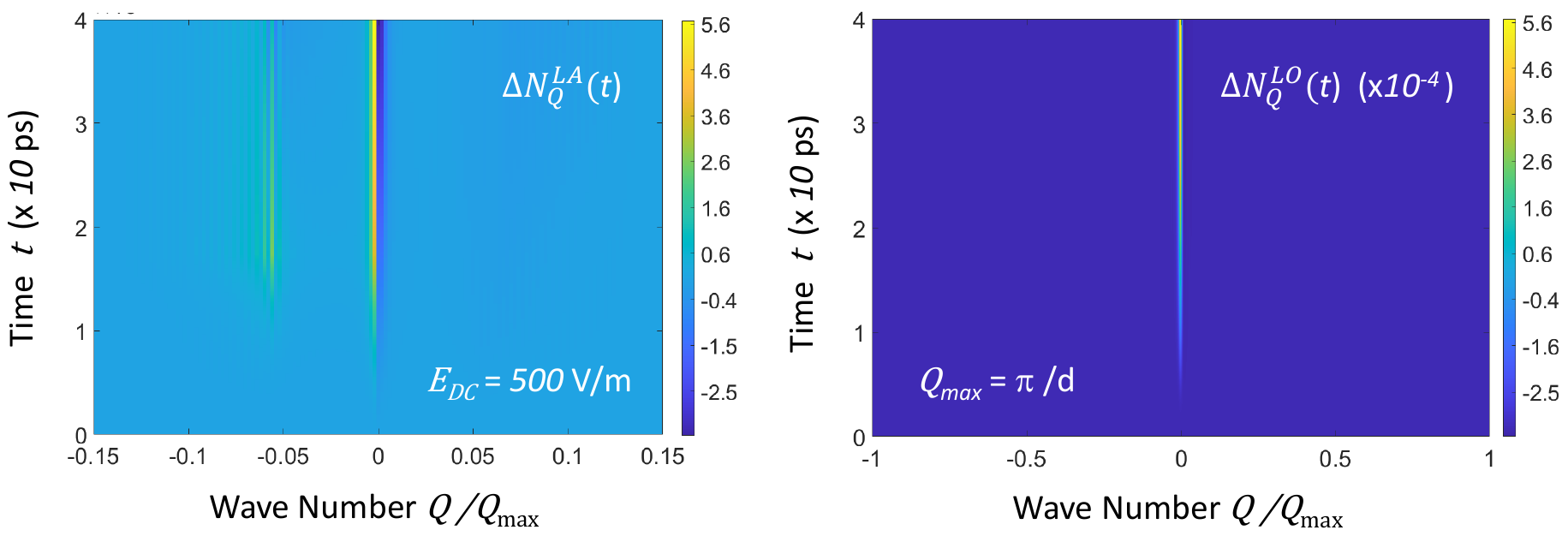}
\caption{Numerically computed time evolution of force-driven non-equilibrium parts $\Delta N^{\rm LA}_Q(t)$ for $LA$-phonons and  $\Delta N^{\rm LO}_Q(t)$ for $LO$-phonons as functions of scaled wave number $Q/Q_{\rm max}$ in the left and right panels, respectively, under $E_{\rm DC}=500\,$V/m.}
\label{nfig9}
\end{figure}

\begin{figure}[htbp]
\centering
\includegraphics[width=0.90\textwidth]{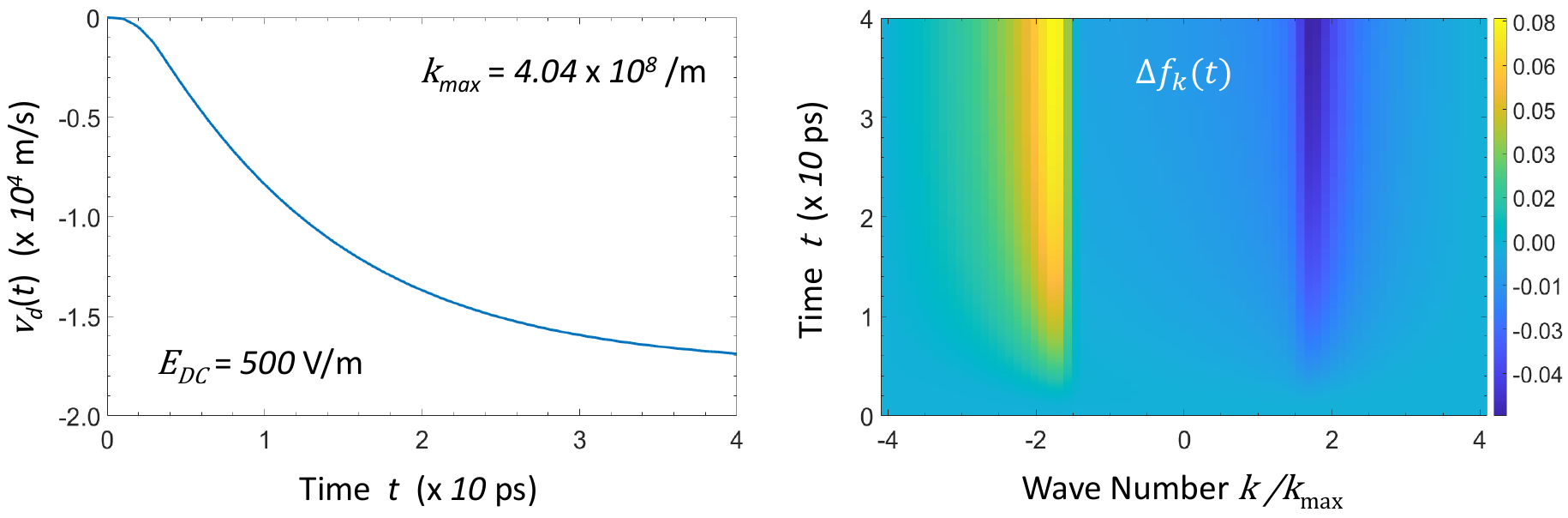}
\caption{Numerically computed time dependence of drifting velocity$\mbox{\boldmath$v$}_d(t)$ for electrons under $E_{\rm DC}=500\,$V/m in the left panel as well as the 
DC-field driven time evolution of non-equilibrium occupation part $\Delta\tilde{f}_k(t)$ of electrons as a function of scaled wave number $k/k_{\rm max}$ in the right panel.}
\label{nfig11}
\end{figure}

Comparisons of $\Delta N_Q^{\rm LA}(t)$ and $\Delta N_Q^{\rm LO}(t)$ at $t=0$ and $t=40\,$ps are selected and presented in Fig.\,\ref{nfig10} for intermediate DC-field strength $E_{\rm DC}=500\,$V/m. Meanwhile, numerically solved time evolution for distribution of non-equilibrium parts, $\Delta N_Q^{\rm LA}(t)$ and $\Delta N_Q^{\rm LO}(t)$, of $LA$- and $LO$-phonons are shown in Figs.\,\ref{nfig9}. 
Furthermore, computed time evolution of non-equilibrium occupation part $\Delta\tilde{f}_k(t)$ for moving electrons is displayed in Fig.\,\ref{nfig11}, along with a time-dependent drift velocity $\mbox{\boldmath$v$}_d(t)$ of drifting electrons under an elevated DC field $E_{\rm DC}=500\,$V/m.
\medskip

We know from the previous discussion on Fig.\,\ref{nfig7} that the Doppler effect on phonon-mode softening takes the form  $\omega^{\rm LA}_Q-\mbox{\boldmath$Q$}\cdot\mbox{\boldmath$v$}_d(t)$, and then, one anticipates that this effect will be strengthened by increasing DC field strength $E_{\rm DC}$ in order to acquire a larger $\left|\mbox{\boldmath$v$}_d(t)\right|$ value. Indeed, as a DC-field is enhanced by an order of magnitude from $E_{\rm DC}=50\,$V/m to $E_{\rm DC}=500\,$V/m, the circled single weak emission peak for $\Delta N_Q^{\rm LA}(t)$ at $E_{\rm DC}=50\,$V/m in the left panel of Fig.\,\ref{nfig7} turns into a mixed emission-absorption peak at $t=40\,$ps, as can be verified from the left panel of Fig.\,\ref{nfig10}. Meanwhile, the single very-sharp emission peak of $\Delta N_Q^{\rm LO}(t)$ at the zone center for $LO$ phonons in the right panel of Fig.\,\ref{nfig7} is still retained but enhanced in strength by one order of magnitude and broadened significantly at the same time, as seen in the right panel of Fig.\,\ref{nfig10} for $E_{\rm DC}=500\,$V/m.
\medskip
 
As DC field in Fig.\,\ref{nfig9} is increased to a relatively high strength $E_{\rm DC}=500\,$V/m, electrons within a wire can move in a regular speed instead of a slow motion. This increases the electron-phonon scattering relative to the situation in Fig.\,\ref{nfig6} by gradually developing a different mixed emission-absorption satellite peak on the $Q<0$ side in non-equilibrium part $\Delta N_Q^{\rm LA}(t)$ for $LA$ phonon distribution. On the contrary, although its magnitude is increased by one order, the time evolution of $\Delta N_Q^{\rm LO}(t)$ for $LO$ phonons still remains very close to the Brillouin-zone center at $Q=0$ even when $E_{\rm DC}$ increases from $50\,$V/m to $500\,$V/m by one order of magnitude.
\medskip

In comparison with the left panel of Fig.\,\ref{nfig8}, the steady-state drift velocity $\mbox{\boldmath$v$}_d(t)=1.8\times 10^3\,$m/s in the left panel of Fig.\,\ref{nfig11} at $t=40$\,ps increases to $\mbox{\boldmath$v$}_d(t)=1.7\times 10^4\,$m/s as $E_{\rm DC}$ changes from $50\,$V/m to $500\,$V/m, indicating a slight deviation from the well-known Ohm's law in this field-strength regime.
From the right panel of Fig.\,\ref{nfig11}, we find that the distribution for non-equilibrium part $\Delta\tilde{f}_k(t)$ of drifting electrons increases in magnitude and expands slightly fast with time $t$, although it still appears as an odd function with respect to electron wave number $k$. The fact that $\Delta\tilde{f}_k(t)\sim 0.08$ indicates that heating of conduction electrons has not played a role yet in this case.
\medskip

\subsection{Doppler Effect of Fast-Moving Electrons on Phonon Scattering}
\label{sec-3-5}

\begin{figure}[htbp]
\centering
\includegraphics[width=0.95\textwidth]{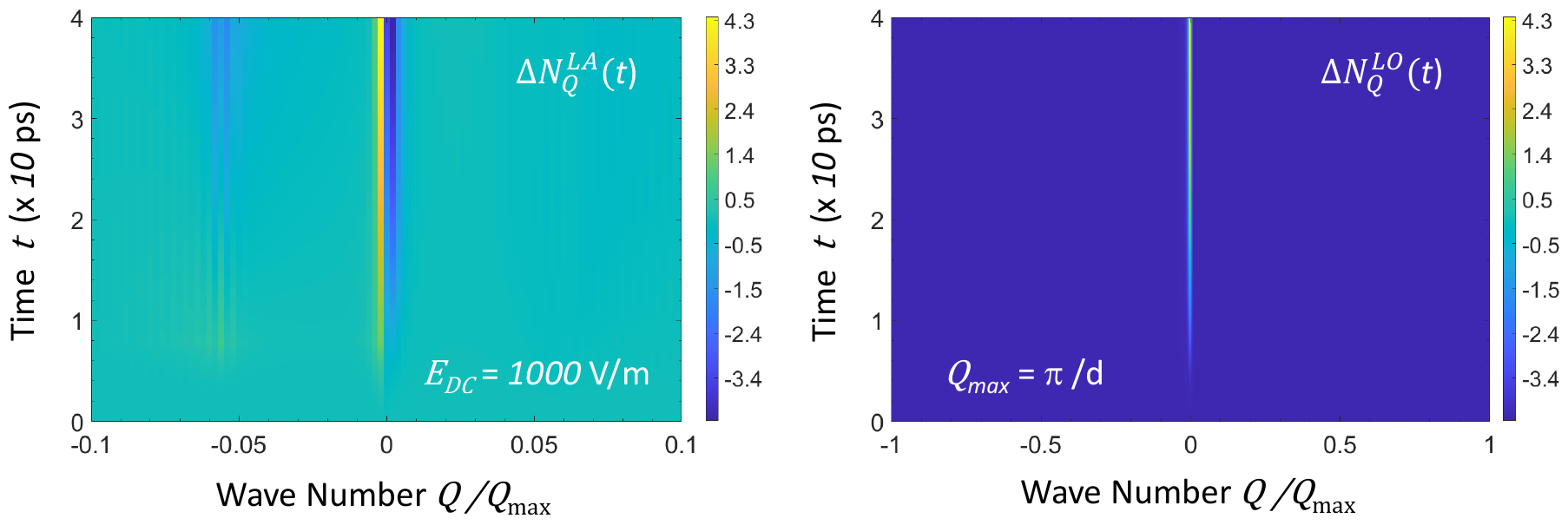}
\caption{Numerically computed time evolution of force-driven non-equilibrium parts $\Delta N^{\rm LA}_Q(t)$ for $LA$-phonons and  $\Delta N^{\rm LO}_Q(t)$ for $LO$-phonons as functions of scaled wave number $Q/Q_{\rm max}$ in the left and right panels, respectively, under $E_{\rm DC}=1000\,$V/m.}
\label{nfig12}
\end{figure}

\begin{figure}[htbp]
\centering
\includegraphics[width=0.85\textwidth]{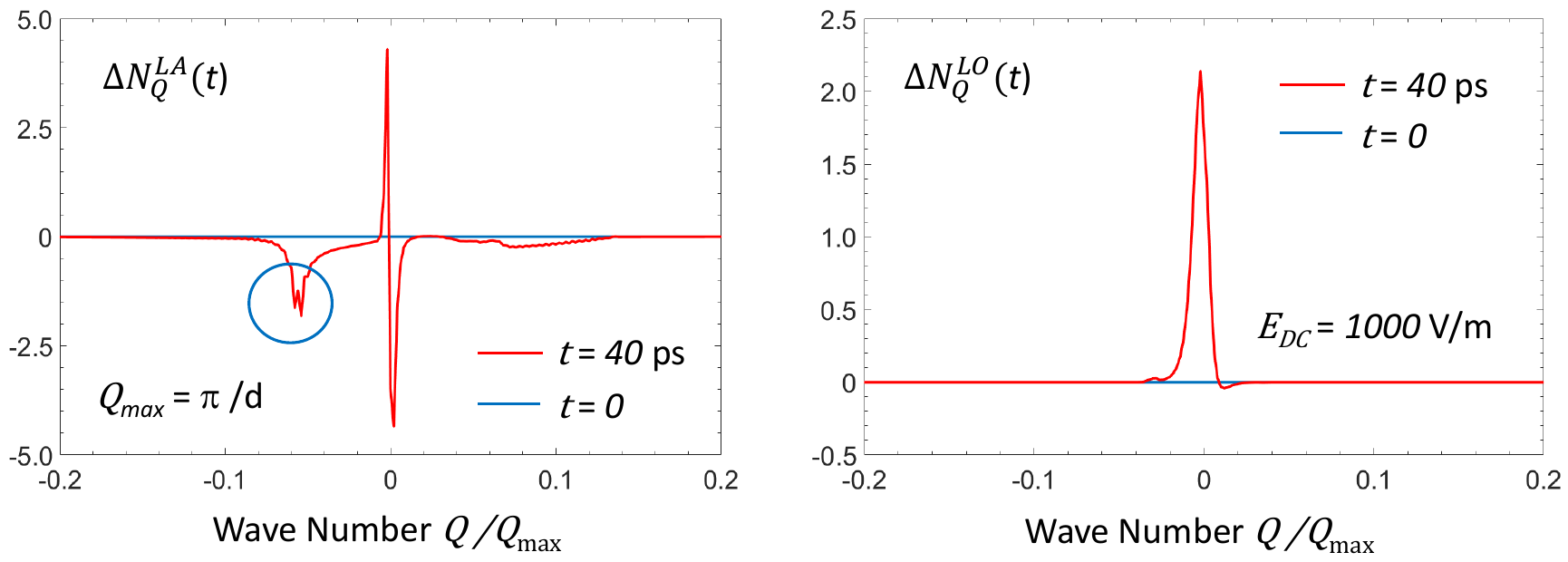}
\caption{Selected comparisons for numerically computed force-driven $\Delta N^{\rm LA}_Q(t)$ for $LA$-phonons and  $\Delta N^{\rm LO}_Q(t)$ for $LO$-phonons as functions of scaled wave number $Q/Q_{\rm max}$ in the left and right panels, respectively, at $t=0$ and $40\,$ps and under $E_{\rm DC}=1000\,$V/m. Here, the circled area highlights the change in Doppler effect from fast-moving electrons on phonon-absorption process only.}
\label{nfig13}
\end{figure}

\begin{figure}[htbp]
\centering
\includegraphics[width=0.90\textwidth]{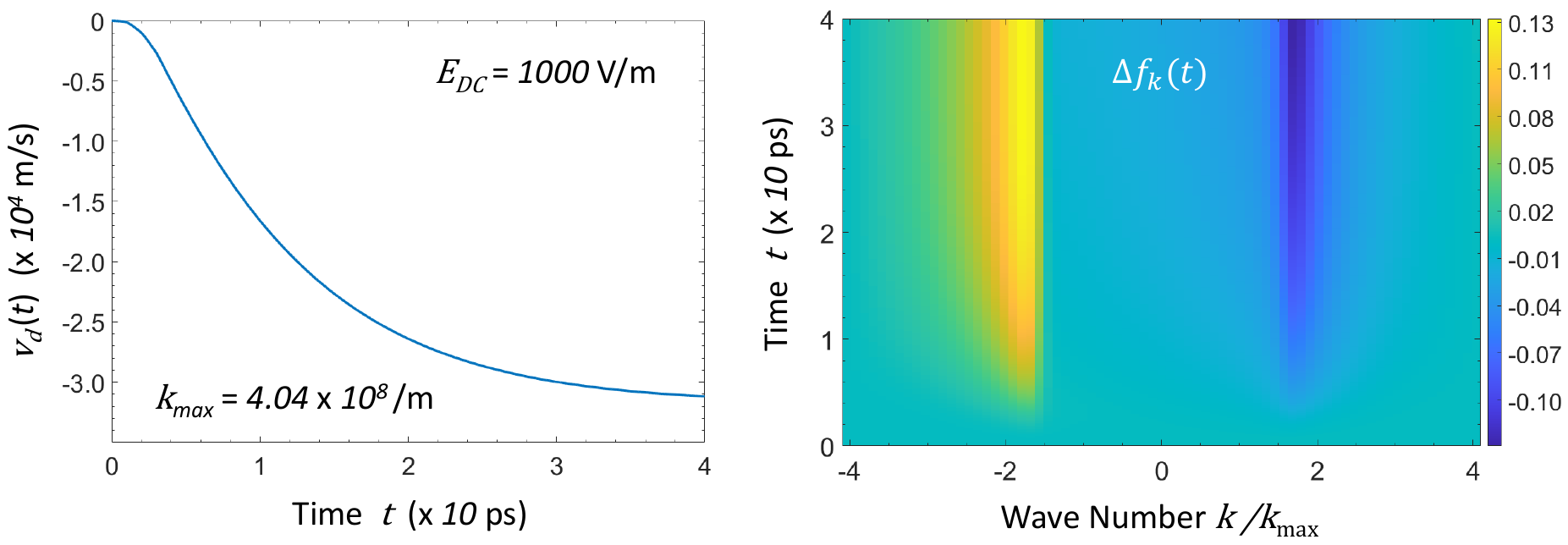}
\caption{Numerically computed time dependence of drifting velocity $\mbox{\boldmath$v$}_d(t)$ for electrons under $E_{\rm DC}=1\,$kV/m in the left panel as well as the DC-field driven time evolution of non-equilibrium occupation part $\Delta f_k(t)$ of electrons as a function of scaled wave number $k/k_{\rm max}$ in the right panel.}
\label{nfig14}
\end{figure}

The numerically calculated time evolution for non-equilibrium distributions $\Delta N_Q^{\rm LA}(t)$ and $\Delta N_Q^{\rm LO}(t)$ of both $LA$- and $LO$-phonons is presented in Figs.\,\ref{nfig12} 
having DC-field strength $E_{\rm DC}=1000\,$V/m for fast-moving electrons.
Meanwhile, comparisons of $\Delta N_Q^{\rm LA}(t)$ and $\Delta N_Q^{\rm LO}(t)$ at $t=0$ and $t=40\,$ps are selected and displayed in Fig.\,\ref{nfig13}. 
Moreover, numerically computed time evolution of non-equilibrium occupation function $\Delta\tilde{f}_k(t)$ for fast-moving electrons is also shown in Fig.\,\ref{nfig14}, along with the time dependence of a drift velocity $\mbox{\boldmath$v$}_d(t)$ under $E_{\rm DC}=1000\,$V/m.
\medskip

In order to explore the limit of a Doppler effect from fast-moving electrons on phonon scattering, we further increase DC field $E_{\rm DC}$ to $1000\,$V/m in Fig.\,\ref{nfig12}, where only a pure negative satellite peak in the $Q<0$ region occurs for $\Delta N_Q^{\rm LA}(t)$ after $t>2.5\,$ps as can be verified by the left panel. This case connects to an absorption-only process for phonons, where the mixed phonon emission exhibited in the left panel of Fig.\,\ref{nfig12} at $E_{\rm DC}=500\,$V/m has been fully suppressed. For $LO$ phonons, $\Delta N_Q^{\rm LO}(t)$ now becomes comparable with $\Delta N_Q^{\rm LA}(t)$ in magnitude.  
However, one still sees a slightly broadened emission peak at the zone center $Q=0$ in the right panel of Fig.\,\ref{nfig12}. Such unique features can be viewed more clearly in two panels of Fig.\,\ref{nfig13} for $LA$ and $LO$ phonons, respectively, by comparing directly corresponding results at initial time $t=0$ and at ending time $t=40\,$ps. 
\medskip

In comparison with $v_d(t)=1.7\times 10^3\,$m/s in the left panel of Fig.\,\ref{nfig11}, the steady-state drift velocity $\mbox{\boldmath$v$}_d(t)$ in the left panel of Fig.\,\ref{nfig14} at $t=40$\,ps increases to $3.1\times 10^4\,$m/s as $E_{\rm DC}$ is doubled  from $500\,$V/m to $1000\,$V/m, indicating an even stronger nonlinear deviation from the well-known Ohm's law in this field range. From the right panel of Fig.\,\ref{nfig14}, we observe dual positive ($k<0$ side) and negative ($k>0$ side) peaks in the distribution for non-equilibrium part $\Delta\tilde{f}_k(t)$ for fast-moving electrons, which implies more electrons have been removed from the $k>0$ states and added to the $k<0$ states with increasing time $t$, but the odd-function feature with respect to electron wave number $k$ remains. The fact that $\Delta\tilde{f}_k(t)\sim 0.13$ is reached indicates that heating of conduction electrons has started already,\,\cite{add-1,add-2} leading to an effective temperature $T_{\rm eff}(t)$ of electrons higher than the initial thermal-equilibrium temperature $T_0$.
\medskip

\subsection{Doppler Effect of Very-Fast-Moving Electrons on Phonon Scattering}
\label{sec-3-6}

\begin{figure}[htbp]
\centering
\includegraphics[width=0.95\textwidth]{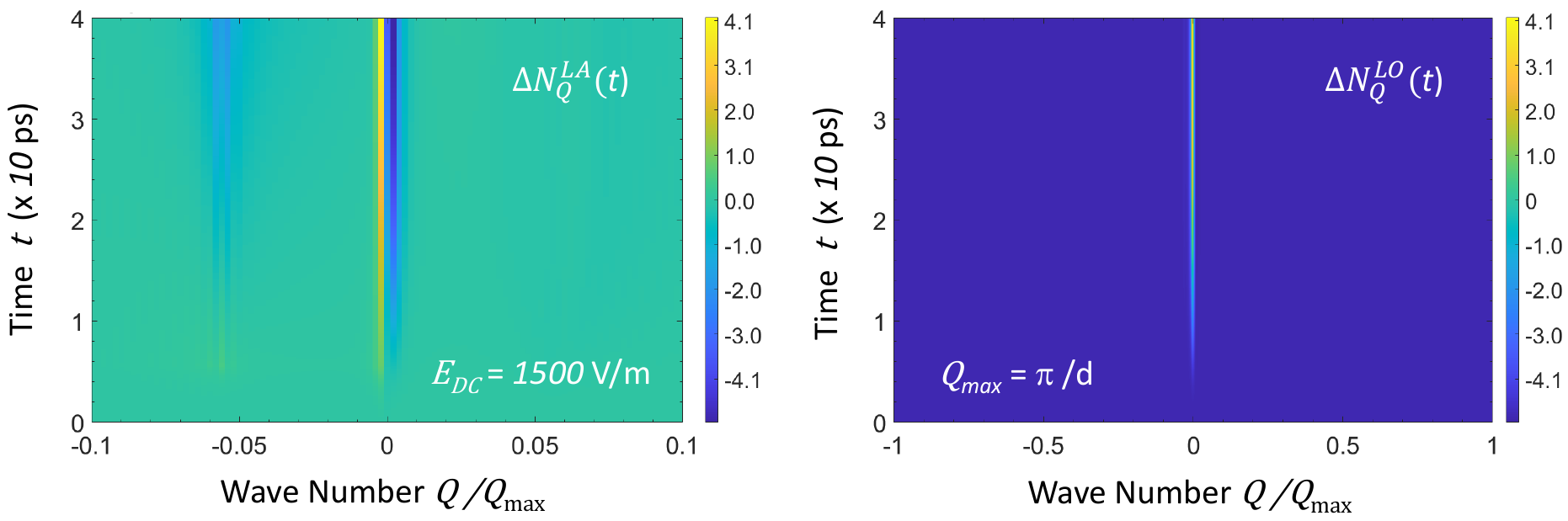}
\caption{Numerically computed time evolution of force-driven non-equilibrium parts $\Delta N^{\rm LA}_Q(t)$ for $LA$-phonons and  $\Delta N^{\rm LO}_Q(t)$ for $LO$-phonons as functions of scaled wave number $Q/Q_{\rm max}$ in the left and right panels, respectively, under $E_{\rm DC}=1500\,$V/m.}
\label{nfig15}
\end{figure}

\begin{figure}[htbp]
\centering
\includegraphics[width=0.85\textwidth]{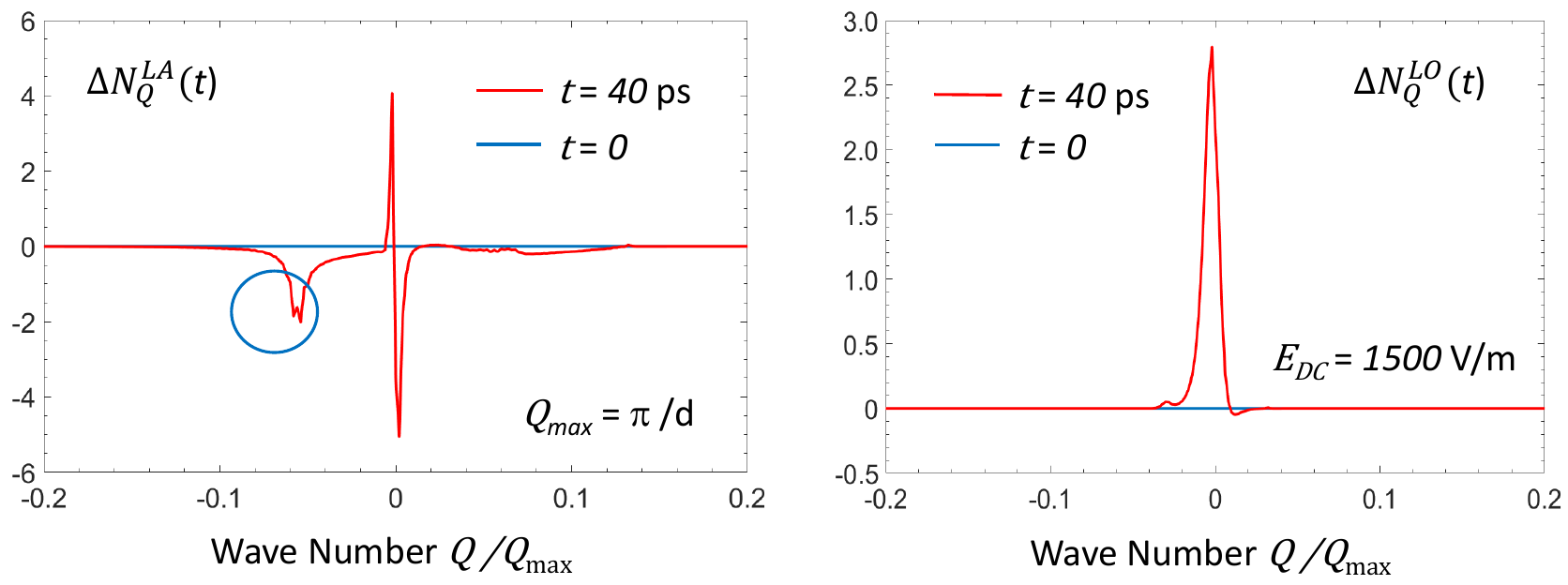}
\caption{Selected comparisons for numerically computed force-driven $\Delta N^{\rm LA}_Q(t)$ for $LA$-phonons and  $\Delta N^{\rm LO}_Q(t)$ for $LO$-phonons as functions of scaled wave number $Q/Q_{\rm max}$ in the left and right panels, respectively, at $t=0$ and $40\,$ps and under $E_{\rm DC}=1500\,$V/m. Here, the circled area highlights the saturation of  Doppler effect from very-fast-moving electrons on phonon-absorption process.}
\label{nfig16}
\end{figure}

\begin{figure}[htbp]
\centering
\includegraphics[width=0.90\textwidth]{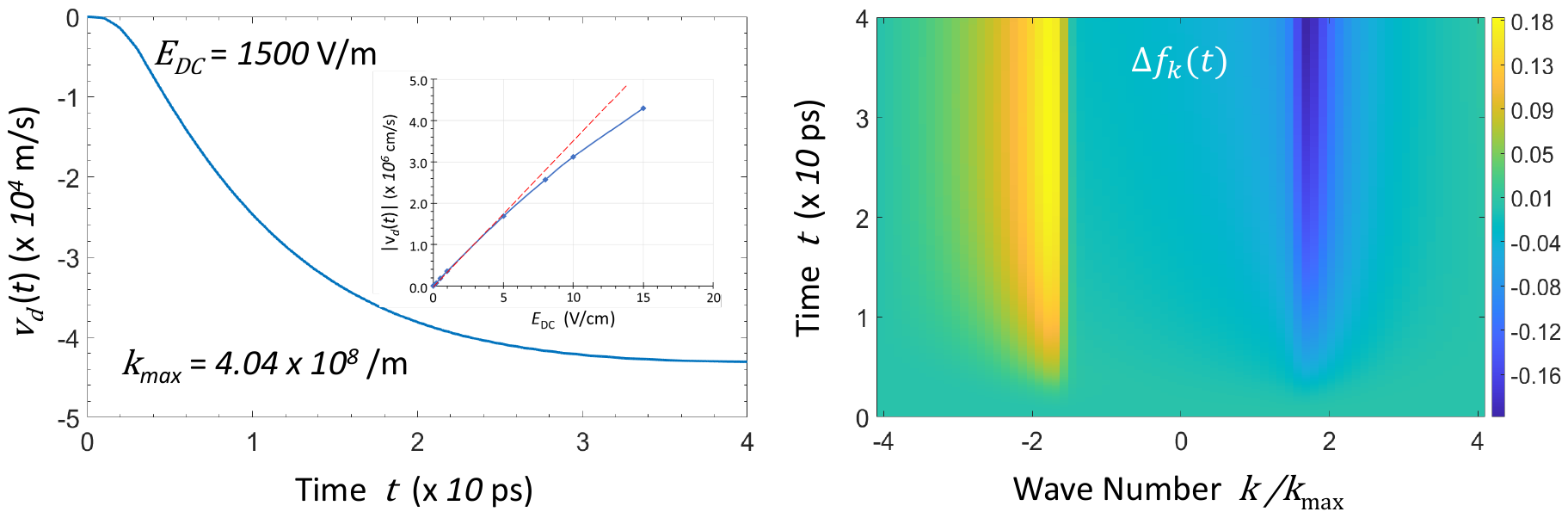}
\caption{Numerically computed time dependence of drifting velocity $\mbox{\boldmath$v$}_d(t)$ for electrons under $E_{\rm DC}=1.5\,$kV/m in the left panel as well as the DC-field driven time evolution of non-equilibrium occupation part $\Delta f_k(t)$ of electrons as a function of scaled wave number $k/k_{\rm max}$ in the right panel. Here, inset in the left panel reveals a non-linear dependence between calculated $\mbox{\boldmath$v$}_d(t)$ at $t=40\,$ps for electrons and applied $E_{\rm DC}$. The linear dependence, or Ohm's law, holds only for low field $E_{\rm DC}\leq 100\,$V/m, as indicated by a dashed straight line. Therefore, the electron mobility becomes field dependent and decreases with DC-field strength in this case.}
\label{nfig17}
\end{figure}

Numerically computed non-equilibrium occupation parts $\Delta N_Q^{\rm LA}(t)$ and $\Delta N_Q^{\rm LO}(t)$ for both $LA$- and $LO$-phonons are presented in Figs.\,\ref{nfig15} 
for high DC-field strength $E_{\rm DC}=1500\,$V/m for very-fast moving electrons.  
Meanwhile, $\Delta N_Q^{\rm LA}(t)$ and $\Delta N_Q^{\rm LO}(t)$ at $t=0$ and $t=40\,$ps are chosen for comparison and displayed in Fig.\,\ref{nfig16}. 
Moreover, calculated time evolution for non-equilibrium occupation part $\Delta\tilde{f}_k(t)$ for moving electrons is also shown in Fig.\,\ref{nfig17}, along with the time dependence of a drift velocity $\mbox{\boldmath$v$}_d(t)$ under $E_{\rm DC}=1500\,$V/m.
\medskip

Even though it is well known that there exists a nonlinear relation between the drift velocity $\mbox{\boldmath$v$}_d(t)$ of electrons and the applied DC-field strength $E_{\rm DC}$, one still hopes to acquire an even larger drift velocity $\mbox{\boldmath$v$}_d(t)$ before its saturation. For this sake, we further lift $E_{\rm DC}$ up to $1500\,$V/m in order to explore an upper limit for self-induced Doppler effect on phonon scattering with very-fast-moving electrons.
\medskip 

By comparing Fig.\,\ref{nfig12} with Fig.\,\ref{nfig15}, we find the time evolution of $\Delta N_Q^{\rm LA}(t)$ in the left panel of Fig.\,\ref{nfig15} still ends with a saturated negative satellite peak in the $Q<0$ region for phonon absorption. However, the time duration required for reaching this saturated-absorption satellite peak from a mixed emission-absorption satellite peak shrunk from $30\,$ps to $20\,$ps. While for $LO$ phonons, one is still left with a broadened emission peak at the zone center $Q=0$ in the right panel of Fig.\,\ref{nfig15} with a bit enhanced strength. 
\medskip

Although the absorption satellite peak gets saturated for $E_{\rm DC}=1500\,$V/m, we still observe a much quickly-expanding dual positive ($k<0$ side) and negative ($k>0$ side) peaks in the time evolution of non-equilibrium part $\Delta\tilde{f}_k(t)$ for very-fast-moving electrons in the right panel of Fig.\,\ref{nfig17}. Here, a bigger peak strength $\Delta\tilde{f}_k(t)\sim 0.18$ is reached for dual peaks of 
$\Delta\tilde{f}_k(t)$. This indicates that even more electrons have been moved from the $k>0$ states to the $k<0$ states within a shorter period of time. 
Meanwhile, the odd-function feature with respect to electron wave number $k$ is kept in a solid way. Additionally, the fact that $\Delta\tilde{f}_k(t)\sim 0.18$ reveals that the heating of conduction electrons has already played a significant role in this case, resulting in a much higher effective temperature $T_{\rm eff}(t)$ for electrons.
\medskip

Here, the extracted steady-state drift velocity $\mbox{\boldmath$v$}_d(t)$ at $t=40\,$ps is plotted as a function of applied DC-field strength $E_{\rm DC}$ in the inset of left panel of Fig.\,\,\ref{nfig17}.
In comparison with $v_d(t)=3.1\times 10^4\,$m/s in the left panel of Fig.\,\ref{nfig14} for $E_{\rm DC}=1000\,$V/m, the steady-state drift velocity $\mbox{\boldmath$v$}_d(t)$ in the left panel of Fig.\,\ref{nfig17} at $t=40$\,ps approaches $4.3\times 10^4\,$m/s as $E_{\rm DC}$ increases to $1500\,$V/m, implying a much stronger nonlinear deviation from the well-known Ohm's law, as shown by a straight dashed line, and accompanied by a reduced mobility with increasing $E_{\rm DC}$ in this field range. 
\medskip 

\subsection{Heat Transport of Force-Driven Non-Equilibrium Phonons Coupled to Drifting Electrons}
\label{sec-3-7}

\begin{figure}[htbp]
\centering
\includegraphics[width=0.40\textwidth]{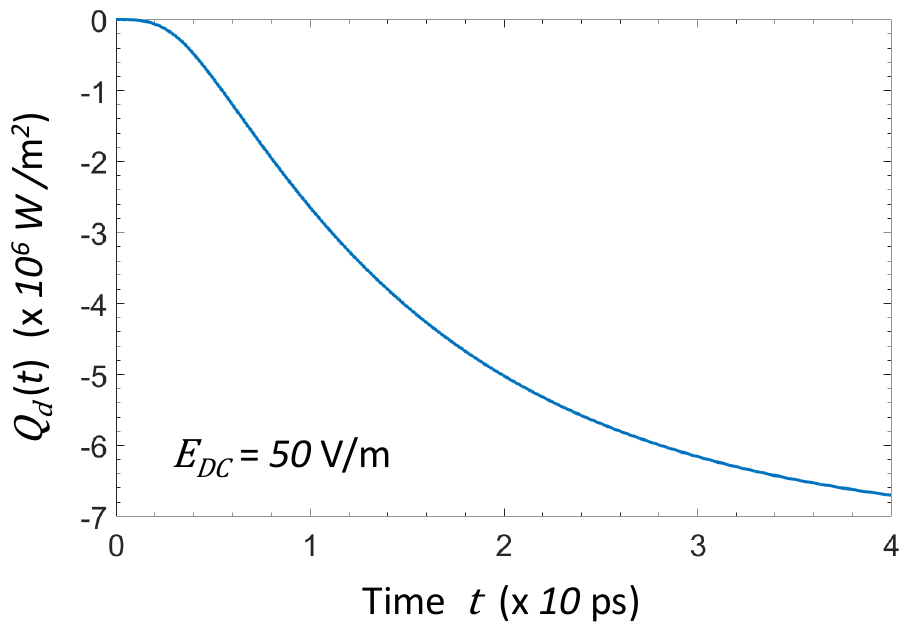}
\caption{Numerically computed time dependence of heat-current density ${\cal Q}_{\rm d}(t)$ from Eq.\,\eqref{diff} due to electron diffusion is presented under an intermediate DC field $E_{\rm DC}=50\,$V/m.}
\label{nfig18}
\end{figure}

\begin{figure}[htbp]
\centering
\includegraphics[width=0.80\textwidth]{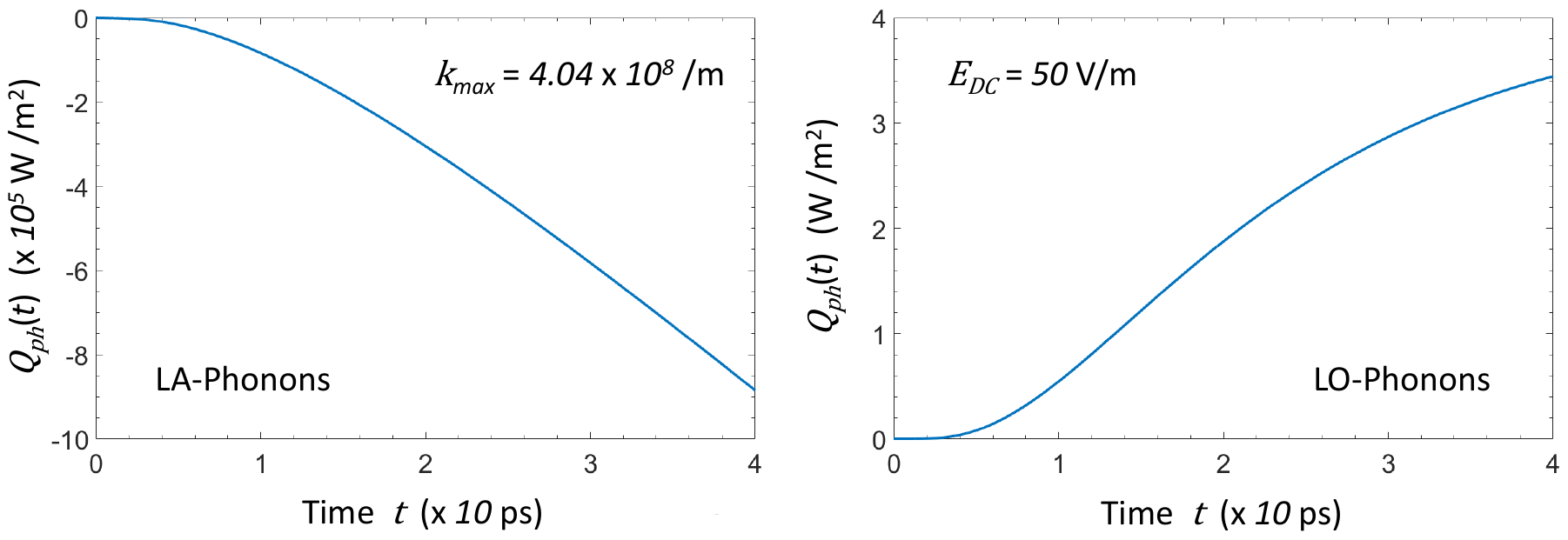}
\caption{Numerically computed time dependence of heat-current densities ${\cal Q}^\gamma_{\rm ph}(t)$ from Eq.\,\eqref{diff} due to transport of force-driven $LA$ (left panel) and $LO$ (right panel) phonons are displayed under an intermediate DC field $E_{\rm DC}=50\,$V/m.}
\label{nfig19}
\end{figure}

\begin{figure}[htbp]
\centering
\includegraphics[width=0.40\textwidth]{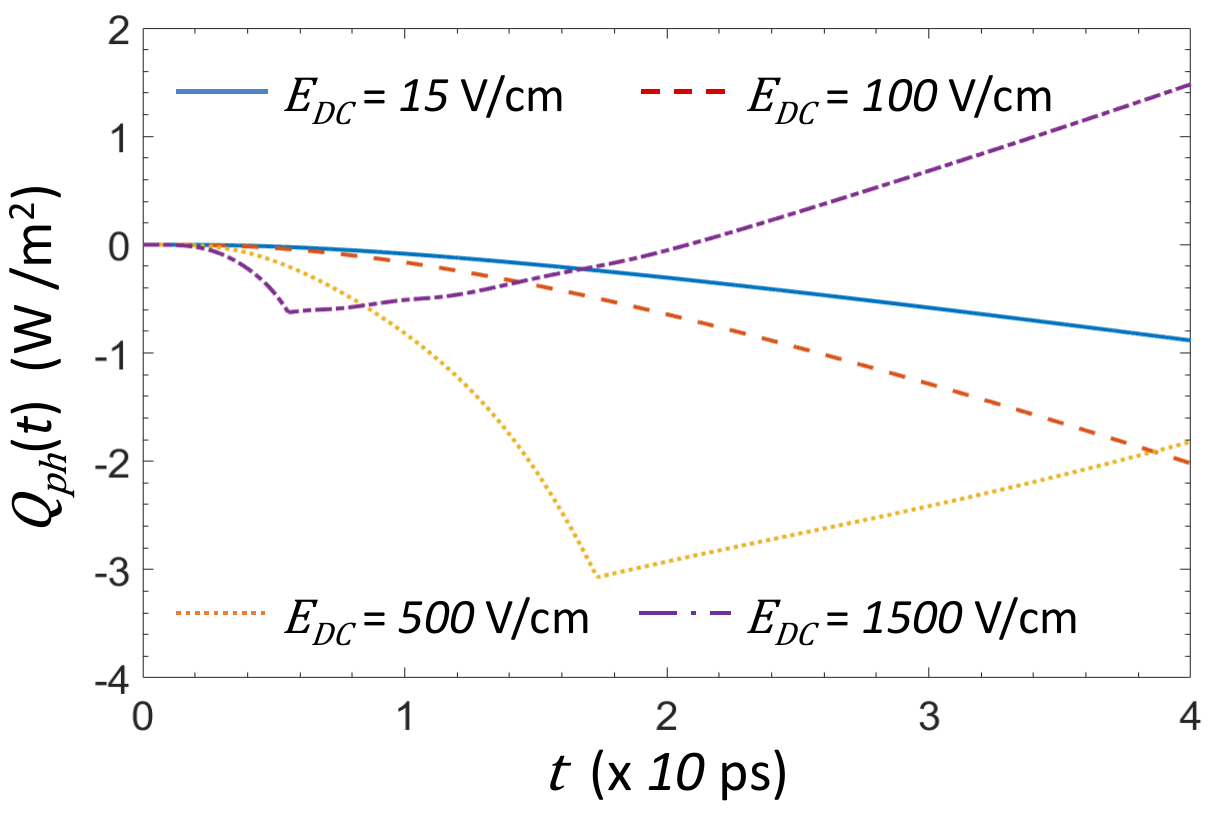}
\caption{Numerically computed time dependence of heat-current densities ${\cal Q}^\gamma_{\rm ph}(t)$ from Eq.\,\eqref{diff} due to transport of force-driven $LA$ phonons scattered by drifting electrons, where different DC fields $E_{\rm DC}=15,\,100,\,500$ and $1500\,$V/m are taken for computations.}
\label{nfig20}
\end{figure}

\begin{figure}[htbp]
\centering
\includegraphics[width=0.85\textwidth]{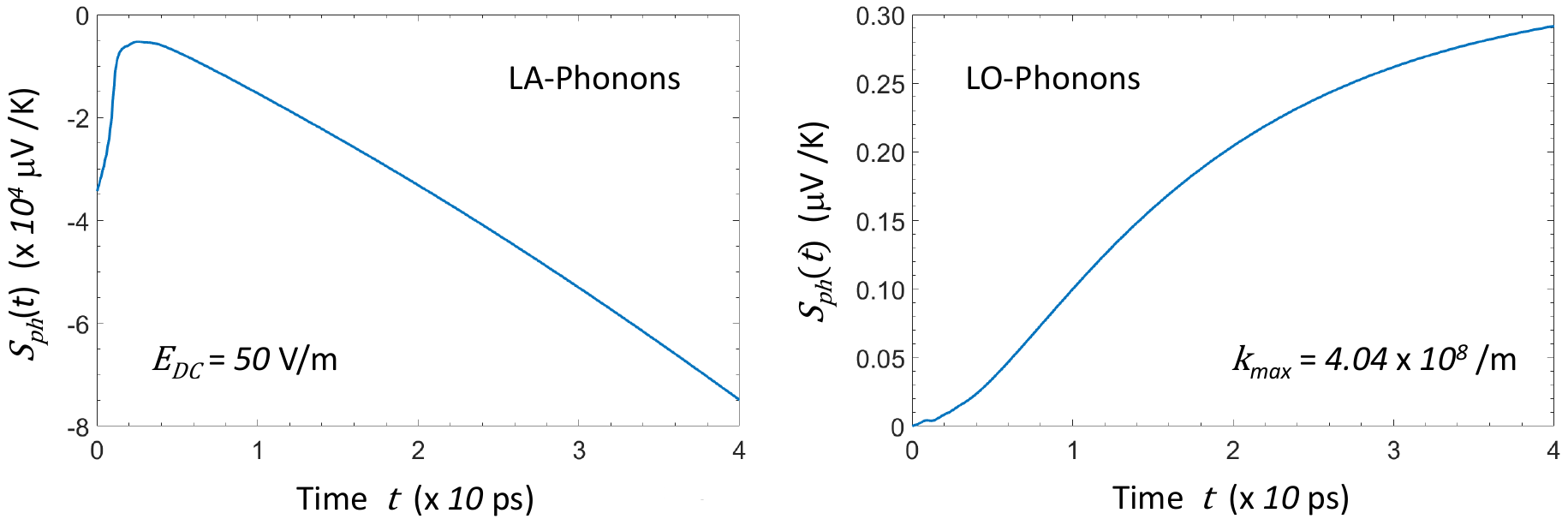}
\caption{Numerically computed time dependence of the phonon-drag thermo-electric power ${\cal S}^\gamma_{\rm ph}(t)$, where a weak DC field $E_{\rm DC}=50\,$V/m is assumed and 
the extracted phonon-drag thermo-electric powers ${\cal S}_{\rm ph}^\gamma(t)={\cal Q}_{\rm ph}^\gamma(t)/T_0J_{\rm e}(t)$ in unit of $\mu V/K$ are presented for $LA$-phonon ($\gamma=LA$ in the left panel) and $LO$-phonon ($\gamma=LO$ in the right panel), respectively.}
\label{nfig21}
\end{figure}

Numerically computed heat-current density ${\cal Q}_{\rm d}(t)$ from Eq.\,\eqref{diff} for electron diffusion is presented in Fig.\,\ref{nfig18}, while the calculated heat-current densities ${\cal Q}^{\rm LA}_{\rm ph}(t)$ and ${\cal Q}^{\rm LO}_{\rm ph}(t)$ from Eq.\,\eqref{drag} for force-driven $LA$- and $LO$-phonons by a thermal field are shown in Fig.\,\ref{nfig19}, where an relatively weak DC field $E_{\rm DC}=50\,$V/m is chosen. 
\medskip

From Fig.\,\ref{nfig18} we see clearly that, under $E_{\rm DC}=50\,$V/m, the negative electron-diffusion heat-current density ${\cal Q}_{\rm d}(t)$, which points to the direction of $\mbox{\boldmath$v$}_d(t)$, quickly reaches its steady-state value at $t=40\,$ps.  
As formulated by Eq.\,\eqref{diff}, ${\cal Q}_{\rm d}(t)$ physically measures a non-equilibrium energy flow carried by drifting electrons relative to their chemical potential $\mu_0$. It is clear from  Eq.\,\eqref{diff} that ${\cal Q}_{\rm d}(t)\equiv 0$ if $\Delta\tilde{f}_k(t)=\Delta\tilde{f}_{-k}(t)$ is satisfied, or equivalently, $\Delta\tilde{f}_k(t)$ is an even function with respect to electron wave number $k$. As $E_{\rm DC}(t)\neq 0$, $\Delta\tilde{f}_k(t)$ becomes nonzero for non-equilibrium drifting electrons, and these electrons in their $k>0$ states will be moved to $k<0$ states due to acceleration by a bias field $E_{\rm DC}$, as seen in the right panel of Fig.\,\,\ref{nfig17} for the time evolution of a pair of positive-negative twin peaks.
\medskip

On the other hand, we display transient features in Fig.\,\ref{nfig19} for non-equilibrium heat-current density $Q^\gamma_{\rm ph}(t)$, resulting from phonon heat transport driven by an external force, for both $LA$ and $LO$ phonons in the left and right panels, respectively. As understood from Eq.\,\eqref{drag}, $Q^\gamma_{\rm ph}(t)$ can be fully determined by an integral over the product of three factors, {\em i.e.\/}, non-equilibrium part $\Delta N^\gamma_Q(t)$ of phonon distribution, phonon energy $\hbar\omega_Q^\gamma$ and phonon velocity $\partial\omega^\gamma_Q/\partial Q$. As seen in the left panel of Fig.\,\ref{nfig6} for $E_{\rm DC}=50\,$V/m, a dual positive-negative peak around the zone center at $Q=0$ implies an effective transferring of $LA$ phonons from $Q>0$ side to opposite $Q<0$ side, similar to electron transferring in the right panel of Fig.\,\ref{nfig17}, and gives rise to $Q^{\rm LA}_{\rm ph}(t)<0$ in the left panel. Since $\partial\omega^{\rm LA}_Q/\partial Q$ is always an odd function of $Q$, according to Eq.\,\eqref{drag}, $Q^{\rm LA}_{\rm ph}(t)$ remains zero if $\Delta N^{\rm LA}_Q(t)=\Delta N^{\rm LA}_{-Q}(t)$ holds, which is obviously contrary to observation in the left panel of Fig.\,\ref{nfig6}. For $LO$ phonons, on the other hand, 
only one narrow positive peak is seen and shifted slightly towards the left side of zone center at $Q=0$, as can be verified in the right panel of Fig.\,\ref{nfig7}. 
This ultimately gives rise to a much smaller but positive heat-current density $Q^{\rm LO}_{\rm ph}(t)$ for $LO$ phonons due to the fact that $\partial\omega^{\rm LO}_Q/\partial Q$ is approximate zero around the zone center.
\medskip

Moreover, a comparison is made in Fig.\,\ref{nfig20} for the time dependence of heat-current density ${\cal Q}^{\rm LA}_{\rm ph}(t)$ with various strengths for a DC field $E_{\rm DC}=15,\,100,\,500$ and $1500\,$V/m, where the appearance of kinks in their time dependence relates to starting an induced phonon absorption of non-equilibrium phonons by enhanced Doppler effect under $E_{\rm DC}=500\,$V/m and $1500\,$V/m in phonon-electron scattering on the negative side of zone center at $\mbox{\boldmath$Q$}=0$, as can be verified by left panels in Figs.\,\ref{nfig13} and \ref{nfig16}. Especially, the negative-value ${\cal Q}^{\rm LA}_{\rm ph}(t)$ is suppressed shortly for $E_{\rm DC}=1500\,$V/m and changes to a positive one after $t>20\,$ps. 
\medskip
 
For practical reason, the extracted phonon-drag thermoelectric powers ${\cal S}_{\rm ph}^\gamma(t)={\cal Q}_{\rm ph}^\gamma(t)/T_0J_{\rm e}(t)$ in unit of $\mu V/K$ are displayed in Fig.\,\ref{nfig21} for $LA$-phonon (left panel) and $LO$-phonon (right panel), respectively, which can be used for characterizing material electric and thermal properties as and thermoelectric performance. Here, the observed features for ${\cal S}^{\rm LA}_{\rm ph}(t)$ in the left panel and ${\cal S}^{\rm LO}_{\rm ph}(t)$ in the right panel follows the heat-current densities ${\cal Q}^{\rm LA}_{\rm ph}(t)$ and ${\cal Q}^{\rm LO}_{\rm ph}(t)$ in the left and right panels of Fig.\,\ref{nfig19}, respectively. The initial rise of ${\cal Q}^{\rm LA}_{\rm ph}(t)$ with time in the left panel is attributed to a flatten starting of $|v_d(t)|\gtrsim 0$ with time, as seen in the left panel of Fig.\,\ref{nfig8}, due to turning on a DC electric field $E_{\rm DC}(t)=\Delta{\cal E}\,\Theta^{(M)}_{\tau_0}(t)$ with a multi-step function $\Theta^{(M)}_{\tau_0}(t)$ given by Eq.\,\eqref{step}.
\medskip

\section{Summary and Remarks}
\label{sec-4}

In summary, we have developed for the first time a self-consistent quantum-kinetic theory in order to study in a non-perturbative way the non-equilibrium coupled phonon/electron transports, as well as mutual interactions between force-driven phonons and drifting electrons, within a quasi-one-dimensional electronic-lattice system. 
From this theory, we have found that after a DC electric field is applied to the system, initial thermal-equilibrium phonons are dragged into motion by drifting electrons 
along the opposite direction of a DC electric field with the help from a momentum-transferring process through electron-phonon scattering. 
Similarly, initial thermal-equilibrium electrons can also be dragged into motion by a force-driven phonon transport based on the same momentum-transferring process
through the phonon-electron scattering in the absence of a DC electric field.
\medskip
 
Interestingly, the Doppler effect resulting from fast-moving electrons in the system is able to play a significant role in modification of electron-phonon scattering by 
softened $LA$-phonon modes, leading to unusual absorption of non-equilibrium force-driven $LA$ phonons away from center of the Brillouin zone at $Q=0$, accompanied by 
a kink in heat-current density of $LA$ phonons as a function of elapsed time in the picosecond regime, including suppression of a negative heat-current density ${\cal Q}^{\rm LA}_{\rm ph}(t)$ and turning its sign after passing through a zero point. 
Meanwhile, the presence of a strong electron-phonon scattering greatly affects the acceleration of electrons by a strong DC-electric field, giving rise to a 
nonlinear transport of electrons in the system beyond well-known Ohm's law applied only to very-weak DC fields.
\medskip

From the application perspective, for a long time,  people have known that thermoelectric generators require a design based on the basic principle discussed in this paper and become all-solid-state devices which do not ask any fluids for fuel or cooling. These unique features make them non-orientation dependent, enabling use in zero-gravity or deep-sea environments. Meanwhile, its solid-state design allows for operation in severe environments. Furthermore, thermoelectric generators have no moving parts, and then, they produce a more reliable device which does not require maintenance for long period of time. Therefore, the durability and environmental stability have made thermoelectrics a favorite for deep space explorers among other applications. In particular, one of the major advantages of thermoelectric generators outside of such specific applications is that they can potentially be integrated into existing technologies to boost efficiency and reduce environmental impact by producing usable power from waste heat.
\medskip

\begin{acknowledgements}
D.H. would like to thank the support from Air Force Office of Scientific Research (AFOSR). D.H. thanks for helpful discussions with Prof. Richard Z. Zhang from Department of Mechanical Engineering of University of North Texas. 
The views expressed are those of the authors and do not reflect the official guidance or position of the United States Government, the Department of Defense or of the United States Air Force.
\end{acknowledgements}
	
\appendix	

\section{Explicit Form of a Mechanical Force Acting on 1D Lattice Phonons}
\label{sec-2.3}

The mechanical-force $Q{\cal P}^{(\tau_0)}_{\rm b}(t){\cal A}d/2$ introduced in Eq.\,\eqref{qdeph} results from the spatial gradient of a position-averaged propagating pressure field $p^{\rm ext}(x,t)$ acting on lattice vibrations within an atomic chain, which originates from a Fourier-transformed force density $f_{\rm ext}(x,t)=\partial p^{\rm ext}(x,t)/\partial x$ and is calculated by

\begin{eqnarray}
\nonumber
f_{\rm ext}(Q,t)&=&\frac{1}{{\cal L}}\,\int\limits_{-{\cal L}/2}^{+{\cal L}/2} dx\,\exp(iQx-i\omega_Lt)\,\left[\frac{\partial p^{\rm ext}(x,t)}{\partial x}\right]=
\int\limits_{-{\cal L}/2}^{+{\cal L}/2} \frac{dx}{{\cal L}}\,\exp(iQx-i\omega_Lt)\,\\
\nonumber
&\times&\sum\limits_{Q'}\,\left(-iQ'\right){\cal P}^{(\tau_0)}_{\rm b}(t)\,\exp(i\Omega_{\rm b}t-iQ'x)
={\cal P}^{(\tau_0)}_{\rm b}(t)\,\exp[i(\Omega_{\rm b}-\omega_L)t]\,\frac{2}{{\cal L}}\,\sum\limits_{Q'}\,\frac{-iQ'}{Q-Q'}\,\sin\left[\frac{(Q-Q'){\cal L}}{2}\right]\\
\nonumber
&=&-i\,{\cal P}^{(\tau_0)}_{\rm b}(t)\,\exp[i(\Omega_{\rm b}-\omega_L)t]\,\sum\limits_{Q'}\,Q'\,{\rm sinc}\left[\frac{(Q-Q'){\cal L}}{2}\right]\approx{\cal P}^{(\tau_0)}_{\rm b}(t)\,\exp[i(\Omega_{\rm b}-\omega_L)t-i\phi_0]\,\\
\label{mech-force}
&\times&\sum\limits_{Q'}\,Q'\,\delta_{Q',Q}
=Q{\cal P}^{(\tau_0)}_{\rm b}(t)\,\exp[i(\Omega_{\rm b}-\omega_L)t-i\phi_0]\ ,\ \ \ \ \ \ 
\end{eqnarray}
where $\phi_0=\pi/2$, $Q{\cal P}^{(\tau_0)}_{\rm b}(t)$ is recognized as the amplitude of a propagating pulse $f_{\rm ext}(Q,t)$. In Eq.\,\eqref{mech-force}, 
${\cal L}$ is the chain length, $\Omega_{\rm b}$ is the blast-wave frequency, $\omega_{\rm L}$ is the usual lattice-vibration frequency\,\cite{callaway}, ${\rm sinc}(x)\equiv\sin(x)/x$, $|Q-Q'|{\cal L}\rightarrow 0$ for very long sound wavelength compared to ${\cal L}$, and $|\Omega_{\rm b}-\omega_{\rm L}|\,t^*\ll 1$ due to very short interaction time $t^*$ of an incident blast wave. Moreover, ${\cal F}_{\rm b}(t)=Q{\cal P}^{(\tau_0)}_{\rm b}(t){\cal A}d/2$ in Eq.\,\eqref{qdeph} represents an external mechanical force associated with an incident blast wave acting on one atom within a two-atom complex unit cell along the chain direction, where ${\cal P}^{(\tau_0)}_{\rm b}(t)=2p_s\exp(-|t-\tau_0|/t^*)\,(1-|t-\tau_0|/t^*)$, 
$p_s$ is the peak amplitude for the pressure of a blast wave peaked at the moment of $t=\tau_0$, and $t^*$ represents the duration time. Furthermore,  
${\cal A}=\pi{\cal R}^2$ is the chain cross-sectional area, and $d=2a$ is the chain-lattice period. Especially, the special case with a spatially-homogeneous ${\cal P}^{(\tau_0)}_{\rm b}(t)\equiv p_0$ $(\mbox{{\em i.e.\/}},  Q\rightarrow 0)$ corresponds to the situation with an isotropic hydro-static pressure, which only affects the elastic moduli and modifies lattice vibrations or energy dispersion of phonons.
\medskip

Physically, we believe that the presence of such a blast wave $(\mbox{{\em i.e.\/}},  Q\neq 0)$ is able to induce a drifting motion of phonons in this system, leading to redistribution of phonons from Brillouin-zone center at $Q=0$ towards two Brillouin-zone boundaries at $Q=\pm\pi/d$ of an atomic chain under a constant temperature $T_0$.


\clearpage
\begin{references}
\bibitem{te-6}W. Li, J. Carrete, N. A. Katcho and N. Mingo, Computer Physics Communications {\bf 185}, 1747 (2014).

\bibitem{te-9}A. Ghukasyan and R. R. LaPierre, Nanotechn. {\bf 32} 042001 (2021).

\bibitem{te-7}H. Haug and S. W. Koch, {\em Quantum Theory of the Optical and Electronic Properties of Semiconductors} (World
Scientific Publishing Co. Pte. Ltd., 2009), 5th ed.

\bibitem{te-10}M. Lindberg and S. W. Koch, Phys. Rev. B {\bf 38}, 3342 (1988).

\bibitem{ziman}J. M. Ziman, {\em Principles of the Theory of Solids} (2nd ed. Cambridge University Press, Cambridge, England, 1972).

\bibitem{new-7}D. Huang, S. K. Lyo, K. J. Thomas, and M. Pepper, Phys. Rev. B {\bf 77}, 085320 (2008).

\bibitem{bookdhh}G. Gumbs and D. Huang, {\em Properties of Interacting Low-Dimensional Systems} (John Wiley \& Sons, 2011).

\bibitem{heat}H. S. Carslaw and J. C. Jaeger, {\em Conduction of Heat in Solids} (Oxford Science Publications (2nd ed., New York, The Clarendon Press, Oxford University Press, 1988). 

\bibitem{te-4}N. Jaziri, A. Boughamoura, J. M\"uller, B. Mezghani, F. Tounsi and M. Ismail, Energy Reports {\bf 6}, 264 (2020). 

\bibitem{te-5}M. A. Zoui, S. Bentouba, J. G. Stocholm and M. Bourouis, Energies {\bf 13}, 3036 (2020).

\bibitem{te-8}J. J. G. Moreno, J. Cao, M. Fronzi, and M. H. N. Assadi, Mater. Renew. Sustain. Energy {\bf 9}, 16 (2020).

\bibitem{r1}S. Ono, Phys. Rev. B {\bf 97}, 054310 (2018).

\bibitem{r21}S. Ono, Phys. Rev. B {\bf 96}, 024301 (2017).

\bibitem{r25}T. Feng and X. Ruan, {\bf 97}, 045202 (2018).

\bibitem{r26}L. Lindsay, A. Katre, A. Cepellotti and N. Mingo, J. Appl. Phys. {\bf 126}, 050902 (2019).

\bibitem{r27}A. J. H. McGaughey, A. Jain, H.-Y. Kim and B. Fu, J. Appl. Phys. {\bf 125}, 011101 (2019).

\bibitem{r3}S. K. Lyo and D. Huang, Phys. Rev. B {\bf 66}, 155307 (2002).

\bibitem{new-2}S. K. Lyo and D. Huang, J. Phys.: Condens. Matter {\bf 16}, 3379 (2004).

\bibitem{iop}E. Bringuier, Eur. J. Phys. {\bf 40}, 025103 (2019).

\bibitem{add-3}D. Huang, A. Iurov, H.-Y. Xu, Y.-C. Lai and G. Gumbs, Phys. Rev. B {\bf 99}, 245412 (2019).

\bibitem{add-4}F. Szmulowicz, H. J. Haugan, S. Elhamri, and G. J. Brown, Phys. Rev. B {\bf 84}, 155307 (2011).

\bibitem{add-31}H. Xie, J. Yan, X. Gu and H. Bao, J. Appl. Phys. {\bf 125}, 205104 (2019).

\bibitem{add-32}X. Gu and R. Yang, J. Appl. Phys. {\bf 117}, 025102 (2015).

\bibitem{add-33}J. Sjakste, K. Tanimura, G. Barbarino, L. Perfetti and N. Vast, J. Phys.: Condens. Matter {\bf 30}, 353001 (2018).

\bibitem{kohn}W. Kohn and J. M. Luttinger, Phys. Rev. {\bf 108}, 590 (1957).

\bibitem{lyo}S. K. Lyo and D. Huang, Phys. Rev. B {\bf 64}, 115320 (2001).

\bibitem{new-1}S. K. Lyo and D. Huang, Phys. Rev. B {\bf 68}, 115317 (2003).

\bibitem{new-3}S. K. Lyo and D. Huang, Phys. Rev. B {\bf 73}, 205336 (2006).

\bibitem{new-4}D. Huang, S. K. Lyo and G. Gumbs, Phys. Rev. B {\bf 79}, 155308 (2009).

\bibitem{add-1}J. R. Gulley and D. Huang, Optics Express {\bf 27}, 17154 (2019).

\bibitem{add-2}J. R. Gulley and D. Huang, Optics Express {\bf 30}, 9348 (2022).

\bibitem{add-30}X. Lu, D. Huang and J. R. Gulley, J. Appl. Phys. {\bf 131}, 073101 (2022).

\bibitem{surfscat}S. Tamura, Phys. Rev. B {\bf 30}, 610 (1984).

\bibitem{surfscat-2}S. Tamura, Phys. Rev. B {\bf 31}, 2574 (1985).

\bibitem{prl}P. Martin, Z. Aksamija, E. Pop, and U. Ravaioli, Phys. Rev. Lett. {\bf 102}, 125503 (2009).

\bibitem{callaway}Joseph Callaway, {\em Quantum Theory of the Solid State} (Academic Press, 1974).

\bibitem{r10}S.Tamara, Phys. Rev. B {\bf 31}, 2574 (1985).

\bibitem{expansion}A. V. Bragas, S. A. Maier, H. D. Boggiano, G. Grinblat, R. Bert\'e, L. S. Menezes, and E. Cort\'es, J. Opt. Soc. Am. B {\bf 40}, 1196 (2023).

\bibitem{ando}N. Nishiguchi, Y. Ando, and M. N. Wybourne, J. Phys.: Condens. Matter {\bf 9}, 5751 (1997).

\bibitem{umklapp}N. Mingo, Phys. Rev. B. {\bf 68} 113308 (2003).

\bibitem{r2}D. Huang, P. M. Alsing, T. Apostolova, and D. A. Cardimona, Phys. Rev. B {\bf 71}, 195205 (2005).

\bibitem{r4}D. Huang and G. Gumbs, Phys. Rev. B {\bf 80}, 033411 (2009).

\bibitem{voigt}Klaus Helbig, {\em Foundations of Anisotropy for Exploration Seismics} (Pergamon, 1994).

\bibitem{hooke}Henry Petroski, {\em Invention by Design: How Engineers Get from Thought to Thing} (Cambridge, Massachusetts: Harvard University Press, 1996).

\bibitem{piezo}A. Manbachi and R. S. C. Cobbold, Ultrasound. {\bf 19}, 187 (2011).

\bibitem{maxwell}John David Jackson, {\em Classical Electrodynamics} (3rd ed., John Wiley \& Sons, 1999).

\bibitem{te-1}T. J. Seebeck, Abhandlungen der K\"oniglichen Preu$\beta$ischen Akademie der Wissenschaften zu Berlin (Verlag von Wilhelm Engelman, 1895).

\bibitem{te-2}J. C. Peltier, Annales de Chimie et de Physique {\bf 56}, 371 (1834).

\bibitem{te-3}W. Thomson, Trans. R. Soc. Edinb. {\bf 3}, 91 (1851).
\end{references}
\end{document}